\documentclass[useAMS,usenatbib,twoside,a4paper,iop,numberedappendix,appendixfloats]{emulateapj} 
\slugcomment{{\sc Accepted for Publication in ApJ}}

\usepackage{graphicx,mathrsfs}
\usepackage{epstopdf}
\usepackage{appendix}
\usepackage{color}
\usepackage{longtable}
\usepackage{array}
 \usepackage{amsmath}

\newcommand{\Msol}{\mbox{$M_{\odot}$}}
\newcommand{\Lsol}{\mbox{$L_{\odot}$}}

\def\deg      {{\ifmmode^\circ\else$^\circ$\fi}}

\shorttitle{Mass-based Classification of Mergers and Clumpy Galaxies}
\shortauthors{Cibinel et al.} 

\begin{document}

\title{A Physical Approach to the Identification of High-$z$ Mergers:  Morphological classification in the Stellar Mass Domain.}
  
\author{{\footnotesize A. Cibinel\altaffilmark{1,2}, 
E. Le Floc'h\altaffilmark{2},
V. Perret\altaffilmark{3},
F. Bournaud\altaffilmark{2},
E. Daddi\altaffilmark{2},
M. Pannella\altaffilmark{4,2},
D. Elbaz\altaffilmark{2},
P. Amram\altaffilmark{5}
and P.-A. Duc\altaffilmark{2}}
}
 \affil{{\scriptsize $^1$Astronomy Centre, Department of Physics and Astronomy, University of Sussex, Brighton, BN1 9QH, UK, \texttt{A.Cibinel@sussex.ac.uk}}} 
\affil{{\scriptsize $^2$CEA Saclay, DSM/Irfu/Service d'Astrophysique, Orme des Merisiers, F-91191 Gif-sur-Yvette Cedex, France}}
\affil{{\scriptsize $^3$Center for Theoretical Astrophysics and Cosmology, Institute for Computational Science \& Physik Institut, University of Z\"{u}rich, 190 Winterthurestrasse, Z\"{u}rich 8057, Switzerland}}
 \affil{{\scriptsize $^4$Universit\"{a}ts-Sternwarte M\"{u}nchen, Scheinerstr. 1, D-81679 M\"{u}nchen}}
 \affil{{\scriptsize $^5$Laboratoire d'Astrophysique de Marseille, Observatoire Astronomique Marseille-Provence, Universit\'{e} de Provence \& CNRS, 2 Place Le Verrier, F-13248 Marseille Cedex 4, France}}

\begin{abstract}
\noindent At $z\gtrsim 1$, the distinction between merging and ``normal" star-forming galaxies based on single band morphology is often hampered by the presence of large clumps which result in a disturbed, merger-like appearance even in rotationally supported disks.
In this paper we discuss how a classification  based on canonical, non-parametric structural indices measured on resolved stellar mass maps, rather than on single-band images, reduces the misclassification of clumpy but not merging galaxies. 
We calibrate the mass-based selection of mergers using the MIRAGE hydrodynamical numerical simulations of isolated and merging galaxies  which span a stellar mass range of $10^{9.8}$--$10^{10.6}\Msol$ and merger ratios between 1:1--1:6.3. These simulations are processed to reproduce the typical depth and spatial resolution of observed Hubble Ultra Deep Field (HUDF) data.
We test our approach on a sample of real $z$\,$\simeq$\,2 galaxies with kinematic classification into disks or mergers and on $\sim$100 galaxies in the HUDF field with photometric/spectroscopic redshift between 1.5\,$\leqslant$\,$z$\,$\leqslant$\,3 and $M$\,$>$\,$10^{9.4} \Msol$. 
We find that a combination of the asymmetry $A_{\rm MASS}$ and $M_{\rm 20, MASS}$ indices measured on the stellar mass maps can efficiently 
identify real (major) mergers with $\lesssim$\,20\% contamination from clumpy disks in the merger sample. 
This mass-based classification cannot be reproduced in star-forming galaxies by $H-$band measurements alone,  which instead result in a contamination from clumpy galaxies that can be as high as 50\%. Moreover, we find that the mass-based classification always results in a lower contamination from clumpy galaxies than an $H-$band classification, regardless of the depth of the imaging used (e.g., CANDELS versus HUDF). 
\end{abstract}

\keywords{high-redshift--galaxies: interactions-- galaxies: irregular--galaxies: structure}

\section{Introduction}
\setcounter{footnote}{5}

Two decades of \emph{Hubble Space Telescope (HST)} optical/near-infrared (NIR) observations -- and more recently also ionized or molecular gas data -- have unveiled a complexity 
of  morphologies in high redshift star-forming galaxies. 
Many $z$\,$\gtrsim$1 galaxies do not display the disk or spheroidal morphology that is observed in the majority of local 
galaxies but are instead characterized by giant, star-forming clumps which dominate the light profiles and result in largely asymmetric 
appearances \citep{Cowie_et_al_1995,Papovich_et_al_2005,Elmegreen_et_al_2007,
Law_et_al_2007,Swinbank_et_al_2010,Forster_Schreiber_et_al_2011,Guo_et_al_2012,
Tacconi_et_al_2013}.
Although such irregular structure has often been associated with mergers events \citep[e.g.,] []{Conselice_et_al_2008,Lotz_et_al_2008a}, 
the use of NIR, integral-field (IFU) spectroscopy has enabled substantial progress in the classification of high redshift galaxies 
\citep[e.g.,] []{Erb_et_al_2004,Shapiro_et_al_2008,Epinat_et_al_2009,Epinat_et_al_2012} and detailed kinematic analysis have revealed
ordered rotational motion in a large number of these visually disturbed galaxies \citep{Genzel_et_al_2006,Bournaud_et_al_2008,
vanStarkenburg_et_al_2008,Forster_Schreiber_et_al_2009}.
The formation of giant clumps in ``normal" disk galaxies is thought to be the outcome of violent disk instability and fragmentation \citep{Noguchi_et_al_1999,Bournaud_et_al_2007,Agertz_et_al_2009,Dekel_et_al_2009,Ceverino_et_al_2010}, developing as a consequence of the high gas fractions that are typical for distant galaxies  \citep{Daddi_et_al_2010,Tacconi_et_al_2010,Magdis_et_al_2012,Saintonge_et_al_2013,Sargent_et_al_2014}.

In the lack of resolved kinematic data, the distinction between 
merging galaxies and clumpy disks remains however an observational challenge.
In this Paper we show how a quantitative classification performed on resolved stellar mass maps, rather than optical or NIR single-band images,
 can help  disentangling the population of truly merging galaxies from that of clumpy disks even without available IFU observations. 
Our approach is motivated by a number of reasons.

Even with the currently available second-generation instruments, obtaining reliable IFU kinematic measurements still requires major telescope 
time investments \citep[see e.g., the KMOS-3D campaign,][]{Wisnioski_et_al_2015}. 
Therefore, techniques which can provide robust proxies for 
the full kinematic informations  are necessary.

Canonical merger classification methods, however, 
suffer from limitations which can hamper the distinction between mergers and ``normal-but-clumpy" galaxies.
The selection of close pairs in the spatial and velocity domain 
\citep[][among the others]{Barton_et_al_2000,Carlberg_et_al_2000,Ellison_et_al_2008,
deRavel_et_al_2009,Kampczyk_et_al_2013,Pipino_et_al_2014}, for example, identifies by definition physically associated systems but it can be biased against very close galaxies 
(``fiber collision") and thus late interaction stages, namely the merger phases in which the distinction between mergers and clumpy disks becomes more hazy. Close kinematic pairs would typically have a large enough separation to allow the individual morphological classification of each galaxy.

Another common way of identifying mergers relies on the degree of irregularity in the light distribution either through  non parametric measures such as the 
concentration, asymmetry, clumpiness (CAS) and  Gini-$M_{20}$ indices \citep[][]{Abraham_et_al_2003,Conselice_2003,Lotz_et_al_2004,Law_et_al_2007,Scarlata_et_al_2007,Conselice_et_al_2008} or through other indicators of the presence of multiple components/tidal interactions 
\citep{Kampczyk_et_al_2007,Bridge_et_al_2010,Kartaltepe_et_al_2012,Lackner_et_al_2014}.
Using hydrodynamical numerical simulations, \citet{Lotz_et_al_2008b} have shown that combinations of non-parametric structural estimators (G-A-$M_{20}$) 
are sensitive to the coalescence phase and thus can be used also for evolved mergers.
As mentioned above, however, at $z>1$ the light profiles are dominated by giant star-forming clumps even in regular disk galaxies and  CAS-like classification schemes applied to single-band optical/NIR images typically fail in distinguishing mergers from non-interacting galaxies \citep[e.g.,] []{Huertas-Company_et_al_2014}.

While the giant $10^{9}\Msol$ clumps contribute to 20\%-50\% of the flux in resolved optical/UV or star formation rate (SFR) maps of $\gtrsim$\,$10^{10}\Msol$ star-forming galaxies, they show a lower contrast with respect to the underlying disks  on stellar mass maps, contributing to $\lesssim 10 \%$ of the total mass budget \citep{Forster_Schreiber_et_al_2011,Wuyts_et_al_2012,Guo_et_al_2012}.
This suggests a potentially lower contamination of falsely identified mergers if measuring the aforementioned non-parametric structural indicators directly on the stellar mass maps instead of the single band images: for clumpy galaxies we expect the stellar mass maps to display a regular, centrally concentrated profile, whereas for merging galaxies multiple components will be present, with no clear central mass concentration.

The exploitation of resolved mass (and also SFR or age) maps has become a common method of investigating the physical properties of low and high-redshift galaxies \citep[see e.g.][]{Welikala_et_al_2008,Zibetti_et_al_2009,Wijesinghe_et_al_2010,Guo_et_al_2012,Wuyts_et_al_2012,Lang_et_al_2014,Tacchella_et_al_2015} but have not been used so far for a quantitative merger classification.
The goal of this paper is to assess the performance of such a classification.
We mainly focus on the testing and calibrating  the proposed mass-based selection of mergers and defer a more in depth discussion on the properties of mass-identified mergers to a forthcoming paper (A. Cibinel et al. in preparation).

We use a set of mergers and isolated galaxies from the MIRAGE simulations \citep{Perret_et_al_2014} to quantitatively determine the efficiency and time scales probed when selecting mergers with mass-based structural parameters. 
Although other studies have investigated the morphology of interacting galaxies in numerical simulations \citep[e.g.,] []{Lotz_et_al_2008b}, 
the use of the MIRAGE sample enables us to make steps forward with respect to these previous analysis.
The MIRAGE simulations suite includes in fact key physical processes   
 that are paramount for the formation and regulation of the giant star-forming clumps 
 and thus naturally reproduces the complexity of clump-dominated morphologies of high redshift galaxies. We then apply our classification scheme on a fiducial sample of 1.5$\leqslant z\leqslant$3 galaxies in the Hubble Ultra Deep Field \citep[HUDF,][]{Beckwith_et_al_2006} and compare the results of our new method with the $H-$band classification and also with previously published kinematic analyses.

Specifically, the paper is organized as follows. We present in Section \ref{sec:Data} the observational data utilized in our analysis,
 the sample basic properties and the generation of the resolved mass maps for the HUDF galaxies.
Section \ref{sec:Sims} presents the MIRAGE simulations and post-processing of the simulation output.
We discuss our definition of a merging galaxy and some caveats regarding the simulations in Section \ref{sec:MergerCons}. 
We provide a summary of the structural measurements performed on both real and simulated galaxies in Section \ref{sec:strcMeasurements}.
In Section \ref{sec:SimulationInsight} we calibrate the classification performed on the mass maps
using the MIRAGE simulations and ancillary data with kinematic information. We then compare in Section \ref{sec:MassLightRealGal} the mass-based classification 
and the standard $H-$band classification for the real HUDF galaxies.
Finally, Section \ref{sec:Conclusions}  summarizes our findings and conclusions.
Considerations about signal-to-noise (S/N) effects and possible systematic biases are presented in Section \ref{sec:biases} and Appendix \ref{appendix:MassMapsReliability}.

All magnitudes are in the AB system and corrected for galactic absorption using the dust maps of \citet{Schlegel_et_al_1998} when necessary.
Throughout the paper we use interchangeably the notation ``$H$-band" to refer to the $HST$/Wide Field Camera 3 (WFC3) F160W filter.
If needed, we also use the abbreviations $b$, $z$ and $Y$ when referring to the $HST$/Advanced Camera for Surveys (ACS) F435W, $HST$/ACS F850LP and $HST$/WFC3 F105W filters, respectively.
Quoted masses assume a Chabrier initial mass function (IMF).
We finally note that we will sometimes use the notation ``clumpy disks" to refer those galaxies that have a clumpy appearance in the $H-$band/optical images but are not classified as mergers with our method.
We stress however that this is not meant to be a quantification of the intrinsic strength of the bulge component in these galaxies.


\begin{figure}
\begin{center}
\includegraphics[width=0.45\textwidth, angle=90]{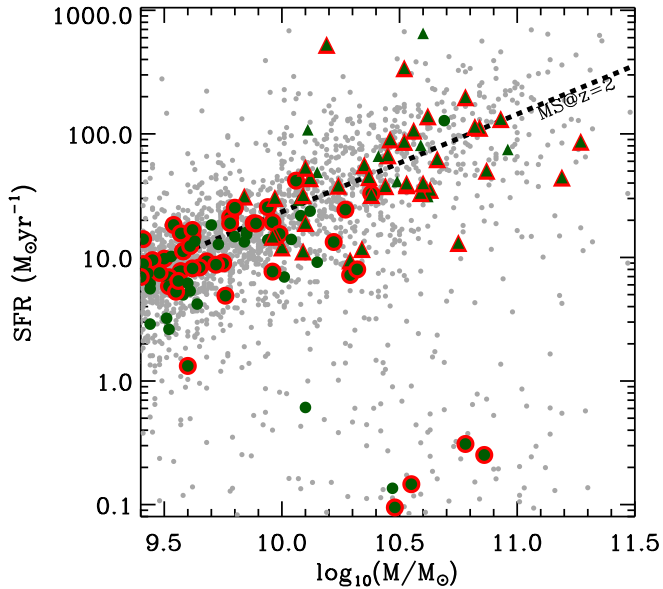}
\end{center}
\caption{\label{fig:f1} Location of the galaxy sample on the mass versus SFR plane. 
The dashed line shows the locus of the $z=2$ main sequence of star-forming galaxies \citep[based on literature compilation in][]{Sargent_et_al_2014} and the  small gray points are all galaxies in the \citet{Guo_et_al_2013} GOODS-S photometric catalog with a photometric redshift 1.5\,$\leqslant$\,$z$\,$\leqslant$\,3. The large green symbols correspond to the initial sample of 132 galaxies studied in the present work, prior to applying the size and magnitude selection of Section \ref{sec:biases}. The final sample of 87 galaxies on which reliable mass maps could be derived after applying this selection is shown with the red symbols.
Triangles indicate IR-based SFR (either from a combination of NUV+monochromatic 24$\mu$m flux density or from a fit to the mid-to-far IR SED).
Circles are instead galaxies without reliable IR photometry and for which the SFR is thus estimated from the dust-corrected UV luminosity.  Galaxies with very low SFR are identified as quiescent based on their $bzH$  or $zYJH$ colors and the upper limits on their IR fluxes. For these galaxies no dust extinction is applied when computing the UV-based SFR and are hence imposed to lie below the main sequence.
The sample considered here is mostly composed of normal (main sequence), star-forming galaxies.}
\end{figure}


\section{Observational Data}\label{sec:Data}

We apply the new classification scheme on a sample of galaxies in the HUDF field which,
 thanks to the availability of extremely deep observations but also medium and shallow coverage over the same area, 
 enables us to generate high accuracy mass maps as well as to assess the impact of S/N on the classification. 
We briefly summarize here the sample selection criteria and the relevant information for the public data sets utilized in this study, 
referring the reader to the original works for further details.

\subsection{Archival $HST$ Observations} \label{sec:archivaldata}

To build the pixel-by-pixel mass maps and perform the analysis described in Section \ref{sec:MassLightRealGal}, we exploited the data from several public campaigns covering the HUDF area with a multi-tiered approach.

For galaxies in the original HUDF ACS field  (3$^{\prime}\times$3$^{\prime}$) we generated two versions of the mass maps using either (1) the deep $HST$/ACS F435W, F606W, F775W and F850LP images from HUDF \citep{Beckwith_et_al_2006} combined with the $HST$/WFC3  F105W, F125W and F160W  images from the CANDELS-Deep survey \citep{Grogin_et_al_2011,Koekemoer_et_al_2011}, or (2) the F435W, F606W, F775W and F850LP observations from the GOODS survey \citep{Giavalisco_et_al_2004} plus the CANDELS-Deep NIR photometry. 

For the sub-set of galaxies in the central 2$^{\prime}\times$2$^{\prime}$ region of HUDF (corresponding to the WFC3 field of view), we also generated a third mass map making use of the extremely deep observations available at all wavelengths: the  
 optical HUDF images and the $HST$/WFC3 F105W, F125W, F140W and F160W coverage from the HUDF09 and HUDF12  surveys \citep{Bouwens_et_al_2009,Ellis_et_al_2013}. We specifically employed the mosaics provided by the HUDF12 team which are combined with  the HUDF09 datasets.
In all cases, CANDELS observations are also used for the F814W filter.


\begin{figure*}
\begin{center}
\includegraphics[width=0.8\textwidth]{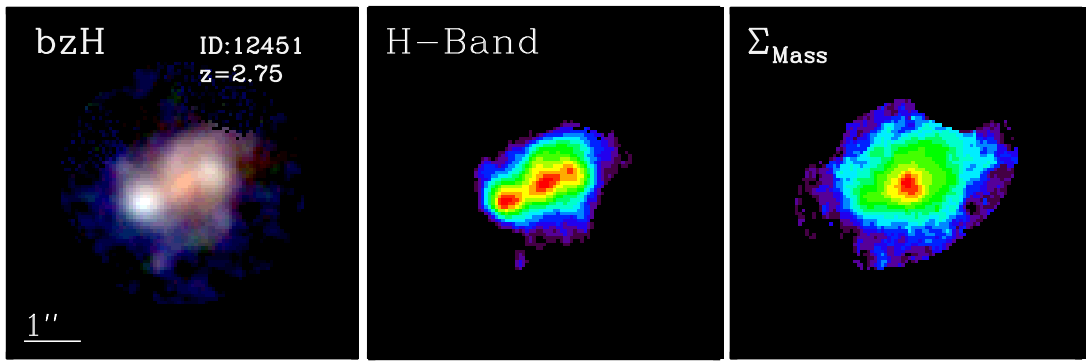}
\end{center}
\caption{\label{fig:f2} Illustration of the mass map obtained in Section \ref{sec:MassMapsReal} for a case example of a galaxy with a merger-like appearance in the optical/NIR images and a disk-like morphology in the mass map (ID 12451 in the \citet{Guo_et_al_2013} catalog, photometric redshift z=2.75). From left to right we present the $bzH$ (F435W, F850LP and F160W) composite image, the $H-$band stamp and the mass map.
In spite of displaying several, equally bright  clumps in the $H-$band or $bzH$ images, the stellar mass map reveals a single galaxy, with a mass profile centrally concentrated at the position of the red (bulge) component in the $bzH$ image.}
\end{figure*}

\subsection{Parent Catalog, Photo-z and Stellar Masses}\label{sec:GalIntegratedProp}

The identification of the galaxy sample and the calculation of photometric redshifts (photo-$z$) are based on the $H$-band selected, 
multi-wavelength catalog published by the CANDELS team in the GOODS-S field \citep[][and references therein]{Guo_et_al_2013}. 
The photometric data available in this compilation consists of imaging in 17 medium and broad-band filters ranging from the $U$-band to the IRAC $8\mu m$ channel.

We derived photo-$z$ and integrated galaxy stellar masses for all galaxies in the \citet{Guo_et_al_2013} sample as described in full details 
in \citet{Pannella_et_al_2014}.
Briefly,  photo-$z$ were estimated from the \citet{Guo_et_al_2013} photometric catalog using the public code EAZY \citep{Brammer_et_al_2008}
and a combination of the standard set of templates from \citet{Whitaker_et_al_2011}. These photo-$z$ reach an accuracy of $\Delta z =|z_{\rm phot}-z_{\rm zpec}|/(1+z_{\rm spec})=0.03$. 
Published spectroscopic redshifts (spec-$z$) are available for about 30\% of our final HUDF sample.\footnote{In particular, we referred to the ESO 
compilation of GOODS/CDF-S spectroscopy.
http://www.eso.org/sci/activities/
garching/projects/goods/MasterSpectroscopy.html
which collects spec-$z$ from several spectroscopic surveys which have covered (also) the HUDF field.}
 For galaxies with a \emph{secure} spectroscopic measurement we considered the spec-$z$ as the final redshift,
  whereas the photo-$z$ was preferred in those cases in which only a tentative (or no spec-$z$) is given. 
We then obtained galaxies stellar masses through fitting of the spectral energy distribution (SED) with \textsc{FAST} \citep{Kriek_et_al_2009},
keeping the redshift fixed and using a set of \citet{Bruzual_Charlot_2003} stellar population models with delayed exponentially declining 
star formation histories. Dust absorption with a maximum of $A_{V}=4$ was allowed in the fitting \citep{Calzetti_et_al_2000}.

\subsection{Sample Selection}\label{sec:sampleSelection}

The initial galaxy sample was extracted from the \citet{Guo_et_al_2013} catalog by selecting galaxies over the HUDF area having a spectroscopic or photometric      
redshift between 1.5\,$\leqslant$\,$z$\,$\leqslant$\,3 and an $H$-band magnitude brighter than $H$\,$\leqslant$\,26 mag.
The redshift selection allows us to probe the rest-frame FUV to optical for all galaxies; 
the luminosity cut is instead applied to ensure that a minimum signal-to-noise is reached in most pass-bands 
and thus a reliable photo-$z$ estimate can be obtained. The exact value of $H$\,$\leqslant$\,26 mag was chosen empirically by requiring that 80\% of all GOODS-S galaxies,
within the same redshift bin as the one here considered, have a S/N$>$3 in at least 10 of the pass-bands which are used in the photo-$z$ calculation. 
This magnitude threshold also ensures that the sample lies comfortably above the 50\% completeness limit of the parent \citet{Guo_et_al_2013} photometric catalog ($H$\,=\,26.6).

The $H-$band magnitude limit translates into a redshift dependent mass completeness threshold. 
For a clear selection of the sample, we hence apply a further cut in stellar mass to include only galaxies above the completeness value.
Given that we are mostly interested in studying the properties of clump-dominated, star-forming galaxies,  
we consider in the following the mass completeness limit for star-forming galaxies.
To derive this threshold we followed the procedure described in e.g., \citet{Pozzetti_et_al_2010}. For each galaxy we estimated the mass $M_{\rm lim}$ that it would have, keeping its mass-to-light ratio constant, if faded to the limiting magnitude  $H=$26. We then calculated, at each redshift, the mass below which lie 90\% of  $M_{\rm lim}$  in the 30\% faintest galaxies --  considered to be representative of the typical $M/L$ of a galaxy close to the magnitude limit. The final completeness limit is set by the highest redshift here considered ($z$=3),  corresponding to a value of $M>10^{9.4}\Msol$ for star-forming galaxies. The equivalent number for quiescent galaxies would be $M>10^{10.2}\Msol$.
 
After also rejecting galaxies which fall too close to the HUDF edges for reliable measurements,
 our initial sample includes 132 galaxies with 1.5\,$\leqslant$\,$z$\,$\leqslant$\,3,   H\,$\leqslant$\,26 and M\,$\geqslant10^{9.4}\Msol$. 
We derived structural parameters for all these galaxies and we provide them in Table \ref{tab:SampleProp}, but the sample is further restricted for our final analysis 
 as a result of the reliability assessment of the stellar mass maps that we present in Section \ref{sec:biases} and which is based
 on tests performed on this initial, larger sample. 
 
While deferring to a forthcoming paper the detailed analysis of the star formation properties of mass-selected mergers, we show in Figure \ref{fig:f1} the position of this initial sample of 132 galaxies on the mass versus SFR plane for illustration purposes.
Even above the mass completeness limit for quenched galaxies, the majority of the galaxies here considered lie on the locus of the so-called main sequence of star formation \citep[e.g.,] []{Brinchmann_et_al_2004,Daddi_et_al_2007,Elbaz_et_al_2007}.
The few galaxies with very low SFR (SFR\,$<$\,1\,$\Msol$\,yr$^{-1}$) have been identified as quiescent from a combination of their  $bzH$ colors or $YHVz$ colors  \citep[see][]{Daddi_et_al_2004b,Cameron_et_al_2011} and the constraints coming from their IR flux upper limits. These galaxies are by definition forced to lie below the main sequence by imposing no dust extinction in the calculation of the UV-based SFR. Our sample is thus representative of the typical population of $z\sim2$ star-forming galaxies.

  A summary of the properties for the full sample of 132 galaxies and the classification into mergers and non-interacting galaxies from the stellar mass map analysis is given in Table \ref{tab:SampleProp}.
 
\subsection{Pixel-by-pixel SED Fitting and Stellar Mass Maps} \label{sec:MassMapsReal}

As a first step for the generation of the resolved mass maps, we registered all the ACS and WFC3 tiles to the same resolution and pixel scale of the $H$-band which has the worst point-spread function (PSF) among the other available filters ($\sim$0.$^{\prime \prime}15$). 
To do so, we created an individual PSF for each filter by stacking several unsaturated stars in the HUDF field and computed the convolution kernels to match the PSF of the $H$-band\footnote{The $HST$ PSF varies slightly across the field of view and this effect could be taken into account by selecting for each object a nearby star instead of using a common PSF for the entire field. However, we estimate that the error introduced by using a single PSF is comparable with the uncertainty associated to noise effects when using individual stars.}. 
From the matched images, we then extracted postage stamps for each galaxy in our sample with a size equal to 3 times the $H$-band Kron radius and, as further justified in Section \ref{sec:mergerDefinition}, we cleaned  from the stamps any nearby companion galaxy with a known spec- or photo-$z$. 

To derive pixel-based stellar masses and the actual mass maps, we extracted pixel-by-pixel SEDs within an elliptical aperture equal to the galaxy $H-$band Kron semi-major axis and fitted them with stellar population models.
Given the relatively low flux in each individual pixel, some degree of smoothing or binning is required to ensure a minimum S/N on most of the filters and thus obtain reliable parameters from the pixel SED fitting.   
Several approaches have been used in the literature to deal with S/N homogenization problems \citep{Sanders_Fabian_2001,Cappellari_Copin_2003,Ebeling_et_al_2006,Wuyts_et_al_2012,Cibinel_et_al_2013b}.
We opted here for the publicly available code \textsc{Adaptsmooth}\footnote{http://www.arcetri.astro.it/$\sim$zibetti/Software/ADAPTSMOOTH.html} developed in \citet{Zibetti_2009} and \citet{Zibetti_et_al_2009}.
Whenever the S/N falls below a given threshold, this algorithm performs an adaptive smoothing of the images by replacing the original pixel values with an average of the galaxy flux over larger and larger circular areas as the S/N decreases. \textsc{Adaptsmooth} features two useful options: (a) the smoothing of several images on the same scale lengths -- necessary to derive self-consistent SEDs -- can be easily performed and (b) the pixel identity is maintained, as opposed to binning schemes in which neighboring pixels are assigned  a common value, effectively grouping them together into a final ``macro pixel".
\begin{figure}

\begin{center}
\includegraphics[width=0.5\textwidth]{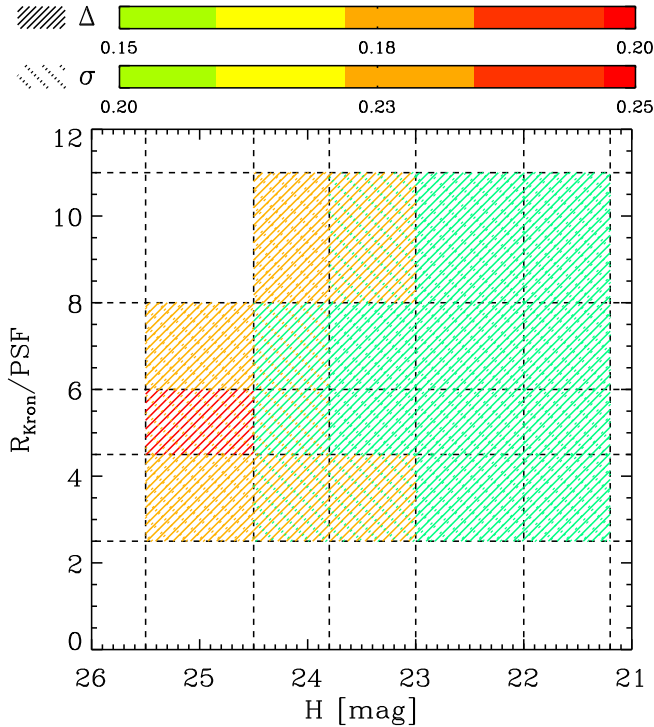}
\end{center}
\caption{\label{fig:f3} 
Results of the tests performed on the toy-model mass profiles to assess the ability of reconstructing the stellar mass distribution.
For all models falling in any region of the the size (Kron radius normalized to PSF) versus $H-$band magnitude plane marked by the horizontal and vertical dashed lines, we calculated the median absolute residual  ($\Delta$) and the dispersion ($\sigma$) between input and reconstructed mass maps.
The region boundaries are selected such to have at least 5 models in each bin.
Areas where $\Delta$ (solid, 45 degrees CW shading) or $\sigma$  (dashed, 45 degrees CCW shading) are low are colored in green, whereas a red color indicates systematic shifts and large scatter around the input models, i.e., a high uncertainty in the derivation of the mass maps. The colorbars at the top of the figure provide the mapping between colors and the absolute values of $\Delta$ and $\sigma$, in dex.
The figure refers to models created to reproduce the typical depth of the optical HUDF and CANDELS-Deep NIR data, which is in between the three combinations of photometry here explored (see Section \ref{sec:archivaldata}).}
\end{figure}

The choice of the reference band(s) defining the smoothing kernels applied by \textsc{Adaptsmooth} is a trade-off between attaining the best S/N 
and maintaining spatial resolution: using the band with the lowest S/N will obviously degrades the resolution, while referring to the one with the highest S/N will likely result in noisy SEDs. Furthermore, also in the light of forthcoming analyses on the comparison between the resolved mass and SFR distribution, we are interested in detecting features such as giant star-forming clumps which may be intrinsically bright in one band but have a smaller flux contrast at longer wavelengths. 
A too broad smoothing on the red band may completely erase these structures in the blue filters.
After testing single or multiple bands smoothing, we found the optimal configuration by running \textsc{Adaptsmooth} on stacked images of all ACS and WFC3 stamps: this ensures that the smoothing is applied on those pixels where the majority of the bands reach a low S/N while preserving the structural variations in the different filters. 
Specifically, we run \textsc{Adaptsmooth} by requiring a minimum S/N\,=\,5 on the stacked images and halting the adaptive smoothing when the averaging area reaches the maximum radius of 5 pixels.
The smoothing pattern thus obtained was then applied to all available bands giving a median S/N\,$\sim$\,5 also on the individual pixels for most filters.
We note that we have tested that our results are not substantially affected by a different choice of the smoothing kernel (e.g., by applying the smoothing on the $H-$band only).

We fitted the adaptively smoothed pixel SEDs with \textsc{LePhare} \citep{Arnouts_et_al_1999,Ilbert_et_al_2006}  using the \citet{Bruzual_Charlot_2003} synthetic spectral library with a \citet{Chabrier_2003} IMF and a delayed exponential star 
formation history, $\psi \propto (t/\tau^2) \exp(-t/\tau)$. The characteristic time scale $\tau$ was let vary between 0.01 and 10\,Gyr in
22 steps and template ages were chosen between 100\,Myr and the age of the Universe at the given redshift. 
We allowed three metallicity values in the fitting ($Z$\,=\,0.2\,$Z_{\odot}$, $Z$\,=\,0.4\,$Z_{\odot}$ and $Z$\,=\,$Z_{\odot}$) and furthermore applied internal dust extinction by assuming a Calzetti law and $E(B-V)$ ranging between 0 and 0.9\,mag.  
We defined our fiducial pixel mass estimate as the median mass from the full probability distribution function from all templates, but our results would remain unchanged if we had used the mass from the best-fit template (i.e., minimum $\chi^2$) instead.
As a validation of the derived mass maps, we verified the consistency between the sum of the pixel-based masses and the integrated galaxy mass in Appendix \ref{sec:Resolved_vs_Integrated}. We find an agreement at the level of $\lesssim$0.1 dex between the two estimates.

An example of the derived mass maps is given in Figure \ref{fig:f2}, where we compare the $bzH$ composite image, 
the $H-$band image and the mass map for a galaxy in our sample (ID 12451). 
We have intentionally chosen a galaxy which displays a different structure in the $H-$band than in the mass map to clearly illustrate how at z\,$>$\,1 
$H-$band light and mass are not equivalent tracers of morphology. This was already pointed out by \citet{Wuyts_et_al_2012} (see for example their Figure 2 which has two galaxies overlapping with our sample in Figure \ref{fig:f11}) and we will further demonstrate it in Section \ref{sec:MassLightRealGal}.

\subsection{Systematic Uncertainties in the Reconstructed Mass Maps}\label{sec:biases}

Numerous studied have shown that the ability of recovering intrinsic galaxy properties from observed flux distributions depends 
strongly on the image S/N, resolution and even on the intrinsic properties of the galaxies themselves 
\citep{Disney_1976,Schweizer_1979,Franx_et_al_1989,Impey_Bothun_1997,Trujillo_et_al_2001,Graham_et_al_2005, Cameron_Driver_2007,Bailin_Harris_2008,Graham_Worley_2008,Maller_et_al_2009,Carollo_et_al_2013a,Cibinel_et_al_2013a}.
Our mass-based measurements will also be affected by similar limitations.
For structural measurements performed on optical images, tests on artificial galaxies have demonstrated that
 it is possible to derive correction schemes that can largely account for the systematic biases in the estimates of galaxy structure -- e.g., galaxy radii and concentrations -- for both local (see \citealt{Cibinel_et_al_2013a} for an application to $z\sim0$ galaxies from the ZENS sample in  \citealt{Carollo_et_al_2013b}) and high redshift galaxies (see \citealt{Carollo_et_al_2013a} for an application to the COSMOS survey, \citealt{Scoville_et_al_2007}).

We do not attempt here a derivation of similar corrections in the mass domain as this would require large suites of
 artificial mass maps and hence significant assumptions on the mass (and dust) distribution in merging and non-merging galaxies. 
We can nonetheless perform some tests which enable us to define the ``boundaries of applicability" of our method, 
i.e. the regimes where we can obtain reliable measurements.

It is clear that for unresolved objects no meaningful mass reconstruction can be performed nor the individual galaxies participating 
in the merger can be identified. Likewise, the reconstruction of the mass distribution becomes more and more difficult as the flux in the pixels reaches the surface brightness limit of the observations.
To derive a global magnitude and size limit below which we cannot reliably derive mass maps, we tested 
our SED-fitting and mass reconstruction technique on a set of artificial galaxies with known mass distribution drawn from our initial sample of 1.5\,$\leqslant$\,$z$\,$\leqslant$\,3. We describe these models in detail in Appendix \ref{appendix:ResolutionLimit}.
For each of the toy galaxies, we compared the mass profile reconstructed following the procedure in Section \ref{sec:MassMapsReal} with the input model and calculated the median of the residuals as well as the typical dispersion around the input model.

The results of the comparison are presented in Figure \ref{fig:f3} where we color code each region of the size versus magnitude plane
according to the mean absolute residual value and the dispersion of all models falling in that specific area: a red/orange color indicates high residuals/scatter in the reconstructed maps, green corresponds to low residuals/scatter.  
The Figure illustrates how the ability of measuring reliable mass maps degrades as galaxies approach the resolution limit or reach low surface brightnesses: 
at sizes $r_{\rm Kron} \lesssim 5 \times$PSF (roughly 15 pixels) and magnitudes $H>24.5$ systematic shifts and/or large deviations from the input model affect 
the measured mass maps.\footnote{The inferred magnitude limit refers to mass maps generated from artificial images matched to the HUDF\,+\,CANDELS-Deep data, 
as described in Appendix \ref{appendix:ResolutionLimit}. For observations at a different depth this limit will scale accordingly.} 

We thus used the thresholds $H$\,$\leqslant$\,24.5 and $r_{\rm Kron}$\,$>$\,5$\times$PSF to select galaxies with reliable mass maps, reducing the sample of $1.5\leqslant z\leqslant 3$ galaxies  to 89 objects. 
Two of these galaxies, ID 13508 and ID 11800, are strongly contaminated by a bright neighbor and have for this reason less robust mass maps. These cases are flagged in Table \ref{tab:SampleProp}  and excluded from the sample used for the analysis in Section \ref{sec:MassLightRealGal}.
We show this final sample of 87 galaxies with red points in Figure \ref{fig:f1}.
Note that the $H$\,$\leqslant$\,24.5 selection is incidentally the same cut that has be applied for a reliable visual morphological classifications on the CANDELS fields \citep{Kartaltepe_et_al_2014}.


 \begin{figure*}[htbp]
\begin{center}
\includegraphics[width=0.47\textwidth,angle=90]{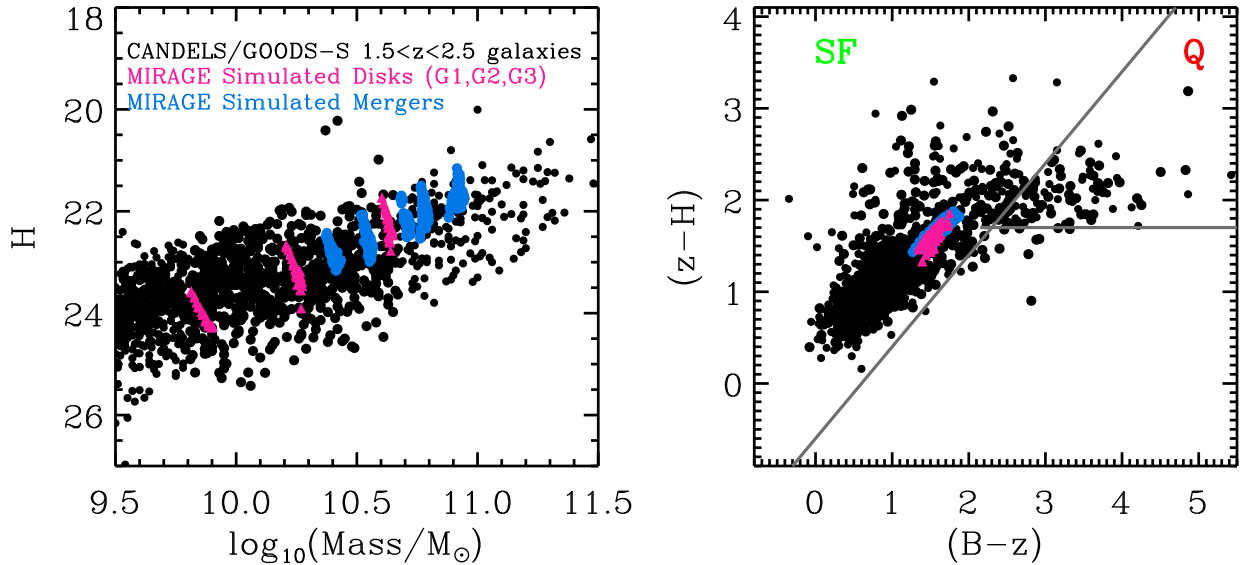}
\end{center}
\caption{\label{fig:f4} \emph{Left:} observed $H-$band magnitude versus mass relation for real galaxies and the simulated MIRAGE isolated disks. With magenta triangles we show  the evolution in mass and luminosity of the isolated disks in the MIRAGE simulation over the  600\,Myr here considered. Blue circles are the MIRAGE simulated mergers.
The black points correspond to real CANDELS/GOODS-S galaxies with 1.5\,$\leqslant$\,$z$\,$\leqslant$\,3. \emph{Right:} $bzH$ (F435W-F850LP and F850LP-F160W) colors for the simulated disks and observed galaxies. The horizontal and slanted lines divide the bzH plane in the locus of z\,$\sim$\,2 star-forming (SF) and quiescent galaxies (Q) following a similar approach as the BzK selection \citep{Daddi_et_al_2004b}. Symbols are the same as in the left panel. The MIRAGE simulations are representative of typical $z\sim 2$, star-forming galaxies. }
\end{figure*}


\section{The MIRAGE Simulated Galaxies}\label{sec:Sims}

\subsection{Description of the Simulations}

The details on the technical aspects of the MIRAGE simulations are presented in \citet{Perret_et_al_2014} and further discussions on the physics implemented in these simulations 
 can also be found  in \citet{Renaud_et_al_2013} and \citet{Bournaud_et_al_2014}. 
Briefly, three closed-box disk models with a bulge-to-total fraction of 8\% in mass and stellar masses of $10^{9.8}\Msol$ (simulation G3 in \citealt{Perret_et_al_2014}),  $10^{10.2}\Msol$  (G2) and $10^{10.6}\Msol$ (G1)  were generated using an adaptive mesh refinement technique with the RAMSES code \citep{Teyssier_2002}. These simulations reach a resolution of 7.3pc at the highest level of refinement and have initial stellar mass particles of 1.2$\times 10^{4}\Msol$ and 1.7$\times 10^{4}\Msol$ in the bulge and disk components, respectively. 
The disk galaxies were evolved in isolation or merged with each other, effectively probing merger ratios of 1:1, 1:2.5 and 1:6.3. In order to construct a representative sample of galaxy mergers, four different orbital parameters were explored for each merger ratio combination, resulting in a total of 20 mergers simulations (plus the 3 isolated disk models).

For each simulation configuration we utilize in the following a set of 16 snapshots separated by 40\,Myr each, covering an epoch from 200\,Myr to 800\,Myr from the simulations initial conditions; for the merger models this corresponds to follow the pre- and post-coalescence  phases for roughly 300\,Myr each (the coalescence time is visually determined in \citealt{Perret_et_al_2014}). 

The MIRAGE simulations feature several aspects which are paramount for a correct description of the ISM physics and thus for a robust comparison with the real data.  First, the high level of grid refinement enables us to resolve Jeans-unstable regions within the disks, consequently we can not only model the disk fragmentation and the formation of large clumps but also properly describe outflows and heating within the giant clumps themselves.

Second, a physically motivated feedback model is implemented in the simulations by coupling standard supernovae feedback \citep{Dubois_Teyssier_2008}  with the novel recipe for photoionization and radiation pressure feedback from OB-stars developed in \citet{Renaud_et_al_2013}. This feedback model reproduces the typical outflows, star formation rates and the stellar population ages ($\lesssim$200\,Myr) observed in real clumps \citep{Bournaud_et_al_2014}.

Third, the simulated disks have an initial total gas fraction of $f_g=65\%$ which is well representative of the observed high molecular fractions in typical  $z>1$, star-forming galaxies \citep[e.g.,] []{Daddi_et_al_2010,Tacconi_et_al_2010} and is essential for the onset of gravitational instability and the generation of the giant clumps.
 
Finally, the MIRAGE simulations were originally designed as a ``numerical counterpart" for the MASSIV galaxy sample \citep{Contini_et_al_2012}, 
and for this reason they are tailored to high-$z$  galaxies in terms of global properties (e.g., their sizes and SFR, see Figures 6 and 9 of \citealt{Perret_et_al_2014}). 


\begin{figure*}[tbp]
\begin{center}
\begin{tabular}{c}
\includegraphics[width=0.6\textwidth,angle=90]{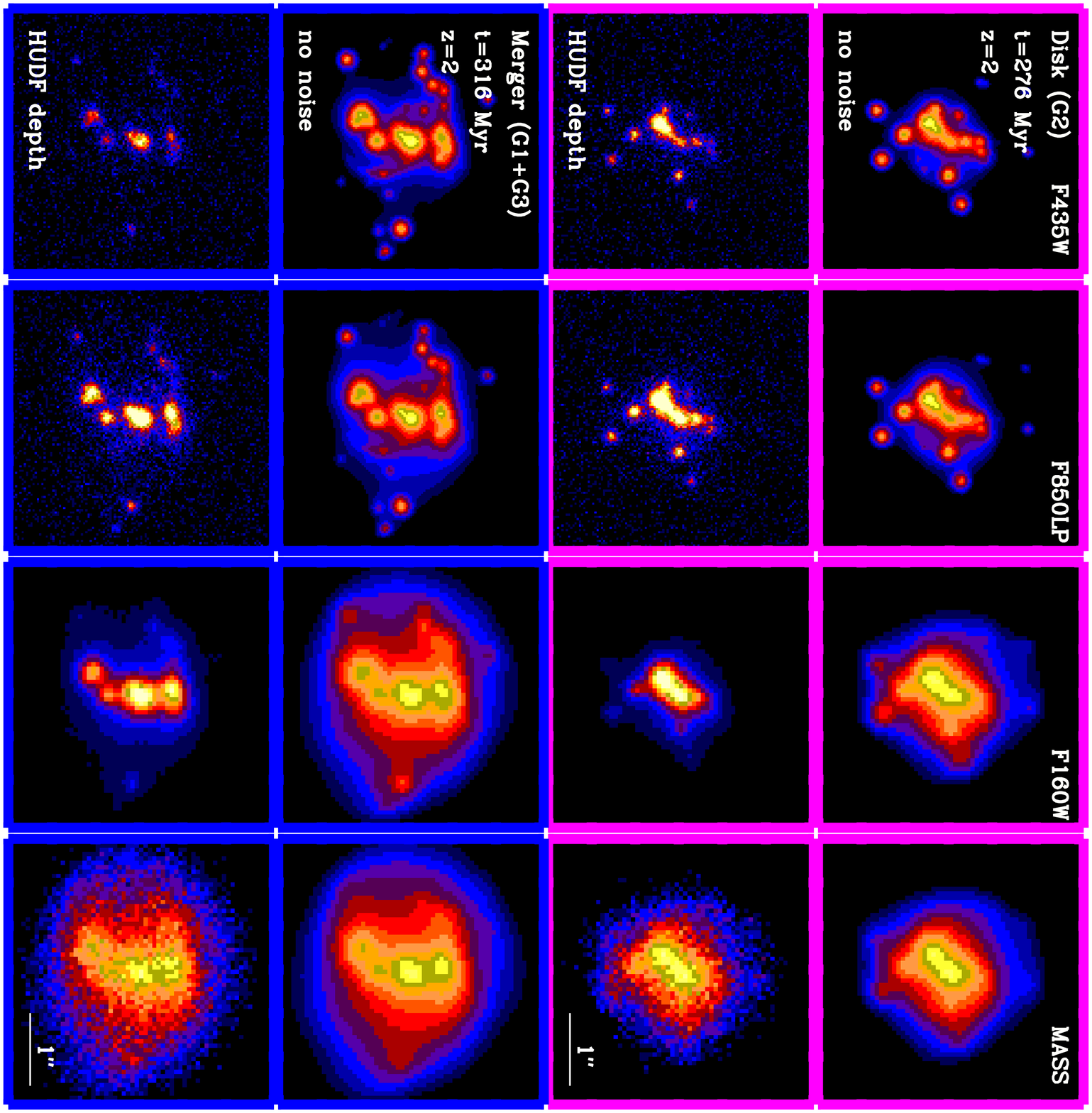}\\
\includegraphics[width=0.16\textwidth,angle=90]{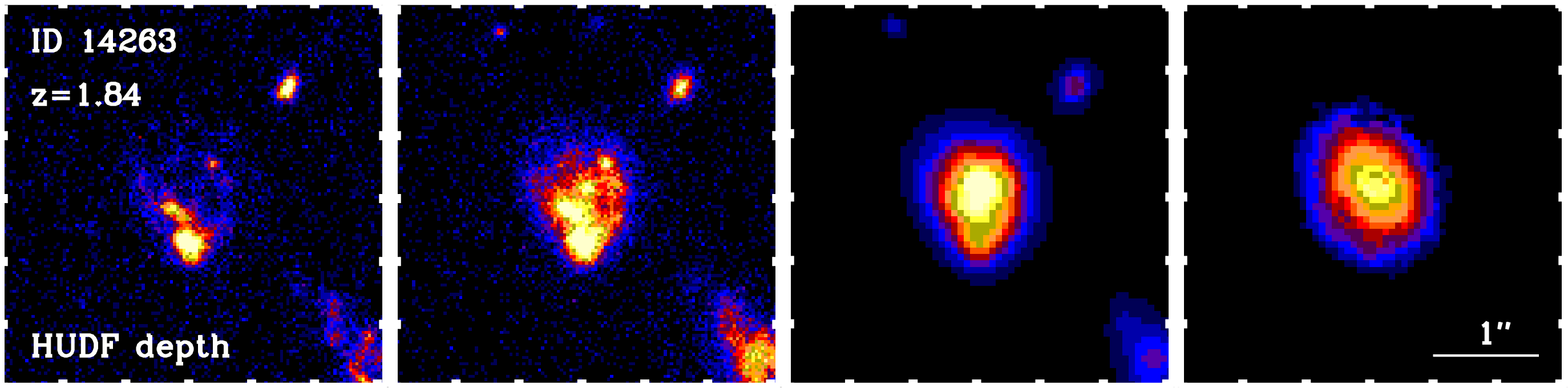}
\end{tabular}
\end{center}
\caption{\label{fig:f5} 
\emph{Upper panels -- magenta frame:} observed-frame mock flux maps and stellar mass density map for one of the isolated disk in the MIRAGE simulations (face-on projection for the isolated disk  `G2') at a time t=276Myr from initial conditions. 
From left to right we show the flux maps in the F435W, F850LP, F160W  filters and the stellar mass map, respectively. 
All images are redshifted and rebinned to the $HST$ pixel scale for an average redshift $z=2$ and convolved to the typical resolution of each band.
In the top row are the original simulation snapshots and the second row presents the same maps once degraded to the typical S/N of the HUDF/HUDF12 observations. 
For the flux maps the color coding shows positive deviations with respect to the mean sky r.m.s. (with the color map ranging from 0 to $20\sigma$).
\emph{Middle panels -- blue frame:} as above but for a simulated merger (minor merger involving the G1 and G3 disks with orbital parameters $\theta_{1}$\,=\,90$^{\deg}$,  $\theta_{2}$\,=\,90$^{\deg}$ and  $\kappa$\,=\,90$^{\deg}$, see \citealt{Perret_et_al_2014}). 
 \emph{Lower panels}: the F435W, F850LP,  F160W images and the derived stellar mass map for a real $z\simeq1.8$ galaxy in the HUDF field are given for comparison with the simulations.
 The MIRAGE simulations well reproduce the wavelength-dependent, complex morphology of real z\,$\sim$\,2 galaxies.}
\end{figure*} 

\subsection{Post-processing of the MIRAGE Output}\label{sec:SimulationPostProcess}

To reproduce the data available for the HUDF galaxies also for the simulated MIRAGE sample, we extracted 
from each simulation snapshot a stellar mass density map and three stellar flux maps, in the $HST$/ACS F435W, F850LP and  $HST$/WFC3  F160W  filters. We used both face-on and edge-on line of sights computed from the angular momentum of the most massive disk, at an initial resolution of 50\,pc. 

The mass maps were simply obtained by projecting the distribution of stellar mass particles in the simulations. 
We instead derived the mock observations by assigning a Starburst99 \citep{Leitherer_et_al_1999} instantaneous burst model with a Salpeter IMF and an effective metallicity $Z=0.004$ to each star particle on the basis of its formation age.
This metallicity corresponds to the integrated value that is expected for a disk galaxy with a central metallicity in agreement with observations of $z\sim2$ galaxies by \citet{Erb_et_al_2006}.
We note here that \emph{no} dust extinction was applied to the simulated fluxes.

To produce flux and mass maps that can be directly compared with observations we: 
\begin{enumerate}
\item Generated observed-frame F435W, F850LP and F160W images by redshifting the Starburst99 SEDs and applying cosmological dimming. We assumed a reference redshift $z$\,=\,2. 
\item Matched the artificial F435W \& F850LP images to the $HST$/ACS resolution ($\sim$\,0$^{\prime \prime}$08)   and the $H-$band and mass maps to the $HST$/WFC3 resolution ($\sim$\,0.$^{\prime \prime}$15).  To do so, we convolved the artificial maps with the PSFs kernels constructed from stars in the HUDF field (see Section \ref{sec:MassMapsReal}).  
\item We also pixelized the simulation images to the ACS and  WFC3 pixel scales (0.$^{\prime \prime}$03  and 0.$^{\prime \prime}$06, respectively), again assuming an average $z$=2. 
\item Finally, we added noise to the flux and mass maps mimicking the typical imaging depth and the uncertainty in the reconstruction of the mass distribution for the real images. 
For the mock $HST$ images, this was done by 
adding poissonian noise (on the maps converted in electron units) and by inserting the simulations into empty sky regions extracted from the real observations.
We used sky regions from the HUDF12 area, when considering the $H-$band mock images, and instead sky areas from the HUDF tiles for the F435W and F850LP filters.
To account for uncertainties on the mass maps, we added a mass-dependent error calculated from the observed mass versus error relation for individual pixels in the sample of real galaxies.
We matched the simulations to the deepest available photometry only, i.e., HUDF\,+\,HUDF12, as we test the effects of different imaging depths on the data itself in Section \ref{sec:MassLightRealGal}.
\end{enumerate}

We then ran \textsc{SExtractor} on the noisy, stellar mock images with similar configuration parameters as those employed for the construction of the GOODS ACS multi-band catalogs in the case of the artificial F435W and F850LP images\footnote{http://archive.stsci.edu/pub/hlsp/goods/catalog\_r2/} and using instead  the ``cold-mode" parameter settings of the \citet{Guo_et_al_2013} CANDELS multiwavelength catalog in the case of the mock $H-$band images.
The \textsc{SExtractor} outputs were used for detecting/deblending the merging galaxies and to define their photometric properties, e.g., sizes or fluxes.

The integrated luminosities and colors of the MIRAGE galaxies resulting from such post-processing are presented in  Figure \ref{fig:f4}, where we compare the simulations with real 1.5\,$\leqslant$\,z\,$\leqslant$\,3 galaxies in the CANDELS/GOODS-S field.
There is a good agreement between the MIRAGE disks or mergers and the real sample of galaxies. At high masses, the MIRAGE models tend to be slightly brighter than observed galaxies, most likely as a consequence of applying no dust extinction to the simulations. Although limited by the intrinsic colors of the Starburst99 templates used to construct the flux maps, the simulated galaxies have also $(b-z)$ and  $(z-H)$ colors that are typical of $z\sim2$ star-forming galaxies.

Examples of the mass and flux maps before and after noise degrading are presented in Figure \ref{fig:f5}, for an isolated disk and a merger simulation. In the same Figure, we also show for comparison a real z\,$\sim$\,2 galaxy extracted from the HUDF field.
The simulated galaxies well reproduce the morphological/structural properties of real observed galaxies and their variation with wavelength of observation.


\section{Caveats and General Comments}\label{sec:MergerCons}

Before proceeding with the analysis, we clarify here some definitions and address caveats which are relevant for the following discussion.

\subsection{Definition of Merger in Our study}\label{sec:mergerDefinition}

As specified in the Introduction, our intent is to optimize the classification for galaxies that are close to the coalescence phase, with less interest to early merger stages which can be identified with other techniques 
(e.g., kinematic pair selection). For this reason, HUDF galaxies in pairs with a sufficient separation to allow distinct identification will be considered as individual objects.

Practically, this means that we perform the structural measurements and classification 
separately for any galaxy appearing as a single entry in the parent \citet{Guo_et_al_2013} catalog, even if the galaxy is in a close pair with another in the catalog.
 Moreover,  this also implies that any galaxy displaying multiple clumps/components which do not appear 
 as individual sources in the $H-$band catalog will be considered as a ``multi-component" single system,
  being it either a clumpy galaxy or a merger (but see next section).

The ability of separating two galaxies in real observations depends on several aspects, such as the intrinsic surface brightness and brightness contrast of the merging galaxies, the signal-to-noise of the images, the size of the galaxies, etc. Hence it is not straightforward to convert the \textsc{SExtractor} deblending threshold into a unique minimum distance between the two galaxies. 
On average, however, the closest separation at which the HUDF 1.5\,$\leqslant$\,$z$\,$\leqslant$\,3 galaxies in the \citet{Guo_et_al_2013} catalog are still deblended is of $\sim$\,10 kpc; 
our merger sample will hence include galaxies at smaller separations.

 Following the same philosophy, we extracted from the simulations a sample of close-to-coalescence mergers  which we refer to as ``pre-coalescence" mergers/snapshots.
In this sample are excluded both pre-merger (i.e. well separated) galaxies\footnote{
In order to follow for the simulation an approach as close as possible to that applied on the real galaxy sample, we relied on the \textsc{SExtractor} output to determine when the two simulated galaxies can be clearly deblended (see Section \ref{sec:SimulationPostProcess} for details of the \textsc{SExtractor} run on the simulations).
 Consistently with the observations, this translates into a maximum separation of about 10 kpc also for the simulated galaxies.}
and post-merger remnants (i.e., snapshots extracted at a time after the coalescence of the two nuclei). In fact, galaxies in such phases  have either to merge yet or have already completed the merger phase and have settle in their final, unperturbed state with no clear structural signatures of the past merger event. In the simulation, and possibly in the observations too, such galaxies are indistinguishable from the isolated galaxies.

Among these pre-coalescence mergers, some reach an almost unperturbed appearance already a $\sim$20 Myr before the coalescence time. Although these simulated galaxies are still nominally undergoing a merger, they also occupy a locus of structural parameters which is overlapping with that of ``normal" galaxies.
To obtain an as pure as possible merger training sample, we excluded these simulations from those employed to calibrate the classification in Section \ref{sec:TestonSims1}. We refer to this clean set of mergers as the ``training-sample".

With such a selection the resulting input training/pre-coalescence sample of simulated  galaxies is roughly equally divided among mergers and isolated disks (82/110 and 96 galaxies respectively) and 70\% of these mergers have a ratio $\leq$2.5, i.e. are major mergers.

\subsection{Keeping Projection Effects under Control}\label{sec:ProjectionEffects}

Obviously, for real data the presence of multiple components which are not deblended into individual sources does not necessarily imply that we are looking at a single clumpy galaxy or that the components are physically participating to a merger.
Especially for those galaxies displaying extreme color gradients, one may worry about chance projections. 
As mentioned in Section \ref{sec:sampleSelection} spectroscopic redshifts are available for only a minority of the sample, hence we cannot rely on the spectral information to confirm associations.

Another way to test this possibility is to compare the source identification/photo-$z$ based on the $H-$band extraction with those obtained at shorter wavelengths to see whether the individual clumps could in principle be identified as single objects located at a different redshift and had simply passed undetected in the $H-$band.
Of course, the advantage of an $H-$band extraction is precisely that -- by probing the rest-frame optical for $z\sim2$ galaxies -- it limits the ``over-deblending" of galaxies which may occur at the rest-frame UV. For this reason we keep the $H-$band source identification as our reference. However, we can use the short wavelength independent measurements to flag  ``bona-fide" multi component galaxies. 

For this purpose we used the publicly available photo-$z$ for the HUDF galaxies from the work of  \citet{Coe_et_al_2006} which are based on an hybrid $i$-band+$BVizJH$ detection.
Specifically, for each galaxy in our sample which displayed several clumps or plausible companions we checked in the \citet{Coe_et_al_2006} catalog whether these would have been identified as separate object and their photometric redshift. Whenever also
the $i$-band extraction was consistent with only a single multi-clump galaxy or the phot-z of the individually extracted clumps/companion was consistent within errors with the redshift of the given galaxy, we considered it as a true multi-component system at best of our knowledge.
Otherwise we flagged this galaxy to remind us that a chance projection cannot be excluded in this case. We highlight such galaxies with a red exclamation mark in Table \ref{tab:SampleProp}. 

\subsection{Limitations of the Simulations}\label{sec:SimsCaveats}

The MIRAGE simulations offer an ideal sample for testing our approach as we argued in Section \ref{sec:Sims}.
Nonetheless, there are also a number of limitations of which we must be aware when comparing with the real data. 

First, in spite of being realistic models of typical high redshift galaxies, they are not \emph{statistically} representative of the full galaxy population  and especially of the relative frequency of normal and merging galaxies. The simulation snapshots are in fact almost equally divided between disks and mergers (and the latter dominated by major mergers) which is not necessarily reflecting reality. 
For this reason, we do not expect that the relative densities of simulated mergers and disks in the structural planes
investigated in the following should reflect the true distribution of real galaxies. 
We will return to this point in Section \ref{sec:SimulationInsight}.

Second, the predicted stellar fluxes for the MIRAGE simulations are not obtained with a self-consistent, 
full radiative transfer treatment and they do not include the effects of dust extinction and scattering which are instead known to be important in real galaxies.  As discussed in detail in the work of \citealt{Lotz_et_al_2008b}, simulated mergers with no dust obscuration appear more concentrated, less asymmetric and have lower $M_{20}$ values than simulations in which dust obscuration is included.
The MIRAGE flux/$H-$band maps are hence likely smoother and may trace the stellar mass more closely than in real high-$z$  galaxies. 
Conversely, the real mass maps may also appear noisier than the simulated maps as a consequence of the extra image processing and SED fitting which was required for the real data (see Section \ref{sec:MassMapsReal} ).

Third, the lifetimes of clumps in simulated galaxies (and hence the observed clumpiness of galaxies) are sensitive to the assumed feedback model \citep[see e.g.,][]{Genel_et_al_2012,Mandelker_et_al_2014,Moody_et_al_2014}. Support for the feedback recipe employed in the MIRAGE suite 
\citep{Renaud_et_al_2013} comes from the fact that the clumps in the simulated disks well reproduce typical stellar ages ($\sim$\,200 Myr) of observed clumps \citep{Bournaud_et_al_2014}. However, our still incomplete understanding of feedback processes could affect the comparison of the simulated galaxies with the real data.

Fourth, the sample of MIRAGE disks consists of purely isolated galaxies obtained from simulations that are lacking a full cosmological context. 
A number of processes acting on large scales could induce a higher clumpiness/asymmetry in the galaxies, including tidal interactions with  massive galaxies or nearby satellites and gas infall through intergalactic streams. Large scale accretion would help maintaining the initial high gas fractions (60\%) for longer times than in the isolated scenario, leading to more prominent instability-induced features (clumps, asymmetry, etc.).
We note however that while fully cosmological simulations would overcome some of the aforementioned limitations,  
the lower resolution imposed by the large cosmological volumes would also most likely result in smoother galaxies than in reality. 

Finally,  a further complication arises from the fact that the gas fraction in the MIRAGE simulations accounts for both molecular and atomic hydrogen.  The study of \citet{Bournaud_et_al_2015} showed that a substantial $\sim$20\,\% of gas in the simulations is found in the atomic phase in moderate-density regions between the clumps and in extended reservoirs. This would lead to an underestimation of the molecular gas fraction, and consequently clumpiness, with respect to observations \citep[e.g.,] []{Daddi_et_al_2010,Tacconi_et_al_2010}. This is witnessed by the too weak CO excitation in these models \citep{Daddi_et_al_2014}.

As a consequence of the above, it is reasonable to expect that the simulations display a somewhat smoother morphology in optical imaging than the real data. We will keep this caveat in mind.


\section{Structural Measurements}\label{sec:strcMeasurements}

\subsection{Non-parametric Morphology} \label{sec:nonparam_indices}


\begin{figure}
\begin{center}
\includegraphics[width=0.48\textwidth,angle=90]{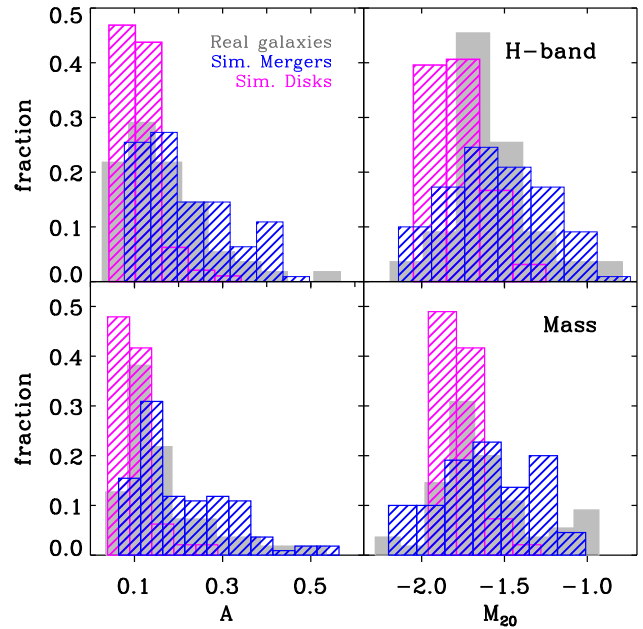}
\end{center}
\caption{\label{fig:f6} Distribution of asymmetry and $M_{20}$ in the MIRAGE simulations (disks=magenta hatched histograms; mergers=blue hatched histograms) and for the sample of real $1.5\leqslant z \leqslant 3$ galaxies in the HUDF field described in Section \ref{sec:sampleSelection} (gray histograms).
The upper row shows the measurements performed on the $H-$band mock and real observations while those obtained on the mass maps are presented in the lower row. 
The simulated galaxies well reproduce the distributions of structural indicators measured on the real HUDF galaxies.}
\end{figure}

On the real as well as simulated single band images and the stellar mass maps, we calculated several non-parametric structural indicators using our own purpose built routines.
 We consider hereafter only  two such indices: 

\begin{enumerate}
\item the asymmetry $A$, which is the normalized residual flux  as obtained  from the difference between the original image and its 180$\deg$-rotated version \citep{Conselice_2003,Zamojski_et_al_2007}. 
         After experimenting with different methods for defining the center used in the asymmetry computation, we chose as our fiducial estimate  
          of $A$ the value calculated with respect to the peak of the emission/mass distribution. 
          We find this option to be most sensitive to multiple components in
          the galaxies and thus to merger features.\footnote{Except for the multiplicity in Appendix \ref{appendix:NoiseEffectsParameters}, all other parameters are referred 
          to the light/mass centroid.}        
\item The normalized second-order moment of the $20\%$ brightest pixels,  $M_{20}$  \citep{Lotz_et_al_2004}, describing the spatial distribution of bright clumps.
\end{enumerate}

The choice of these two indices is justified in detail in Appendix \ref{appendix:NoiseEffectsParameters}, where we also describe the other indicators that were explored. Briefly,  we found that combination of $M_{20}$ and $A$ indices is optimal in terms of both the capability of separating merging from normal galaxies and the (in)sensitivity to S/N effects.

It is known that dust extinction and young star-forming regions affect short wavelength morphologies \citep[e.g.,] []{Bohlin_et_al_1991,Giavalisco_et_al_1996,Windhorst_et_al_2002,Papovich_et_al_2003}. For this reason we performed our measurements on the $H-$band only for the real sample of HUDF galaxies (this is also the reference band used in recent morphological studies on the GOODS/CANDELS fields, e.g., \citealt{Huertas-Company_et_al_2014,Kartaltepe_et_al_2014}).
In the case of the MIRAGE galaxies, on the other hand, we computed the structural parameters also for the artificial F435W and F850LP images for self-consistency check and comparison purposes with the $H-$band and mass maps. 
We measured the indices within the \textsc{SExtractor} Petrosian semi-major axis of each filter and within the Petrosian semi-major axis of the $H-$band in the case of the mass maps.\footnote{The choice of the Petrosian aperture is also motivated in Appendix \ref{appendix:NoiseEffectsParameters}.}


\begin{figure*}
\begin{center}
\includegraphics[width=0.9\textwidth]{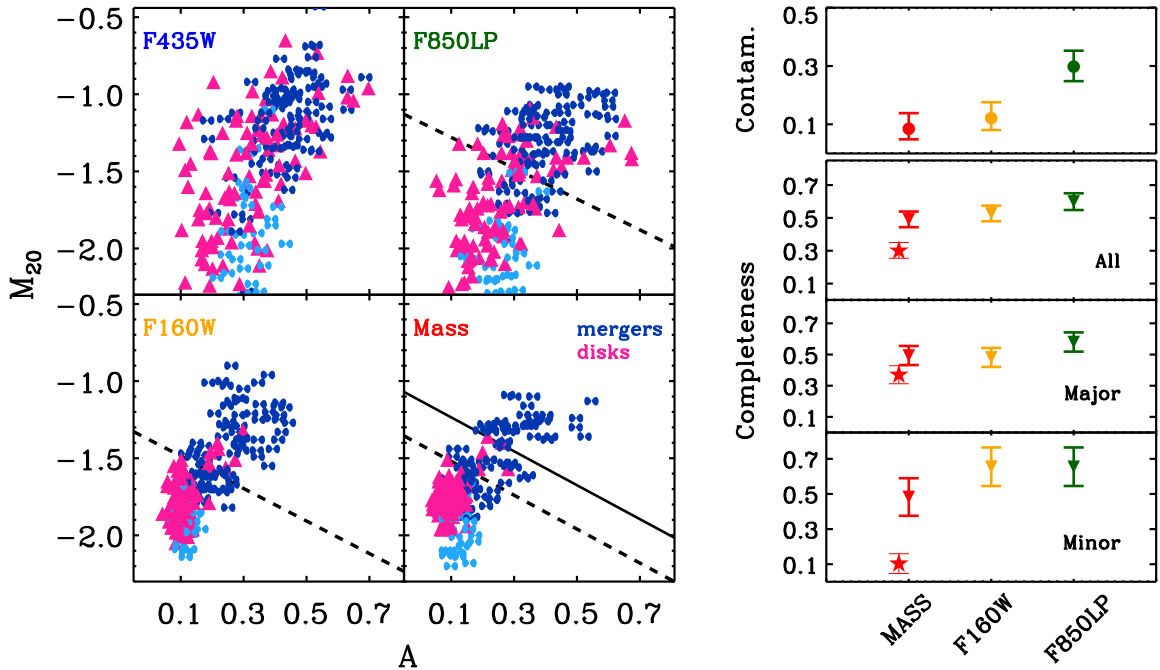}
\end{center}
\caption{\label{fig:f7}  \emph{Left panel block:} relation between  $A$ and $M_{20}$, for the sample of MIRAGE simulated mergers and isolated disks. From top to bottom and left to right, the structural indices are measured on the artificial $HST$ F435W, F850LP, F160W images and on the mass map as indicated by the figure labels. Isolated disks are shown with magenta triangles, while the dark blue points correspond to the training sample of mergers.
The light-blue symbols are mergers that are nominally in the  pre-coalescence sample but have already reached a visually unperturbed appearance, these galaxies are excluded from the SVM training set of mergers.
The dashed lines highlight the maximum margin classifier separation between mergers and disks obtained with the SVM approach.
The solid line in the lower right panel indicates the classification criterion for mass-selected mergers of Equation \ref{eq:eq1}.
 \emph{Right panel block}: 
for the snapshots falling above the SVM dashed lines, we plot in the top most panel the contamination from isolated disks  and in the lower three panels the completeness level for mergers of all mass ratios and for major mergers (ratios 1:1 and 1:2.5) or minor mergers (1:6.3) only. Green, orange and red colors show the values obtained when using $A$ and $M_{20}$ measured on the artificial F850LP images, F160W images and mass maps, respectively.   The red stars indicate the completeness in the mass-based merger selection defined by the condition in Equation \ref{eq:eq1} (shown by a solid line in left panels). The completeness levels are always calculated with respect to all pre-coalescence mergers (i.e., including the light blue points). Errors on the contamination and completeness values are calculated from Poissonian statistics.}
\end{figure*}

Figure \ref{fig:f6}  shows the distribution of $A$ and $M_{20}$ measured on the $H-$band and mass for the real HUDF galaxy sample (gray histograms) in comparison with the distributions that are obtained on the MIRAGE isolated and interacting galaxies (magenta and blue histograms, respectively).
Although the exact shape of the observed gray histograms reflects the relative abundance of mergers and normal galaxies which is not necessarily matched in the simulations,  the MIRAGE and observed galaxies clearly span the same region of the structural parameter space. 
This confirms the reliability of the simulations in probing the morphology of z\,$\sim$\,2 star-forming galaxies and enables us to use them for calibrating the classification in the next Section.
  
\subsubsection{Uncertainties of the Structural Indices} \label{sec:errors on Parameters}

For the real HUDF galaxies,  we estimated the errors on the structural indices as follows.
We resampled  the mass maps and $H-$band images 101 times by replacing each pixel value with a new value randomly extracted within the mass and flux uncertainties, assuming a Gaussian distribution. 
We then recalculated the structural indices each time and used the median value and the 16th and 84th percentiles of all the 101 realizations 
as our final estimates of the non-parametric indices and the associated errors.
Using the median over the resampled maps rather than the directly measured indices has the advantage of down-weighting  the impact of isolated extreme pixels with large uncertainties.

\subsection{Visual Multiband and Mass Morphologies} \label{sec:visualClass}

As complementary information to the quantitative structural measurements, we also visually inspected all galaxies in the HUDF sample. We performed the visual classification independently on composite $bzH$ images and on the mass maps. Each galaxy was assigned to one of the following three broad classes of morphology: 

\begin{enumerate}
\item ``Compact galaxies". These galaxies are either consistent with a spheroidal morphology with no signatures of perturbations or 
too compact for detecting any structural feature.
\item ``Disks", characterized by a regular, centrally symmetric  light or mass profile for which an underlying disk-like  morphology is discernible;
\item ``Multi-component/disturbed" galaxies, which present several peaks/clumps and/or lopsided distribution of light or mass with no clear disk or spheroidal morphology.
\end{enumerate}


\section{Calibration of the Mass-based Classification}\label{sec:SimulationInsight}

We now turn to the main goal our paper, i.e., to verify whether a mass-based classification can improve the identification of merging galaxies.
We start  by testing and calibrating the method on the simulated MIRAGE galaxies. 

\subsection{Comparison with Single-band Classification in the Simulations}\label{sec:Mass_vs_LightSims}

For the MIRAGE mergers and the isolated disks, we show in the left panels of Figure \ref{fig:f7}
the relations between  $M_{20}$ and  $A$ as measured on either the three optical/NIR bands F435W, F850LP and F160W or on the mass maps (we will use the subscript ``H-BAND" or ``MASS" to differentiate $H-$band or mass-based indices from now on).
The F435W results are presented for a comprehensive comparison with the other bands and the mass maps, but a separation between mergers and disks  on the artificial F435W images is very difficult as in this case the two populations clearly overlap in almost the entire parameter space.
Therefore, we do not consider the F435W band for the quantitative analysis described in the following.

On the F850LP, F160W and mass panels  of Figure  \ref{fig:f7} we ran a support vector machine (SVM) algorithm and identified the best dividing line between simulated mergers and disks (dashed line in the figure). We remind that for this calculation we excluded from the training sample of mergers any pre-merger galaxy, merger remnants  and mergers which have reached an unperturbed state (light points in Figure \ref{fig:f7}).  Our method is hence optimized to select galaxies that are clearly displaying interaction features.
For any given sample of snapshots falling above the dashed lines in Figure \ref{fig:f7}, we then estimated the  ``contamination"  and the ``completeness" of the selected merger sample -- i.e., the fraction of misclassified isolated disks among all selected snapshots and the fraction of all pre-coalescence merger galaxies which are correctly classified. 
The results of these calculations are summarised in the right panels of Figure  \ref{fig:f7} where we plot contamination (uppermost panel, circles) and completeness (lower three panels, triangles). We use a red symbol when the classification is performed on the mass maps, orange when it is based on the F160W images  and green for the F850LP filter.

\begin{figure*}
\begin{center}
\begin{tabular}{cc}
\includegraphics[width=0.46\textwidth]{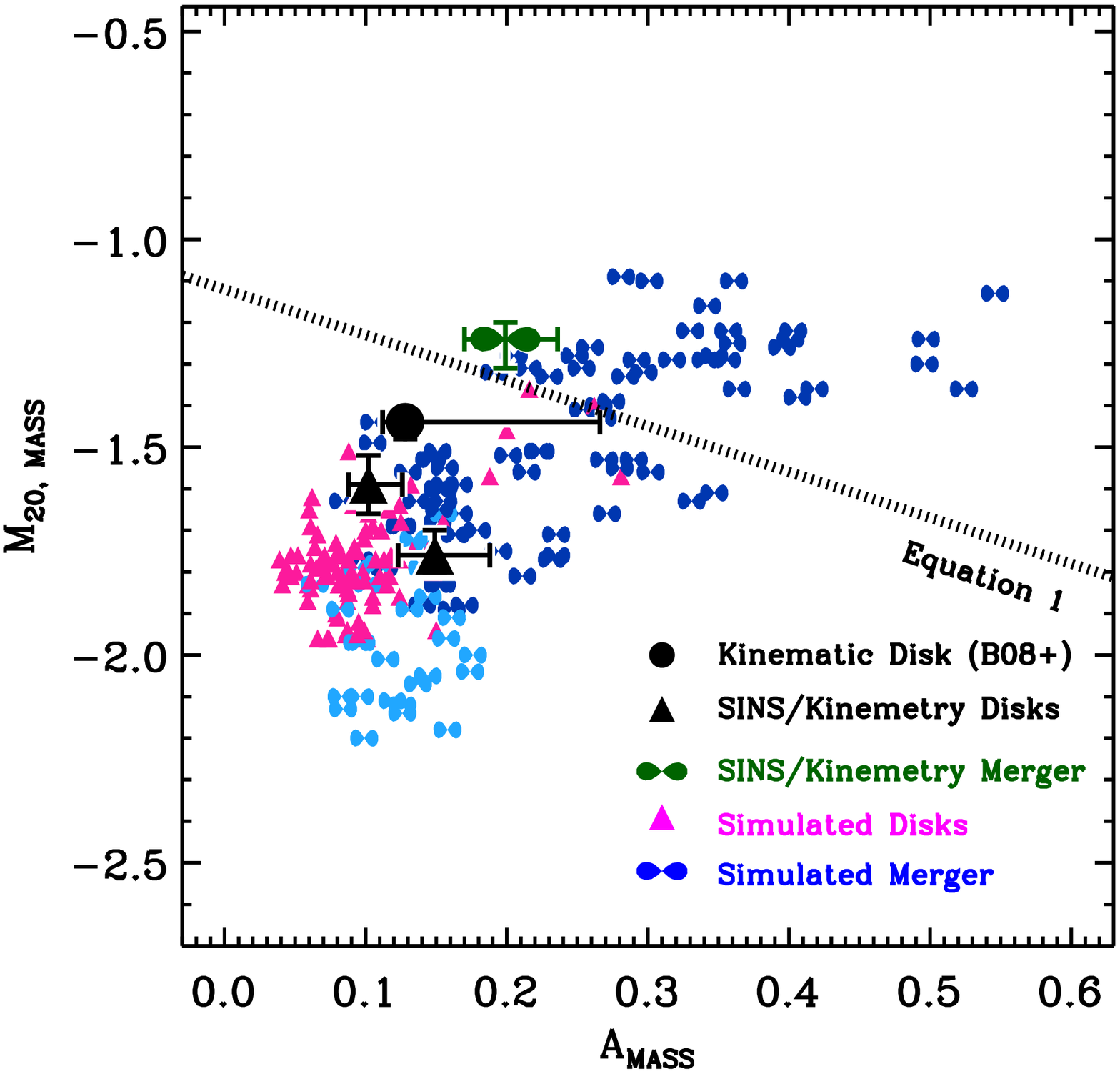} & \includegraphics[width=0.5\textwidth]{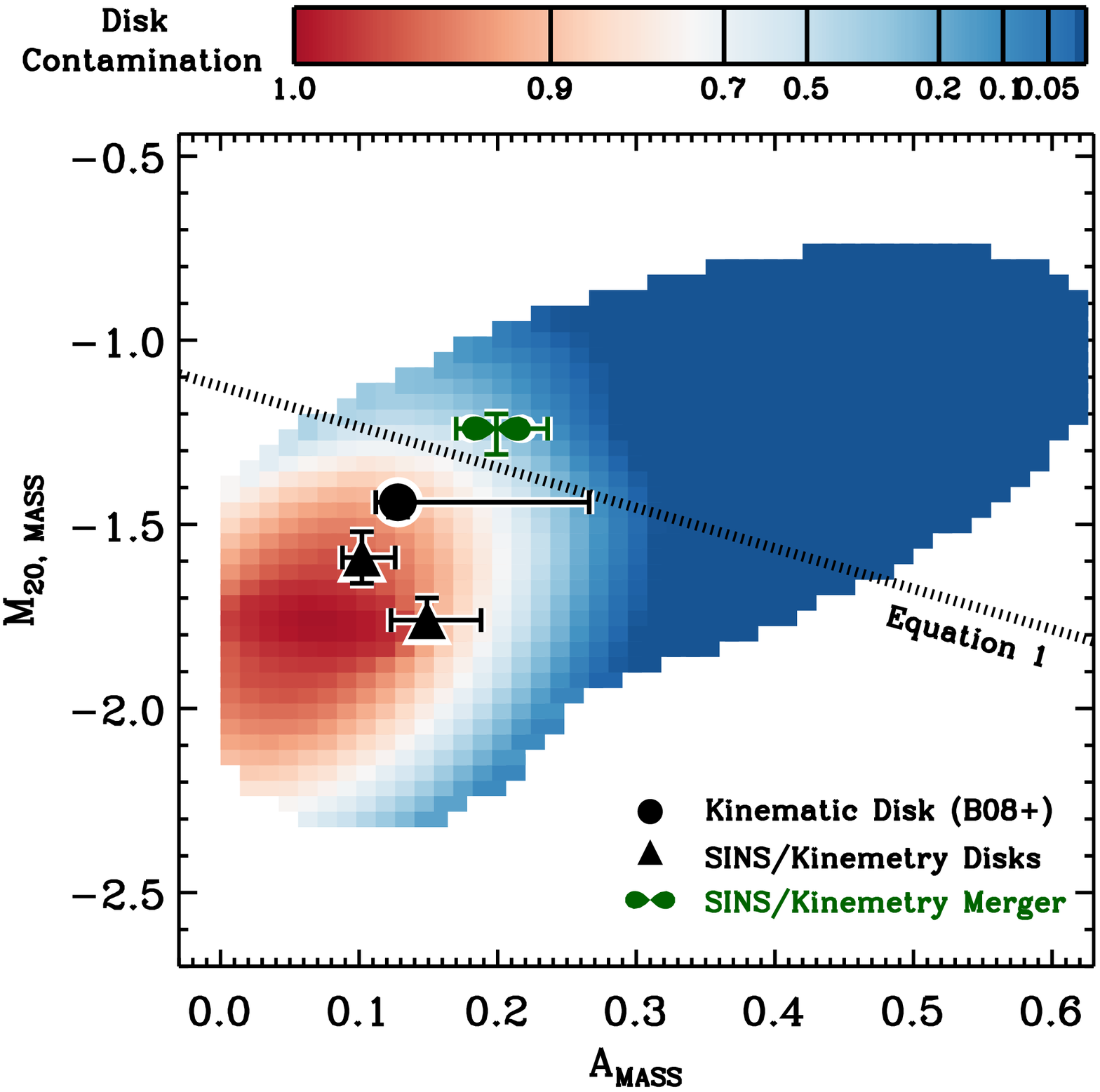} \\
\end{tabular}
\end{center}
\caption{\label{fig:f8}\emph{Left}: location of real galaxies with known kinematic classification on the  $M_{\rm 20, MASS}$ and $A_{\rm MASS}$ plane, in comparison with the results from the MIRAGE simulations.
The simulated isolated disks are shown as magenta triangles and the blue symbols are simulated mergers, as in Figure \ref{fig:f7} 
(light blue corresponds to pre-coalescence mergers with already a regular, unperturbed appearance).
Real observed galaxies that have publicly available kinematic classification are plotted as large points with error bars: in black for kinematically confirmed disks and in green for the known merger.
The sample of observed galaxies with kinematic information is composed by the subset SINS galaxies with kinemetry classification in the GOODS-S area and by the clumpy disk in the study of \citet{Bournaud_et_al_2008} (see Section \ref{sec:SINSMergers} for details.)
\emph{Right:} as on the left panel, but now the color-coding provides the level of contamination from isolated disks (i.e., the fraction of disk snapshots over all snapshots) in any given area of the $M_{\rm 20, MASS}$ and $A_{\rm MASS}$. The shading is obtained by modeling the disk and merger populations with a gaussian mixture and assuming a total merger fraction of 10\% (see text in Section \ref{sec:SINSMergers}).
The division into mergers and disk galaxies based on the simulations and Equation \ref{eq:eq1} (dotted line) is in very good agreement with the position of real kinematically confirmed disks and mergers.}
\end{figure*}

There is a clear trend of decreasing contamination from clumpy disks going from the $z$- to $H$-band and mass maps, supporting our approach.  Although in the simulations the difference between $H-$band and mass is not statistically significant, this trend is reinforced by the results in Figure \ref{fig:Ap2}, showing that the contamination is minimized if the classification is performed on the mass maps also for other combinations of structural indicators.
 We emphasize here that as a consequence of all limitations listed in Section \ref{sec:SimsCaveats}, the simulated $H$-band images likely trace the mass distribution more closely than in real galaxies in which, e.g., patchy dust obscuration or higher gas fractions, will increase the galaxy clumpiness and worsen the ability of separating mergers from disks. 
We hence expect that the differences between the mass-based and flux-based classification should be more pronounced in real data. This hypothesis is supported by the observational results discussed in Section \ref{sec:MassLightRealGal}.

We also note that the mass-based selection of mergers reaches a completeness level that is comparable to the $H-$band for simulated major mergers, whereas it is somewhat lower for the minor mergers. This is not surprising since, by definition, the mass-based classification identifies galaxies with large mass contrasts/asymmetries and hence is less sensitive to minor mergers.  In the simulations, some minor mergers can still be detected in the optical/NIR bands as a consequence of SFR enhancements. Again, in real observations the classification of these minor mergers will be further complicated by the effects discussed above.  


\begin{figure*}[htbp]
\begin{center}
\includegraphics[width=0.7\textwidth,angle=90]{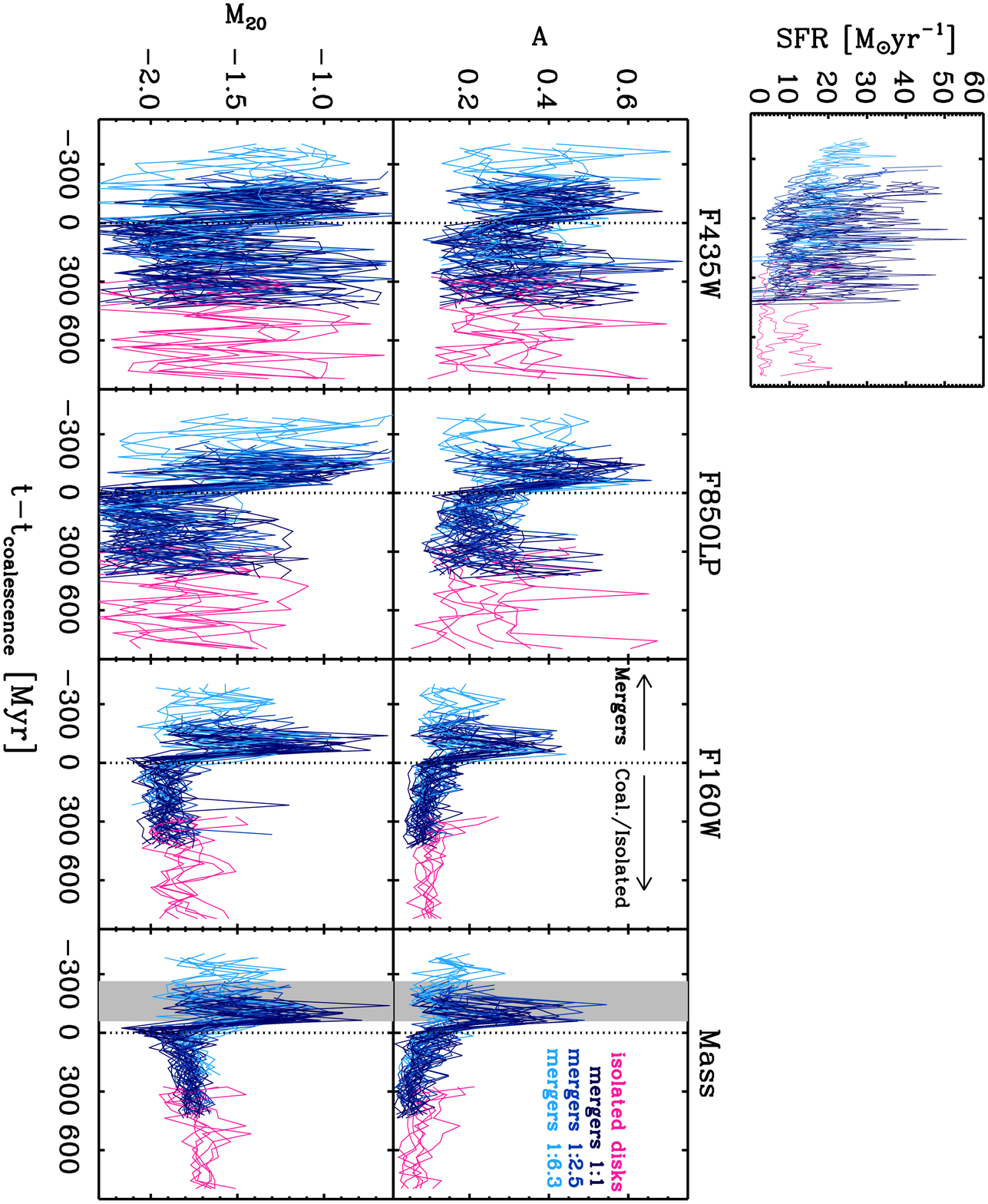}
\end{center}
\caption{\label{fig:f9} Evolution of the Asymmetry and $M_{20}$ indices in the MIRAGE simulated disks and mergers. From left to right the four columns show the results obtained from the mock observations in the F435W, F850LP,  F160W filters and on the mass maps.
The upper-most, isolated panel shows for reference the SFR histories of the simulated galaxies.
Merger simulations are shown as blue lines of increasingly darker shades with increasing merger ratio (see figure legend) while the magenta lines correspond to the isolated disks.
The x-axis is the simulation time with respect to the coalescence time (shown with thin dotted line a $t=0$), negative times hence indicate an on-going merger while for positive $t$ values the two galaxies have already merged into a single object. No coalescence time exists for the isolated disks which are simply shown over the time range considered. Note that the pericenter time for simulated mergers is at about $t=-200$. 
The shaded gray area in the right panels indicates the time scale over which the simulated mergers satisfy the condition in Equation \ref{eq:eq1} between the $M_{\rm 20, MASS}$ and $A_{\rm MASS}$ indices and thus would be classified as mergers according to that criterion.}
\end{figure*}

\subsection{Quantitative Definition of Mergers}\label{sec:TestonSims1}

We use the results of the SVM partition to define a quantitative criteria for selecting mergers based on the position in the $M_{\rm 20, MASS}$ and $A_{\rm MASS}$ plane.
We concluded in Section \ref{sec:SimsCaveats} that  an exact one-to-one match between the relative numbers of disks and mergers  in the simulations and the real data is not expected. 
For this reason we have refrained from performing any fine-tuning of the coefficients in  the relation between $M_{\rm 20, MASS}$ and $A_{\rm MASS}$  such as to optimize the completeness versus contamination level. This would in fact depend on the input merger fraction in the simulations.
We rather follow the conservative approach of identifying 
a locus in the mass-derived $M_{\rm 20, MASS}$ versus $A_{\rm MASS}$ plane which -- unless real galaxies have extreme mass distributions that are not reproduced by current models -- should be populated by mergers only.
This is of course still model dependent, but less affected by the specific choice of the merger fraction in the simulations.

 \begin{figure*}
\begin{center}
\includegraphics[width=0.9\textwidth]{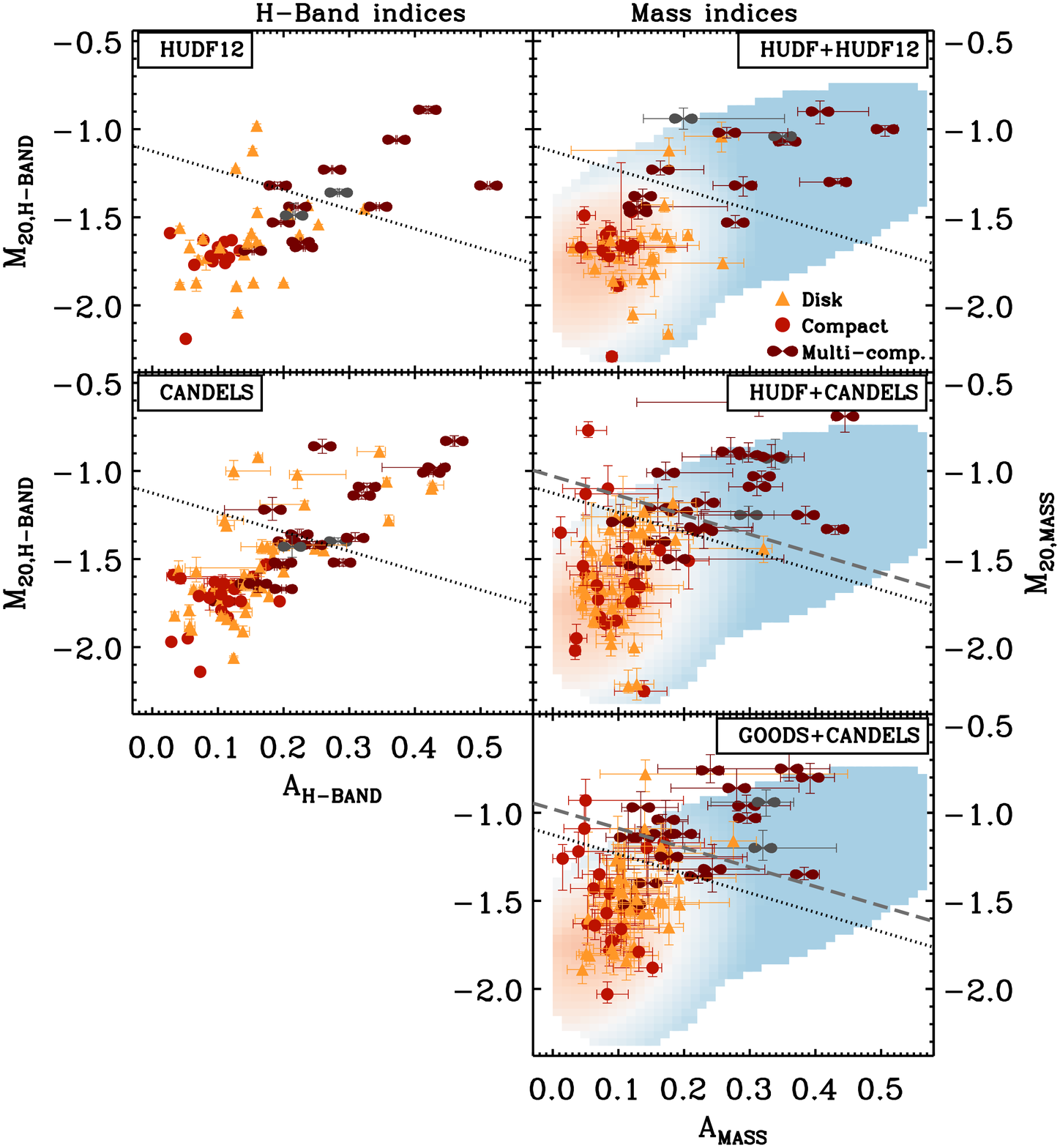}
\end{center}
\caption{\label{fig:f10}$M_{20}$ versus asymmetry plane for the sample of 87 1.5$\leqslant$\,z\,$\leqslant$3 galaxies in the HUDF area with reliable mass maps. In the left panels we present the structural indices measured on the $H-$band images while on the right panels are those obtained on the stellar mass maps. From top to bottom measurements performed on images of increasingly shallower depth are presented, as highlighted by the figure legend. The deepest HUDF\,+\,HUDF12 are available only for 50 of the 87 galaxies; the points in the upper-most panels are hence a sub-set of those in the other planes.
The different symbols correspond to the broad visual morphological classes in which the sample is divided in Section \ref{sec:visualClass}: triangles\,=\,disk galaxies; circles\,=\,compact/smooth galaxies; lemniscate\,=\,multi-component/disturbed galaxies. Grey symbols highlight the two galaxies for which the mass map reconstruction is uncertain due to large contamination from companion galaxies. 
The dotted lines indicate the relation  $M_{\rm 20, MASS}>-1.1\times A_{\rm MASS}-1.12$ dividing simulated mergers and disk galaxies in Equation \ref{eq:eq1}. The dashed lines are the depth-adjusted criteria of Equation \ref{eq:eq2}. 
In the right-hand panels, the background shaded areas reproduce the disk contamination fraction presented in Figure \ref{fig:f8} for simulations matched to the HUDF\,+\,HUDF12 depth.
As illustrated by the several orange triangles in the left panels, merger samples selected using  $A_{H-BAND}$ and $M_{20, H-BAND}$ cuts are substantially contaminated by visually clumpy galaxies that however have smooth, disk-like mass profiles.}
\end{figure*}

Our simulation-justified dividing line between mergers and normal galaxies is hence given by the following relation:

\begin{equation}\label{eq:eq1}
M_{\rm 20, MASS}>-1.1\times A_{\rm MASS}-1.12 \, ,
\end{equation}
and it is obtained by simply applying a shift to the best SVM line in the $M_{\rm 20, MASS}$ versus $A_{\rm MASS}$ plane of Figure \ref{fig:f7} such to exclude  the most extreme disks.
This relation is shown with a solid line in Figure \ref{fig:f7}.

The completeness level over all pre-coalescence merger snapshots (including the unperturbed light-blue points in Figure \ref{fig:f7}) reached with the selection of Equation \ref{eq:eq1} is shown with a star symbol in the lower-right panel in Figure \ref{fig:f7}. As a consequence of our rather conservative choice of the dividing line in Equation \ref{eq:eq1}, the mass-based classification results in a very high purity of the selected mergers at the expense of the completeness of the merger sample. On the MIRAGE simulated data, we estimate a completeness of up to $\sim$\,40\% for major mergers (ratios 1:1 to 1:2.5) and up to 10\% completeness for the minor mergers when applying Equation \ref{eq:eq1}. 

 When dealing with real data, it is useful to introduce, together with the binomial classification of Equation \ref{eq:eq1}, also a more probabilistic description of the $M_{\rm 20, MASS}$ versus $A_{\rm MASS}$ plane which could be used to assign a merger likelihood on individual galaxies. For this reason using the results from the simulations, we also derived a continuous parametrization of the contamination from clumpy disks over that plane.
  This was obtained by modeling the number densities of the mergers and disk snapshots with a Gaussian mixture approach and calculating the fraction of isolated disks over all snapshots in any given region.
As we discussed in detail in Section \ref{sec:SimsCaveats}, the relative number of mergers and disks in the simulation is almost 1:1, whereas the major merger fractions reported in the literature at the redshift considered here are of order of 10\% \citep[e.g.,] []{Bluck_et_al_2009,Newman_et_al_2012,Williams_et_al_2011,Man_et_al_2014}. To derive a realistic value of the relative abundance of disk and merger snapshots in $M_{\rm 20, MASS}$ versus $A_{\rm MASS}$ space, we hence rescaled the number counts of simulated mergers such as to reach a total merger fraction of 10\%.\footnote{This is simply done by down-weighting the merger counts by the factor $w_m=(f_m\times N_{d})/((1-f_m)\times N_m)$ such that $f_m\equiv 0.1 =w_m\times N_m/(N_d+w_m\times N_m)$. In the above, $N_d=96$ and $N_m=88$ are the original number of simulated disks and mergers and $f_m$ is our target fraction of 10\%.} 
We show this probabilistic mapping of  disk contamination in the $M_{\rm 20, MASS}$ versus $A_{\rm MASS}$ plane in the right-hand panel of Figure \ref{fig:f8}. This information can be used to refine the classification on a case by case analysis.

\subsection{Validation on Real Galaxies with Kinematic Classification} \label{sec:SINSMergers}

A kinematic classifications into rotationally supported disk or merger is currently available for some $z>1$ galaxies 
\citep[e.g.,] []{Epinat_et_al_2009,Bournaud_et_al_2008,Forster_Schreiber_et_al_2009} which can thus be used to
further validate the calibration of the classification in the $M_{\rm 20, MASS}$ versus $A_{\rm MASS}$ plane obtained from the MIRAGE simulations.
For this test we consider galaxies with a similar optical+NIR wavelength coverage as our HUDF sample.

In particular, a subset of the galaxies in the SINS survey \citep{Forster_Schreiber_et_al_2009} are located in the GOODS-S field and thus stellar mass maps can be built using the GOODS\,+\,CANDELS data.
 Specifically, we refer to those SINS galaxies which have a kinemetry  classification in Table 9 of \citet{Forster_Schreiber_et_al_2009}.
There are three such galaxies falling in the CANDELS/GOODS field: K20-ID6, K20-ID7, K20-ID8 which were originally extracted from the sample of \citealt{Daddi_et_al_2004a}. K20-ID6 and K20-ID8 are classified as disks, while the nature of K20-ID7 is less clear as it has been classified as merger by the kinemetry analysis in \citet{Forster_Schreiber_et_al_2009} or as a disk displaying disturbed rotation in \citet{Tacchella_et_al_2015}. We label it as ``perturbed/merger" to indicate that it is not a regular disk.
For another galaxy that is included in our sample of galaxies with HUDF\,+\,HUDF12 coverage (CANDELS ID 15011 or UDF ID 6462)
a kinematic analysis based on SINFONI data has been published in the study of \citet{Bournaud_et_al_2008}.
 As discussed in that work, in spite of its disturbed appearance this galaxy clearly displays a rotational motion in the $H_{\alpha}$ velocity field.
 We hence include this object in the sample of kinematically classified disks. We performed our mass map analysis on these galaxies and compare in Figure \ref{fig:f8} their location with the expectations from the simulations.
 
Inspecting Figure \ref{fig:f8}, we find a very good agreement between our simulation-based partition of the $M_{\rm 20, MASS}$ versus $A_{\rm MASS}$ plane and the location of the observed galaxies: all kinematically confirmed disks lie below the line defined in Equation \ref{eq:eq1} in the disk-dominated locus and the kinematic merger/perturbed galaxy is on top of the merger locus, on a region where the disk contamination is expected to be  $\lesssim$\,20\% at the HUDF depth.

We note that the real kinematically confirmed disks display somewhat higher $A_{\rm MASS}$ and $M_{\rm 20, MASS}$ values than the bulk of simulated MIRAGE disks.
 As mentioned in Section \ref{sec:SimsCaveats}, a lower gas fraction in the simulations with respect to real galaxies or the lack of dust could be responsible for the difference. 
Furthermore, we discuss in Section \ref{sec:HvsMassHUDF} that a lower S/N causes a shift to higher $M_{20}$.  Small modifications to Equation \ref{eq:eq1} should be applied when using data that is shallower than HUDF, as for the SINS galaxies.  When accounting for this effect, we still find a good agreement between the kinematic and mass classification, with galaxies K20-ID6, K20-ID8 and ID 15011 falling in the disk-dominated region and the perturbed galaxy K20-ID7  on the transition region between disks and mergers, consistently with its ambiguous kinematic classification.

\begin{figure*}[htbp]
\begin{center}
\includegraphics[width=0.8\textwidth]{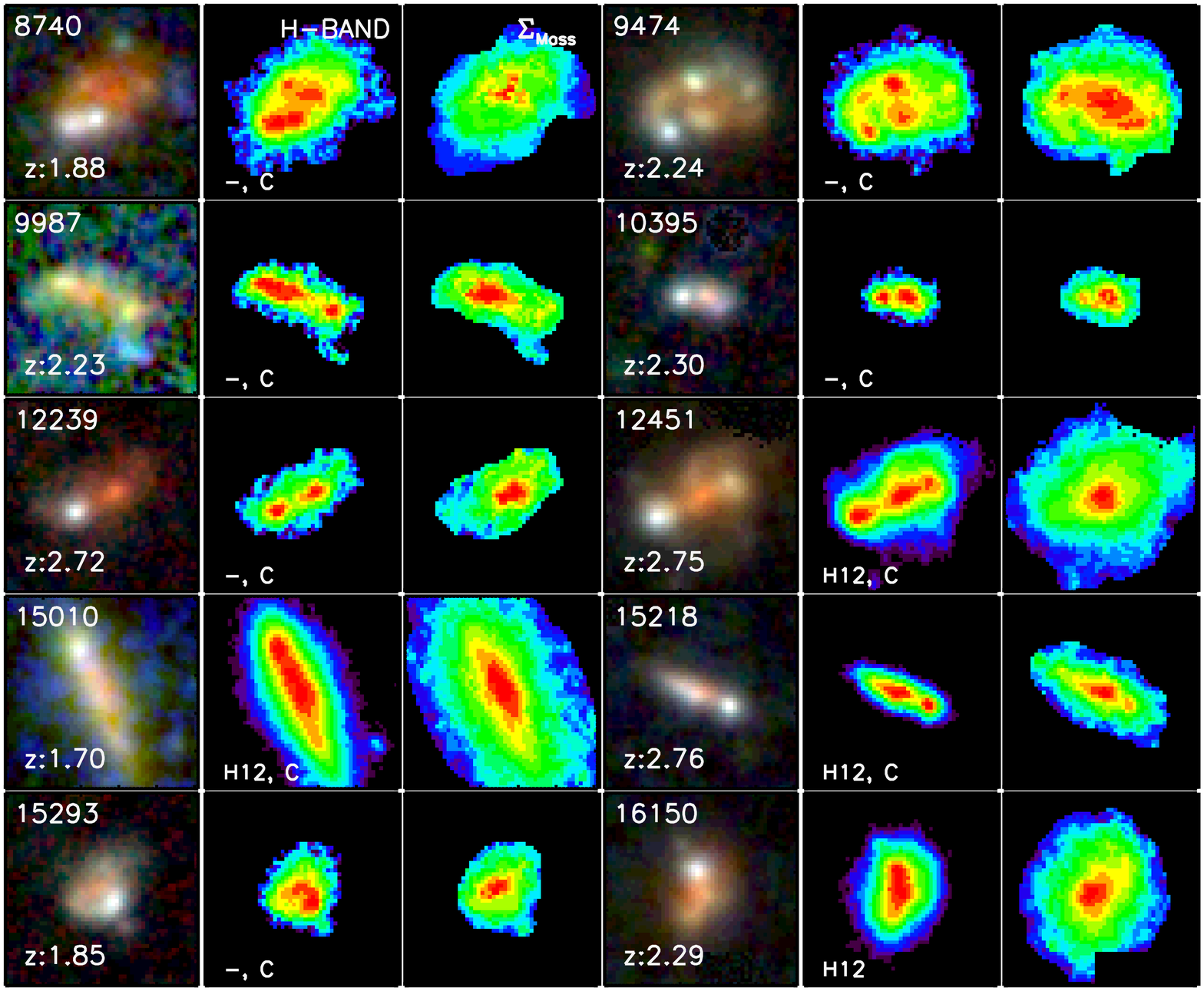}
\end{center}
\caption{\label{fig:f11} Composite $bzH$ image, $H-$band stamp and mass map (from left to right) for  galaxies which have a sufficiently asymmetric appearance in the $H-$band to be classified as mergers on the basis of their $A_{H-BAND}$ and $M_{20,H-BAND}$ values, but are instead visually and quantitatively classified as normal, disk-like galaxies based on their mass distribution and the $A_{\rm MASS}$ and $M_{\rm 20, MASS}$ indices -- see filled orange triangles above the dotted line in the lower left panel of Figure \ref{fig:f10}. We mark the $H$-band stamps with ``C" or ``H12" depending on whether the galaxy satisfies the criteria for being a merger when using the CANDELS-Deep $H-$band photometry or the HUDF12 images (or both). A ``-" sign highlights those galaxies for which only CANDELS observations are available. All images are 3$^{\prime\prime}$ wide. When available, the stamps and mass maps from the HUDF/HUDF12 imaging are presented; HUDF/CANDELS-Deep images are shown in the other cases.}
\end{figure*}

\subsection{Probed Timescales and Merger Ratios}

We can also use the simulations to convert the structural selection into a typical observability window during which mergers are identified as such by the mass-based classification. 
Figure \ref{fig:f9} shows the evolution in time of the structural parameters in the MIRAGE simulated disks 
and mergers. As a further illustration of the results of Section \ref{sec:Mass_vs_LightSims}, we present the time variations for indices measured on the stellar mass maps  as well as for those derived on the three optical/NIR bands.
The different shades of blue illustrate the evolution of mergers with different mass ratios (increasing merger ratio for darker shades), 
whereas the magenta lines are for the isolated disks.  For  interacting galaxies we trace the parameters 
from an early merger phase at 50Myr before pericenter -- when the two galaxies are still separated  and hence the 
structural indices are measured on the primary only -- to after the complete fusion between the two galaxies.
The peak observed in most of the indices coincides with the pre-coalescence phase that we are targeting with our selection; clearly it becomes more distinguishable from other phases when moving from the optical to the NIR and the mass maps.

The shaded gray areas in the right panels of the figure highlight the time interval over which 90\% of the mergers identified by  the selection in Equation \ref{eq:eq1} are found: following our classification criteria we are able to identify mergers for about 200\,Myr, i.e. between 300\,Myr and 50\,Myr before coalescence. 
This time scale is consistent with those reported by, e.g., \citet{Lotz_et_al_2008b,Lotz_et_al_2010} for optically selected mergers, although these authors find a wide range of observability time-scales  (between $\sim$0.1 and 1\,Gyr) depending on the orientation, type of mergers as well as on the structural indicators used to detect the merger features.

Finally, in the top panel of Figure \ref{fig:f9} we show for comparison the evolution of the SFR in the simulated galaxies 
(see also Figure 7 of \citealt{Perret_et_al_2014}). Interestingly, the variation in the structural parameters occurring during the merger event is not associated to a clear SFR enhancement in the simulations.
\citet{Perret_et_al_2014} list several plausible causes for the lack of the starburst phase, including the interplay between the high gas fractions and the feedback treatment in the simulations.
A detailed study of the star formation histories in the MIRAGE simulations is beyond the scope of this paper (and is also affected by  uncertainty in the models), the information is presented for completeness here but we will not further speculate on the links (or absence thereof) between the structural properties and star formation activity of the simulations.


\begin{figure*}
\begin{center}
\includegraphics[width=0.7\textwidth,angle=90]{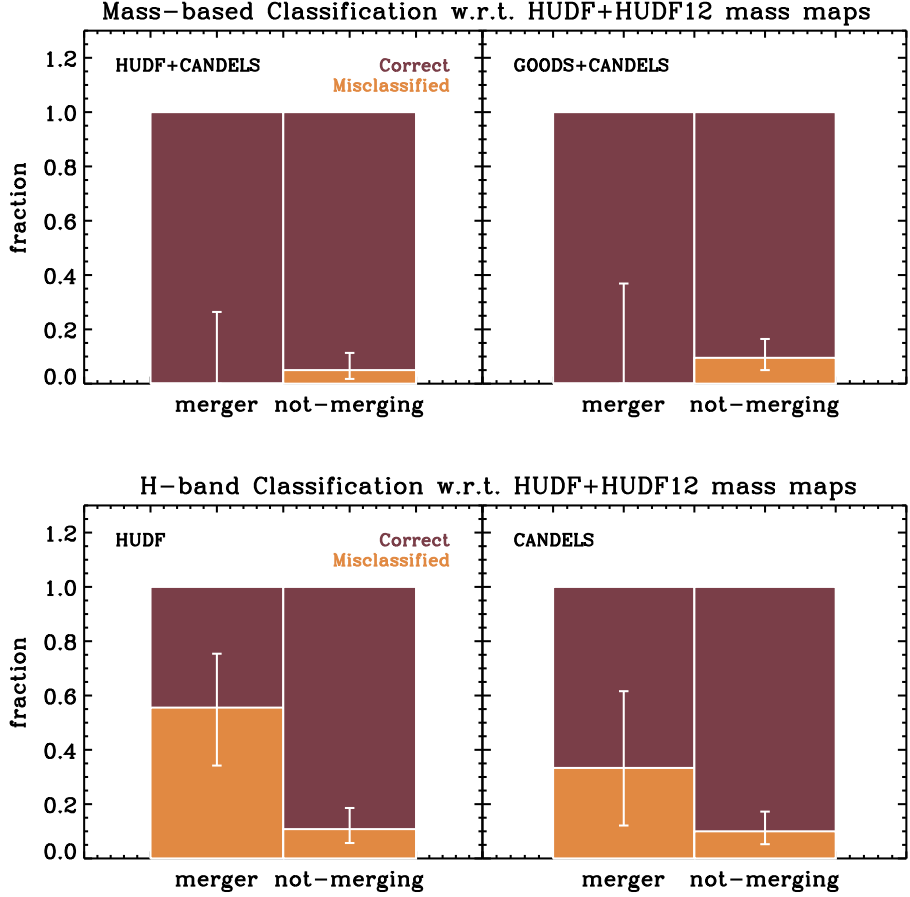}
\end{center}
\caption{\label{fig:f13}\emph{Top:} comparison between the classification based on $A_{\rm MASS}$ and $M_{\rm 20, MASS}$ calculated on the stellar mass maps constructed with deep HUDF\,+\,HUDF12 photometry and the classifications which are instead obtained from mass maps derived from the shallower combination of HUDF\,+\,CANDELS-Deep (left) or GOODS\,+\,CANDELS-Deep imaging (right). 
For galaxies classified as either non-interacting or mergers according to Equation \ref{eq:eq2}  when using the HUDF\,+\,CANDELS or GOODS\,+\,CANDELS data, we show: with a dark histogram the fraction that have a consistent classification at the HUDF\,+\,HUDF12 depth and with a light histogram the fraction that is misclassified at the shallower depths, namely  the fraction of
HUDF\,+\,CANDELS-Deep or GOODS\,+\,CANDELS-Deep merger candidates that are instead classified as disks from the HUDF\,+\,HUDF12 maps, and vice versa.
For each class, the light and dark histograms sum up to one. White error bars indicate Poissonian uncertainties on the measured fractions.
\emph{Bottom:} As above but for the classification performed on the $H-$band images. In particular, we plot the fraction of galaxies classified as mergers or disks based on the $A_{H-BAND}$ and $M_{20,H-BAND}$ measured on the HUDF12 (left) or CANDELS-Deep images (right) that have a consistent or a different classification when using $A_{\rm MASS}$ and $M_{\rm 20, MASS}$ from the  HUDF\,+\,HUDF12 mass maps. 
At any depth, the classification based on measurements performed on the $H-$band results in a twice as large contamination of clumpy galaxies in the mergers sample than the mass-based classification.}
\end{figure*}


\section{Mass versus Light: Classification for Real HUDF Galaxies at Different Depths}\label{sec:MassLightRealGal}

Building on the results from the previous section, we now apply the mass-based classification and compare the outcome with the results from the $H-$band images for the 87 galaxies with reliable mass maps in the HUDF area (i.e., the 89 galaxies with \textsc{SExtractor} kron radii larger than 5\,$\times$ the PSF FWHM and magnitudes $H\leqslant$24.5 minus the two galaxies with mass maps affected by strong contamination from neighboring galaxies, see Section \ref{sec:biases}).

For the galaxies in the HUDF area, the availability of multi-depth data allows us to verify the reliability of the mass-based classification at different S/N.
Specifically,  we consider mass maps derived  from the three combinations of optical plus NIR imaging described in Section \ref{sec:MassMapsReal}: (1) GOODS  and CANDELS-Deep  data, (2)  HUDF plus CANDELS-Deep, and (3) the deepest maps obtained from the HUDF12+HUDF observations. 
Of the 87 galaxies here considered,  50 galaxies have coverage at all depths, while for the other 37 only the first two versions of the mass maps could be derived. 
For comparison with the mass maps results, we also consider structural parameters measured on the $H-$band images at the HUDF12 or CANDELS-Deep depths.

In Figure \ref{fig:f10} we present the location of the HUDF galaxy sample in the $M_{20}$ versus Asymmetry plane, either derived from the $H-$band images (left panels)  or the stellar mass maps (right panels). 
From top to bottom we show measurements performed on images of increasingly shallower depths.
The dotted lines in the figure indicate the dividing locus of Equation \ref{eq:eq1} and the different symbols highlight the visual types of Section \ref{sec:visualClass}.
In the following we will regard galaxies with a visual class for the mass maps that is inconsistent with the quantitative criterion on $A_{\rm MASS}$ and  $M_{\rm 20, MASS}$  as failures of the mass-based classification, i.e., as contaminants in the merger sample. In some cases such galaxies display genuine color gradients which could justify the mass asymmetries, but we will take the conservative approach of assuming they are the result of noise in the mass maps.\footnote{Note that we also implicitly assume here that  there are no systematic biases in the derivation of the mass maps,  which could generate artificial multi-component, merger-like features in high S/N maps.}

\subsection{(Falsely) $H-$band Selected Mergers}\label{sec:HvsMassHUDF}

We start the analysis on the real galaxies by identifying in Figure \ref{fig:f10} those galaxies which would be classified as mergers based on the irregularity of their $H-$band images.
For a consistent comparison with the selection based on the mass map, we consider as mergers candidates those galaxies that satisfy Equation \ref{eq:eq1} also for the $H-$band measurements, i.e., galaxies above the dotted lines in Figure \ref{fig:f10}.  
Other criteria are often used in the literature when using  $M_{20}$ and asymmetries derived from optical images.
At low redshift (z$<$\,1) a cut at $A$\,$>$\,0.3-0.35 is typically applied to select major mergers \citep{Conselice_2003,Lotz_et_al_2008b} but for  z\,$>$\,1.5 galaxies a value of $A$\,$\gtrsim$\,0.2 is preferred to account for the effects of the morphological $k$-correction, the decrease in the image resolution and surface brightness dimming \citep{Conselice_2003,Conselice_et_al_2005}. 
The distribution of irregular galaxies in Figure 10 of \citet{Scarlata_et_al_2007} would also suggest a division at $M_{20,H-BAND}$\,$>$\,-1.7 to  identify mergers candidates.
We have checked (and it can be verified by inspecting Figure \ref{fig:f10}) that the result here presented would not  change if these other criteria are applied.

More than 90\% of the galaxies that satisfy Equation \ref{eq:eq1} in the $H-$band were assigned to the class of ``multi-component/disturbed" galaxies also by the visual $bzH$ morphological analysis. This means that these galaxies would be classified as mergers by both a multi-wavelength visual inspection and a quantitative structural analysis on the $H-$band. 
Conversely, it can be seen that several of the merger candidates lying above the dotted line in the $H-$band planes are  visually classified as normal disk galaxies based on their mass map appearance (orange filled triangles in the left panels of Figure \ref{fig:f10}). These galaxies have accordingly low asymmetry and $M_{20}$ indices measured on the mass map (see right panels of Figure \ref{fig:f10}). At both the HUDF12 and CANDELS-Deep depth, about $\sim$ 50\%  of those galaxies with  large enough $A_{H-BAND}$ and  $M_{20,H-BAND}$ to be classified as mergers in the $H-$band have a mass profile consistent with that of a normal disk galaxy and fall below Equation  \ref{eq:eq1} when the mass-based indices are used.
To illustrate these differences between mass and $H-$band classification, we present  in Figure  \ref{fig:f11}  the stamp images for  those galaxies with  $A_{H-BAND}$ and  $M_{20,H-BAND}$ as high as those of mergers (using either the HUDF12 or CANDELS photometry) but a disk-like mass profile. 
In spite of displaying composite $bzH$ and $H-$band morphologies dominated by multiple clumps and asymmetric light distribution which could be suggestive of a merger event, all such galaxies have a regular, centrally concentrated distribution of mass, typically associated with a red nucleus.

\subsection{Mass-selected Mergers}\label{sec:HUDFMergers}

Focusing on the right panels of Figure \ref{fig:f10}, we can instead study the results of the measurements performed on the mass maps. 
At the HUDF\,+\,HUDF12 depth -- top most panel -- we find an excellent agreement between the  locus of mergers expected from the analysis on the simulations (see light shaded area and dotted line) and the visual classes for the real data. 
A comparison with the other right-hand panels, however,  reveals some important facts:  not surprisingly,  the separation between galaxies visually classified as disks/compact and galaxies with visually disturbed morphologies  becomes less clear as the S/N of the mass maps decreases. At the GOODS\,+\,CANDELS-Deep depth, the distribution of galaxies in the  $A_{\rm MASS}$ and  $M_{\rm 20, MASS}$  plane is noticeably more clustered than for the structural parameters derived on the HUDF\,+\,HUDF12 mass maps.

It also appears rather clearly from  Figure \ref{fig:f10} that the bulk of real galaxies shifts toward higher $M_{\rm 20, MASS}$ values as a consequence of the decrease in S/N. This is particularly evident when comparing the data points with the shaded locus defined by the simulations.
We showed in Appendix \ref{appendix:NoiseEffectsParameters} that, while the Asymmetry is almost insensitive to the image noise,  some dependence of the $M_{20}$ on the S/N  is expected and variations of the order of 10\% are measured for  $M_{20}$ indices derived at different depths. 
This suggests that the criterion in Equation \ref{eq:eq1}, which was derived for simulations reproducing the HUDF\,+\,HUDF12 depth, should be revised as an ``adaptive" threshold varying with the images S/N. 
Furthermore, we note that at the shallower depth a few merger candidates appear in a region of very low asymmetry ($A_{\rm MASS}<0.1$) but very high $M_{\rm 20, MASS}$ ($M_{\rm 20, MASS}>-1$) where virtually no simulated mergers nor galaxy visually classified as merger are found.  We suspect these galaxies to be affected by measurement errors.

We use the above empirical findings to provide refined criteria to robustly select mergers at the various depths: 

\begin{flalign}\label{eq:eq2}
M_{\rm 20, MASS}&> -1.1\times A_{\rm MASS}+
\begin{cases}
-1.12  \quad H+H12  \\
-1.00  \quad H+C  \\
-0.98  \quad G+C  \\
\end{cases}
\nonumber \\ 
A_{\rm MASS}&>0.1\, ,
\end{flalign}

where ``H+H12", ``H+C" and ``G+C" stand for the mass maps derived from the HUDF\,+\,HUDF12, HUDF\,+\,CANDELS-Deep and GOODS\,+\,CANDELS-Deep photometry, respectively.
These new relations where obtained by adding to the locus of Equation \ref{eq:eq1} the average offsets in $M_{\rm 20, MASS}$ measured in Figure \ref{fig:Ap2}. The new selection limits are shown with dashed lines in Figure \ref{fig:f10}.

By selecting as merger candidates those galaxies that satisfy the generic  Equations \ref{eq:eq1}  we find a total of  9, 21 and 26 mergers  for the HUDF\,+\,HUDF12, HUDF\,+\,CANDELS-Deep and GOODS\,+\,CANDELS-Deep combinations, respectively.
Of these 2, 6 and 10 have a visual class that is inconsistent with the quantitative analysis on the mass maps, i.e they have been assigned an either ``compact" or ``disk" class in the visual inspection of the mass maps.
 The use of the fixed selection of Equation \ref{eq:eq1} would hence result in an increase of the contamination from $\sim 20\%$ for the HUDF\,+\,HUDF12 maps to $40\%$ for the GOODS\,+\,CANDELS-Deep combination.
Conversely, using the relations in Equation \ref{eq:eq2} we find 9, 15 and 15 mergers candidates with a total of 2, 3 and 4 galaxies visually classified as non interacting, namely a roughly constant  contamination of $\sim 20\%$, regardless of the image depth.

What should be stressed here is that applying the `adaptive' criteria of Equation \ref{eq:eq2} also to the H-Band would not reduce the contamination of clumpy disks: as it is clear from the bottom left panel of Figure \ref{fig:f10}, these clumpy disks are distributed  everywhere in the  $M_{20,H-BAND}$ and $A_{H-BAND}$  plane and there is no selection that would substantially reduce the contamination.

As a validation of the latter statement we considered the 50 galaxies for which all three versions of the mass maps are available and, for each combination of the photometric data, we independently identified the samples of non-interacting galaxies and mergers, even when accounting for errors (i.e. we selected as merger candidates those galaxies that have error bars above the relations in Equation \ref{eq:eq2}). We did so  for both the $H$-band and mass structural indices.
We then assumed that the classification performed using the deep HUDF\,+\,HUDF12 mass maps is the ``correct" one and we calculated for the mergers and disks identified at the other depths or in the $H-$band the fraction that have a consistent classification  -- i.e. mergers or non-interacting galaxies classified as such also on the deep mass maps --  and the fraction of galaxies which are instead misclassified as a consequence of the lower S/N -- i.e. galaxies classified as mergers that are identified as not-merging on the deep HUDF\,+\,HUDF12 maps or  $H-$band  images, and vice versa.

We show the results of this calculation in Figure \ref{fig:f13}. The number of identified mergers is small and thus the error bars relatively large, 
however the figure shows that by applying the relations in Equation \ref{eq:eq2} the fraction of ``misclassified mergers" in the mass selected sample is consistent with being $\sim$\,20\% for  mass maps obtained from both the HUDF\,+\,CANDELS-Deep and GOODS\,+\,CANDELS-Deep combinations.  Conversely, for the structural indices  $A_{H-BAND}$ and $M_{20,H-BAND}$  the contamination of false mergers remains as high as $\sim$\,50\% for both the HUDF and CANDELS observations, even when imposing the condition in Equation \ref{eq:eq2}.
As witnessed by the increase in the fraction of galaxies that are misclassified as disks in Figure \ref{fig:f13}, the drawback of the refined selection is that a larger number of mergers are missed from the selection at the shallower depths.

Summarizing the above findings, the availability of deep photometry is certainly necessary for maximizing the accuracy and completeness of the mass based classification.  Nonetheless, applying small corrections to our simulation-motivated selection of mergers reliable mergers samples can be identified also on medium depth images, albeit reaching a lower completeness.
Most notably, independently of whether Equations \ref{eq:eq1} or the refined relation in \ref{eq:eq2} are applied to the $A_{H-BAND}$ and $M_{20,H-BAND}$ indices, we find the $H-$band classification results in a twice as large contamination of clumpy disks in the merger sample at any depth here considered. 

\begin{figure*}[htbp]
\begin{center}
\includegraphics[width=0.85\textwidth]{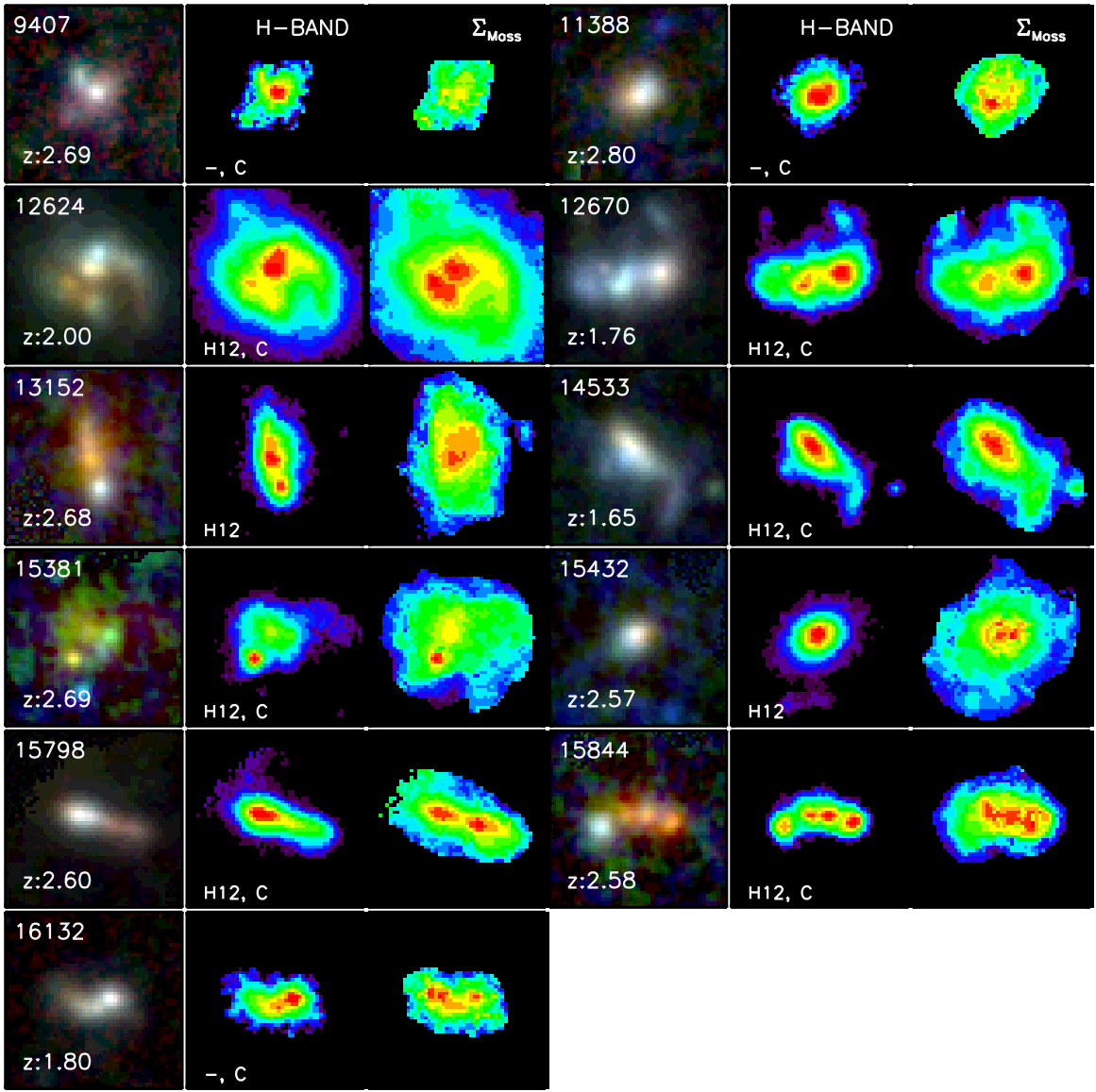}
\end{center}
\caption{\label{fig:f12}
As in Figure \ref{fig:f11}, but in this case showing  galaxies which satisfying the relations in Equation \ref{eq:eq2}  even when accounting for errors in the measurements and hence are candidate mergers according to our definition. }
\end{figure*}

We identify our final sample of merger candidates among the HUDF sample as those galaxies that satisfy the selection criteria in Equation \ref{eq:eq2} even when accounting for the error in the measurements (i.e., have error bars above those relations). The classification is performed on the deep HUDF\,+\,HUDF12 mass maps when available and on the HUDF\,+\,CANDELS-Deep maps for those galaxies with no HUDF12 coverage. 
 We find a total of 11 such candidates over the 87 galaxies with reliable mass maps.  For the latter, we show in Figure \ref{fig:f12}  the $bzH$, $H-$band and mass stamps.
Among this sample of mergers are included one galaxy with an inconsistent visual class and also one object that has been flagged as possible chance projection in Section \ref{sec:ProjectionEffects} (ID 15844, but note that even excluding the secondary blue clump this galaxy would still be classified as a merger), suggesting again a 20\% contamination level.


\subsection{Mergers Missed in the $H-$band}

Thus far we have focused on those galaxies which are misclassified as mergers in the $H-$band because of the presence of bright clumps. 
It is however interesting to test whether also the opposite occurs, i.e., whether there are galaxies classified as mergers in the structural analysis performed on the mass maps that instead appear smooth(er) in the $H-$band, for example as a result of dust lanes or intrinsically high mass-to-light ratios.
Figure \ref{fig:f13} suggests that this happens for roughly 10\% of the ``$H$-band smooth" galaxies.
In our sample of 11 mass-selected merger candidates we find five galaxies (IDs  9704, 11388, 12624, 14533 and 15432) that fall below the dotted lines in Figure \ref{fig:f8} when using $M_{20,H-BAND}$ and $A_{H-BAND}$ and hence would not be classified as mergers in the $H-$band.

To understand the origin of this discrepancy, we generated new versions of the mass maps for these galaxies obtained by applying no dust extinction during the pixel-by-pixel SED fitting. This gives us an indication of whether the mass asymmetry is driven by the dust correction applied to the templates.
IDs 9704 and 14533 display significant asymmetries also in these mass maps, indicating that the classification is independent of the extinction law assumed. IDs 12624, 11388 and 15432  instead would not satisfy our selection criteria with dust-free mass maps.  We note that both IDs 12624 and 15432  are detected in the FIR with total IR luminosities $L_{IR}>10^{12}\Lsol$, implying a substantial amount of dust in these galaxies leading us to favor the results obtained including dust corrections.

While it is clear that additional, e.g., kinematic, information is required for a complete characterization of the population of ``mass-smooth and H-clumpy" galaxies, the available data suggest that most of these galaxies have distinct substructure and that dust obscuration can explain some of the differences between the $H$-band and mass morphology.

\section{Summary and Conclusions} \label{sec:Conclusions}

Using a sample of about 100 HUDF galaxies with $1.5\leqslant z \leqslant3$, we have compared a morphological classification of merging galaxies based on non parametric structural indices derived on resolved stellar mass maps with a canonical classification obtained from $H-$band measurements.  We tested this approach using photometry at the different depths available on the HUDF area (GOODS, CANDELS and HUDF/HUDF12) and, performing tests on artificial stellar mass distributions, we derived luminosity and size limits for which reliable mass maps can be obtained. 
The selection of mergers in the mass domain is calibrated using a sample of isolated and merging galaxies from the MIRAGE hydrodynamical simulations which have been post-processed and analyzed to closely reproduce the observational data.

We summarize our findings as follows:

\begin{enumerate}
\item as also discussed in previous works \citep[][]{Wuyts_et_al_2012}, the stellar mass maps morphologies cannot be reproduced by $H-$band data alone which often display merger-like features even for galaxies with perfectly smooth, disk-like mass profiles. 
Although some of these galaxies display a red nucleus indicative of a central mass concentration, even a combined visual inspection of multiple bands is still affected by the presence of star-forming clumps and does not provide a full proxy for the mass distribution. As a consequence of the above, we quantify that merger samples identified on the basis of asymmetry/irregularity in the $H-$band images can have a contamination from clumpy galaxies as high as 50\%. 
Our analysis also suggests that differences between the mass- and $H$-band selection may originate from a population of galaxies with smoother $H$-band images than mass distribution, possibly as a result of dust extinction.

\item On both the data and the simulations, we find that a combination of $A_{\rm MASS}$ and $M_{\rm 20, MASS}$ measured on the stellar mass maps is instead most cleanly separating major mergers from isolated, clumpy galaxies. 
We use the results from the simulations and the analysis on the real HUDF galaxies to provide in Equations \ref{eq:eq1} and \ref{eq:eq2} quantitative criteria to separate the galaxy population into mergers and disks using mass maps derived at the HUDF\,+\,HUDF12, HUDF\,+\,CANDELS-Deep and GOODS\,+\,CANDELS-Deep depths.
Applying these criteria, we estimate that chance projections or the scattering of clumpy disks in the merger sample by noise in the mass maps result in a contamination of roughly 20\%.
When applied to galaxies with available kinematic data, our $A_{\rm MASS}$ and $M_{\rm 20, MASS}$ based classification results in morphologies that are consistent with the kinematic classes.

\item The ability of the mass-based $A_{\rm MASS}$ and $M_{\rm 20, MASS}$ indices to select true mergers is not significantly affected by a moderate decrease in the S/N of the parent images if the conditions in Equation \ref{eq:eq2} are used. Conversely, the $H$-band classification result in a twice as large contamination from clumpy disks, independently of the criteria that are applied.  

\item  From the analysis of the MIRAGE simulated mergers we estimate
the proposed $A_{\rm MASS}$, $M_{\rm 20, MASS}$ selection should be sensitive to major mergers between 300 Myr and 50 Myr before coalescence.

\end{enumerate}

Based on the above results, we thus suggest that to identify major mergers
a classification in the mass domain rather than from optical/NIR images should be preferred and performed whenever possible.
Obviously, our technique is only sensitive to merger phases in which the
perturbation in the mass profiles is measurable (roughly up to 300\,Myr before coalescence as specified above) and, as also a consequence of imposing a high purity in the resulting merger sample, is strongly biased against earlier merger stages.
A combination with other techniques (e.g., close kinematic pair selection) would be hence necessary for a full census of merging systems.

Finally, it is possible that variations in the gas fraction of galaxies and the actual masses of giant clumps could introduce some redshift dependent scaling of the threshold here derived, moving the locus of clumpy disks toward slightly higher $A_{\rm MASS}$ and/or $M_{\rm 20, MASS}$ values at higher redshift. However, we expect that these variations would be comparable to the uncertainties in the mass maps and the derived structural parameters and hence would not strongly affect the proposed classification for sufficiently large samples.

\section{Acknowledgements}

A.C. thanks P. Hurley, M. T. Sargent and V. Strazzullo for useful discussions and suggestions.
We also thank the anonymous referee for valuable comments which improved the manuscript. 
We acknowledge financial support from the Swiss National Science Fundation (A.C., Project PBEZP2\_137312), from the Agence Nationale de la Recherche  (A.C., E. l. F., contract \#ANR- 12-JS05-0008-01)  and from the E. C. through an ERC grant (A.C., F. B., StG-257720).
The simulations used in this work were performed on GENCI resources at the Tres Grand Centre de Calcul (project 04-2192) and at the LRZ SuperMUC facility under PRACE allocation number 50816.
This work is based on observations taken by the CANDELS Multi-Cycle Treasury Program with the NASA/ESA HST, which is operated by the Association of Universities for Research in Astronomy, Inc., under NASA contract NAS5-26555.
This work has made use of the adaptive smoothing code adaptsmooth, developed by Stefano Zibetti and available at the URL http://www.arcetri.astro.it/~zibetti/Software/ADAPTSMOOTH.html
 \\



\appendix

\section{Reliability of the Mass Maps and Derived Parameters}\label{appendix:MassMapsReliability}

\subsection{Comparison with Integrated Masses}\label{sec:Resolved_vs_Integrated}

\begin{figure*}[htbp]
\begin{center}
\includegraphics[width=0.45\textwidth]{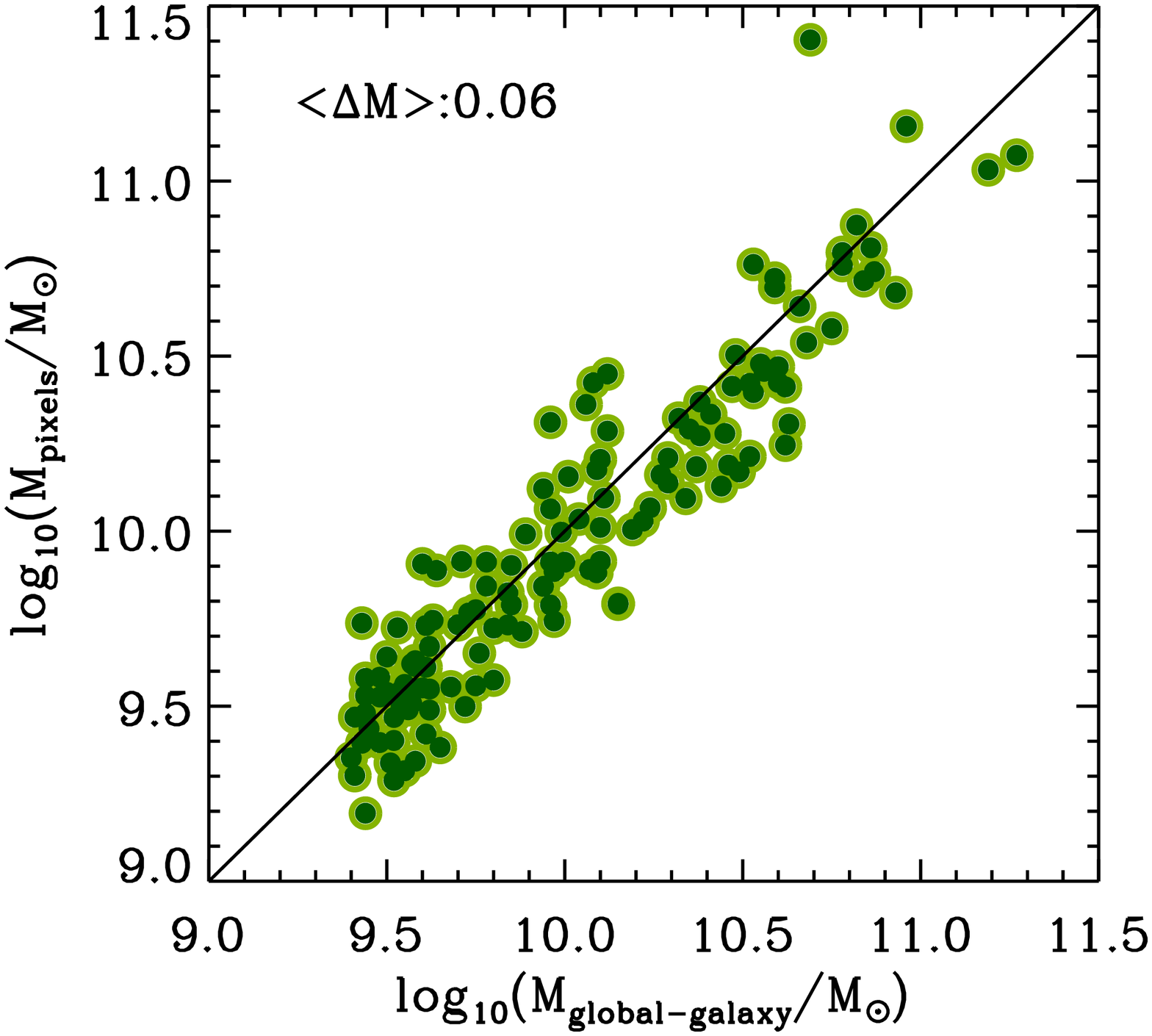}
\end{center}
\caption{\label{fig:Ap1} For the 132 galaxies with $ 1.5 \leqslant z \leqslant 3$ presented in Section \ref{sec:sampleSelection},  we show the comparison between the  galaxy mass obtained from SED fitting to the integrated photometry ($M_{global-galaxy}$) and the sum of the masses in each individual pixel in the mass maps ($M_{pixels}$). On the top right corner of the Figure we provide the median difference between the two estimates for the entire sample of galaxies here considered.}
\end{figure*}

As a basic consistency check of the mass maps  for the sample described in Section \ref{sec:sampleSelection}, we show in Figure \ref{fig:Ap1} the comparison between the sum of the masses in each individual pixel of the galaxy and the total galaxy mass that is obtained from the integrated photometry. Although for highly obscured or strongly star-forming galaxies differences between the two estimates can be expected on an object by object basis due to the patchy distribution of dust, strong disagreement between the resolved and integrated estimates would be an indication of biases in the pixel-by-pixel SED fits. We instead find a very good agreement between the pixel-based total mass and the whole galaxy  mass, with a median difference which is less than 0.1\,dex.


\subsection{Description of the Models used for Testing the Mass Reconstruction}\label{appendix:ResolutionLimit}

As mentioned in Section \ref{sec:biases}, the resolution and noise of the parent observations set major limitations for the estimate of the stellar mass maps.
In order to derive the size and magnitude limits for which a reliable mass map can be obtained, we tested the pixel-based SED fitting on a sample of artificial galaxies with known mass distribution. To perform the test on model galaxies which are representative of the typical real galaxies, the models were drawn from the observed sample as follows. 

For each real HUDF galaxy in the initial sample presented in Section \ref{sec:sampleSelection}, we generated a toy mass distribution characterized by a S\`ersic profile with a total
mass equal to the galaxy integrated mass and structural parameters determined by the best-fit $H-$band GALFIT \citep{Peng_et_al_2010} model.
To densely populate the magnitude-size plane, we also generated extra models by adding random perturbations on the original galaxies parameters. 
We then imposed each pixel in the artificial galaxy to have an SED equal to the observed best-fit SED for the entire galaxy (including a uniform extinction for all pixels) and used this SED to predict the pixel fluxes.

From these flux maps we generated artificial $HST$ images  which we pasted into blank sky regions to mimic the typical S/N properties reached by the real galaxies with  HUDF optical plus CANDELS-Deep NIR coverage. This depth is in between the two extreme combination of imaging (deep HUDF\,+\,HUDF12 or shallow GOODS\,+\,CANDELS-Deep) used for creating the mass maps (see Section \ref{sec:archivaldata}).
Specifically, we created stamps at the HUDF depth for the F435W, F606W, F775W and F850LP filters, while the F814W, F105W, F125W and F160W images were matched to the CANDELS-Deep depth.  
All images were degraded to the resolution of the F160W filter as in the real sample. 

We finally computed ``observed" mass maps form these artificial images as described in Section \ref{sec:MassMapsReal}, i.e.,  by running \textsc{Adaptsmooth}, extracting pixel-based SEDs and fitting them with \textsc{LePhare} and used these ``observed" mass maps for our tests in Section \ref{sec:biases}. 
Although these simulations are undoubtedly a simplification of reality which is complicated by pixel-by-pixel variations of the SED, dust content, etc.  they provide us with a measure of systematic biases in the mass estimation.

\subsection{The Choice of Classification Parameters: Discriminating Power and S/N Effects}\label{appendix:NoiseEffectsParameters}

Together with the $M_{20}$ and $A$ indices described in Section \ref{sec:nonparam_indices}, we also evaluated the following structural indicators:

\begin{enumerate}
\item The concentration $C$ defined as the logarithmic ratio of the radii containing 20\% and 80\% of the total flux  $C = 5 \log (r_{80}/r_{20})$ \citep{Bershady_et_al_2000,Conselice_2003}.
\item The Gini coefficient \citep{Abraham_et_al_2003,Lotz_et_al_2004}, which describes the uniformity of the flux distribution on a scale between 0 (all pixels with equal flux) and 1 (all flux in just one pixel).
\item And finally the multiplicity $\psi$  that quantifies the presence of multiple components through a comparison between the original image and a resampled version in which the pixels are re-arranged in  decreasing flux order from the brightest pixel \citep{Law_et_al_2007}. For this latter parameter we  slightly modified the original definition proposed in \citet{Law_et_al_2007} to account for ellipsoidal light/mass distribution.
\end{enumerate}

\begin{figure*}
\begin{center}
\includegraphics[width=0.83\textwidth]{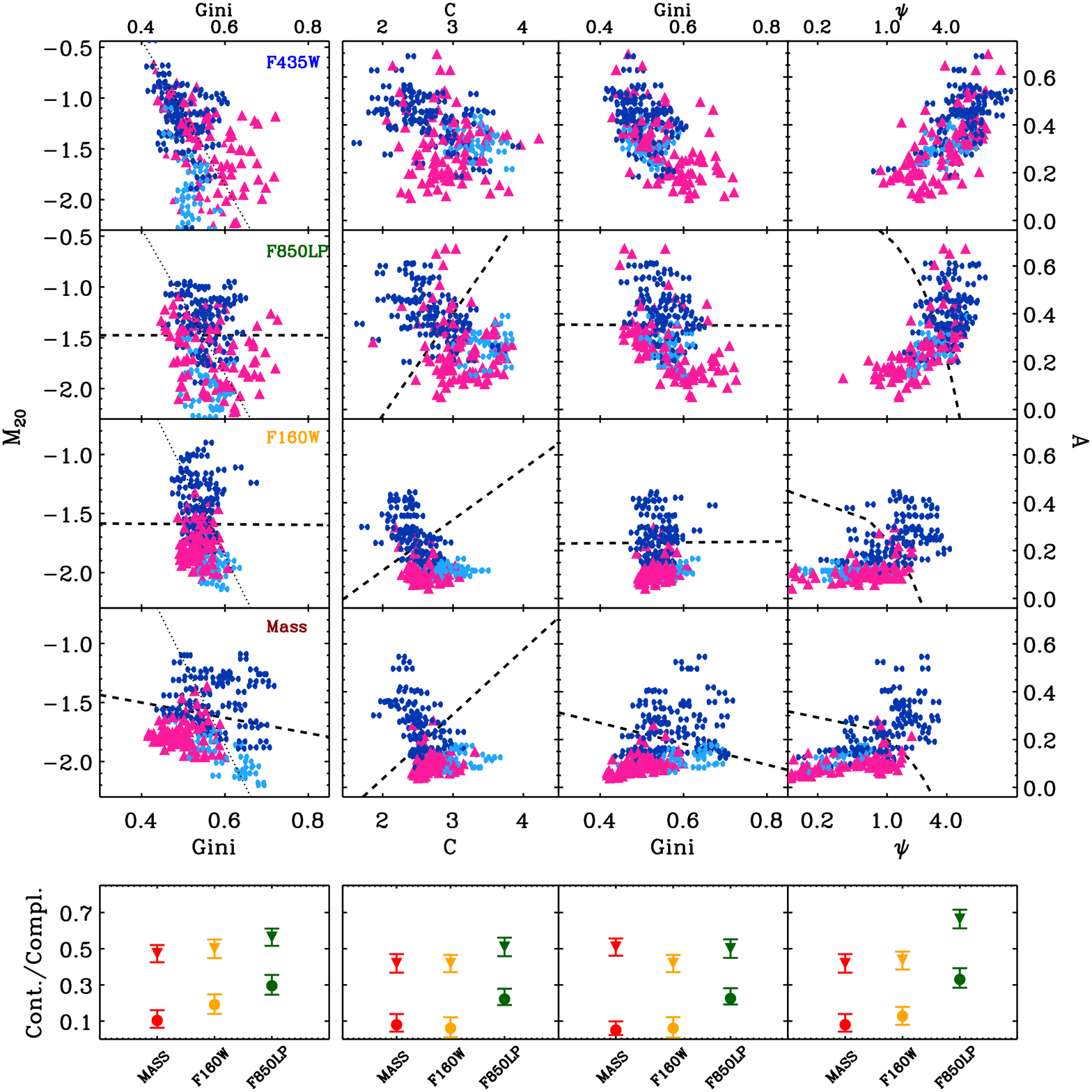}
\end{center}
\caption{\label{fig:f7_Appendix} As in Figure \ref{fig:f7} but this time also considering the structural parameters $C$, Gini and $\psi$. \emph{Top panel block:} Relation between different structural indices  for the sample of MIRAGE simulated mergers and isolated disks. From top to bottom measurements are performed on the artificial $HST$ F435W, F850LP, F160W images and on the mass map as indicated in the labels.
The dashed lines highlight the best dividing relation between mergers and disks (maximum margin classifier) obtained with a SVM approach.
 In the Gini-$M_{20}$ plane, we also show for reference the merger threshold of \citet{Lotz_et_al_2004} with a thin dotted line. Our results remain unchanged if using this relation instead of that derived from the SVM algorithm.
 \emph{Lower panel block}: 
For each of the selection criteria in the upper panels, we plot the contamination from isolated disks in the merger sample (snapshots falling above dotted line) and the completeness in the selected sample of mergers. All mergers independently of their ratio are here considered. Colors and symbols are as in Figure \ref{fig:f7}.}
\end{figure*}

We present in Figure \ref{fig:f7_Appendix}  the analogous of Figure \ref{fig:f7} but this time considering planes that include also $C$, Gini and $\psi$.
On one hand, the figure reinforces the results of Sections \ref{sec:SimulationInsight} and \ref{sec:MassLightRealGal}: even for these different combinations of structural indicators, the contamination in a mass-based selection of mergers is smaller than that obtained with measurements performed on the F160W and F850LP images.
On the other,  it justifies the choice of $M_{\rm 20, MASS}$--$A_{\rm MASS}$ as our fiducial combination: it can be noticed in fact that these other sets of indices result in either a higher contamination or lower completeness with respect to a classification based on $M_{\rm 20, MASS}$--$A_{\rm MASS}$ (see lower panels of Figure \ref{fig:f7_Appendix} and Figure \ref{fig:f7}). 
An exception is the Gini$_{\rm MASS}$--$A_{\rm MASS}$ combination which produces comparable results to those obtained for  $M_{\rm 20, MASS}$--$A_{\rm MASS}$. We show in the following that the Gini coefficient is however more sensitive to the choice of the aperture used in the calculation and the noise in the image than the $M_{20}$ or $A$ indices and hence a less stable structural measurement. 

\begin{figure*}[htbp]
\begin{center}
\includegraphics[width=0.95\textwidth]{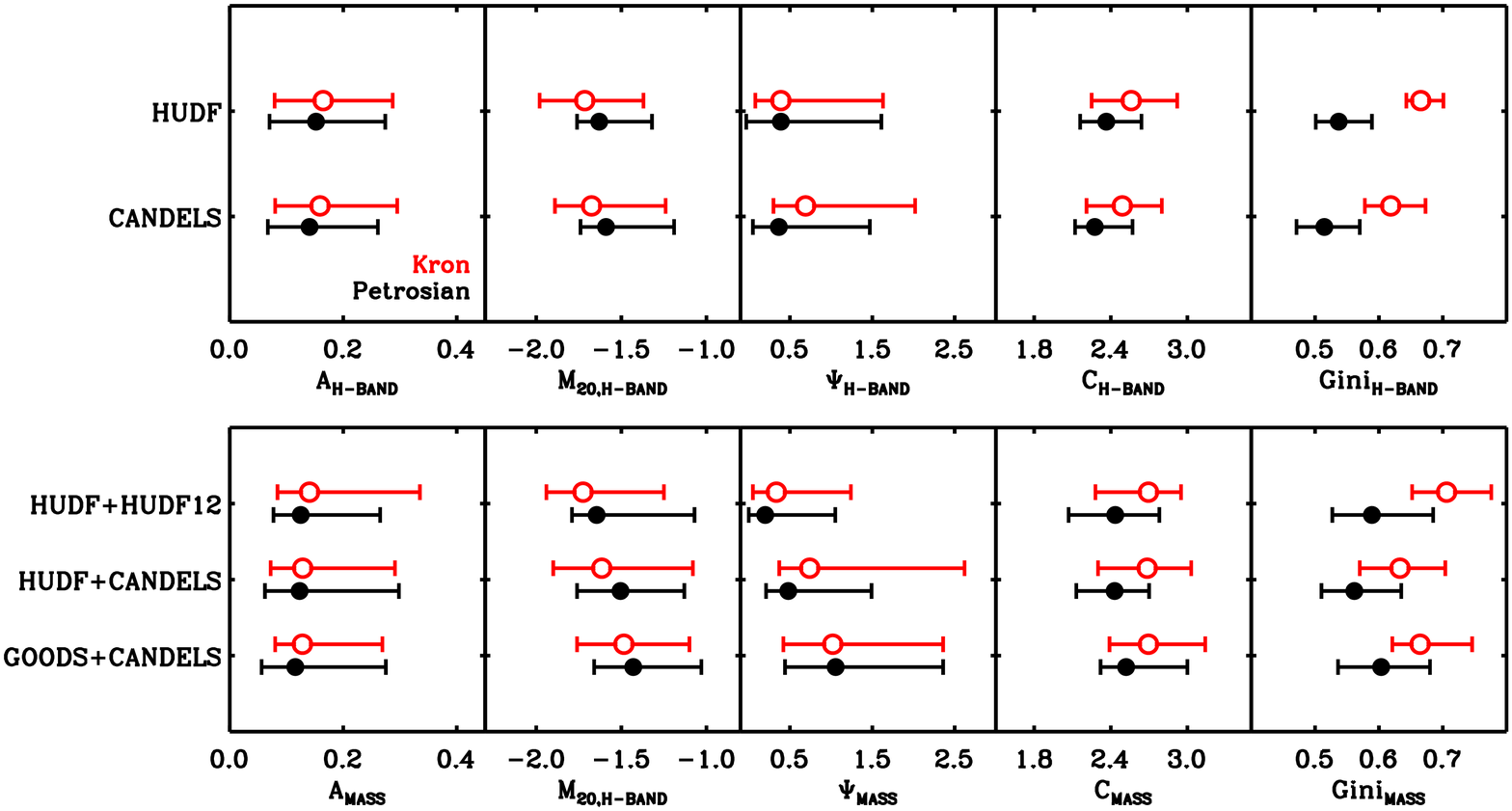}
\end{center}
\caption{\label{fig:Ap2} Comparison between the structural indices ($A$, $M_{20}$,  $\Psi$, $C$ and Gini coefficient) measured on images of different depths and within different apertures. Points in the figure correspond to the median values of the structural indices over all 50 galaxies that have  $r_{kron}>5\times$PSF , $H\leqslant$24.5 and  coverage in the HUDF, CANDELS as well as GOODS imaging.
The black and red points are the medians obtained within either a Petrosian or a Kron elliptical aperture, respectively. The error bars indicate the 16th and 84th percentiles of the distributions. \emph{Top panels:}  structural indices derived from the $H-$band images extracted from the HUDF  and CANDELS fields, as shown in the y-axis label. 
\emph{Bottom panels:} structural indices measured on the mass maps obtained from a combination of HUDF\,+\,HUDF12, HUDF\,+\,CANDELS-Deep and GOODS\,+\,CANDELS-Deep photometry, as shown in the y-axis label.}
\end{figure*}

This is illustrated in Figure \ref{fig:Ap2}  where study the variation of the structural indices with image depth and aperture size used in the calculation.
We only discuss the comparison for the structural indices measured  on the mass maps obtained from the combinations of HUDF\,+\,HUDF12, HUDF\,+\,CANDELS-Deep and GOODS\,+\,CANDELS-Deep photometry. Nonetheless, we also present the results for the indices derived on the $H-$band at HUDF or CANDELS depth for completeness of information.
Specifically, we considered  those galaxies (50) with $r_{kron}>5\times$PSF and $H\leqslant$24.5 that have coverage in the HUDF, CANDELS as well as GOODS fields and calculated the structural indices at all depths, within either a Kron or Petrosian aperture.
The use of two different apertures allows us to further test the impact of the noise in the images: the Kron aperture is usually larger than the Petrosian radius\footnote{We note that the Petrosian aperture considered here is the actual Petrosian radius, i.e., not the default \textsc{SExtractor} aperture equal to 2$\times R_{petrosian}$.} (see Table \ref{tab:SampleProp}) and hence includes a higher number of low flux pixels; differences between the two estimates can be used to assess the stability of the measurement. This is also the reason why we opted for a Petrosian aperture in the calculation of the parameters in Section \ref{sec:nonparam_indices}. 
The points in Figure \ref{fig:Ap2} show the median values of the structural indices over all the 50 galaxies here considered.

A reliable measure of structure should show as little variation as possible with both imaging depth and aperture size in Figure \ref{fig:Ap2}.
 The asymmetry $A_{\rm MASS}$  is in this sense very robust with median values that remain almost identical at all depths and for both apertures.
This should be compared, for example, with the results for the Gini coefficient in the right-most column which clearly shows a strong dependence on the aperture size with differences up to about 30\%.
As already pointed out in the work of \citet{Lisker_2008}, this is a consequence of the fact that Gini becomes a tracer of the noise in the images rather than galactic structure for large apertures. 
The other parameters, $M_{\rm 20, MASS}$, $\Psi_{\rm MASS}$ and $C_{\rm MASS}$, all show some dependence on the depth of the images used for constructing the maps, shifting towards lower concentrations/higher
clumpiness when using the GOODS\,+\,CANDELS-Deep photometry instead of the deep HUDF\,+\,HUDF12 imaging. The largest variations are measured for the multiplicity $\Psi_{\rm MASS}$ (more than 50\% change among the various cases), whereas the $M_{\rm 20, MASS}$ coefficient varies by $\leqslant 15\%$.

In combination with Figure \ref{fig:f7}, these findings justify our choice of  $A_{\rm MASS}$- $M_{\rm 20, MASS}$ in Section \ref{sec:TestonSims1} as the most powerful and less noise- or aperture-dependent parameter set for the discrimination of merging and clumpy galaxies.


\clearpage
\LongTables
\begin{deluxetable*}{cccccccccccc}
\tablewidth{0pt}
\tabletypesize{\small}
\tablecaption{Classification of HUDF galaxies\label{tab:SampleProp}}
\tablehead{
\colhead{ID} & \colhead{BzH} & \colhead{H-band} &\colhead{Mass}  & \colhead{RA[J2000]} & \colhead{DEC[J2000]} & \colhead{z} & \colhead{$M_{\rm 20, MASS}$} & \colhead{e\_$M_{\rm 20, MASS}$} & \colhead{$A_{\rm MASS}$} & \colhead{e\_$A_{\rm MASS}$} & \colhead{Class} }
\startdata
 8740 &
 \includegraphics[width=1cm,height=1cm]{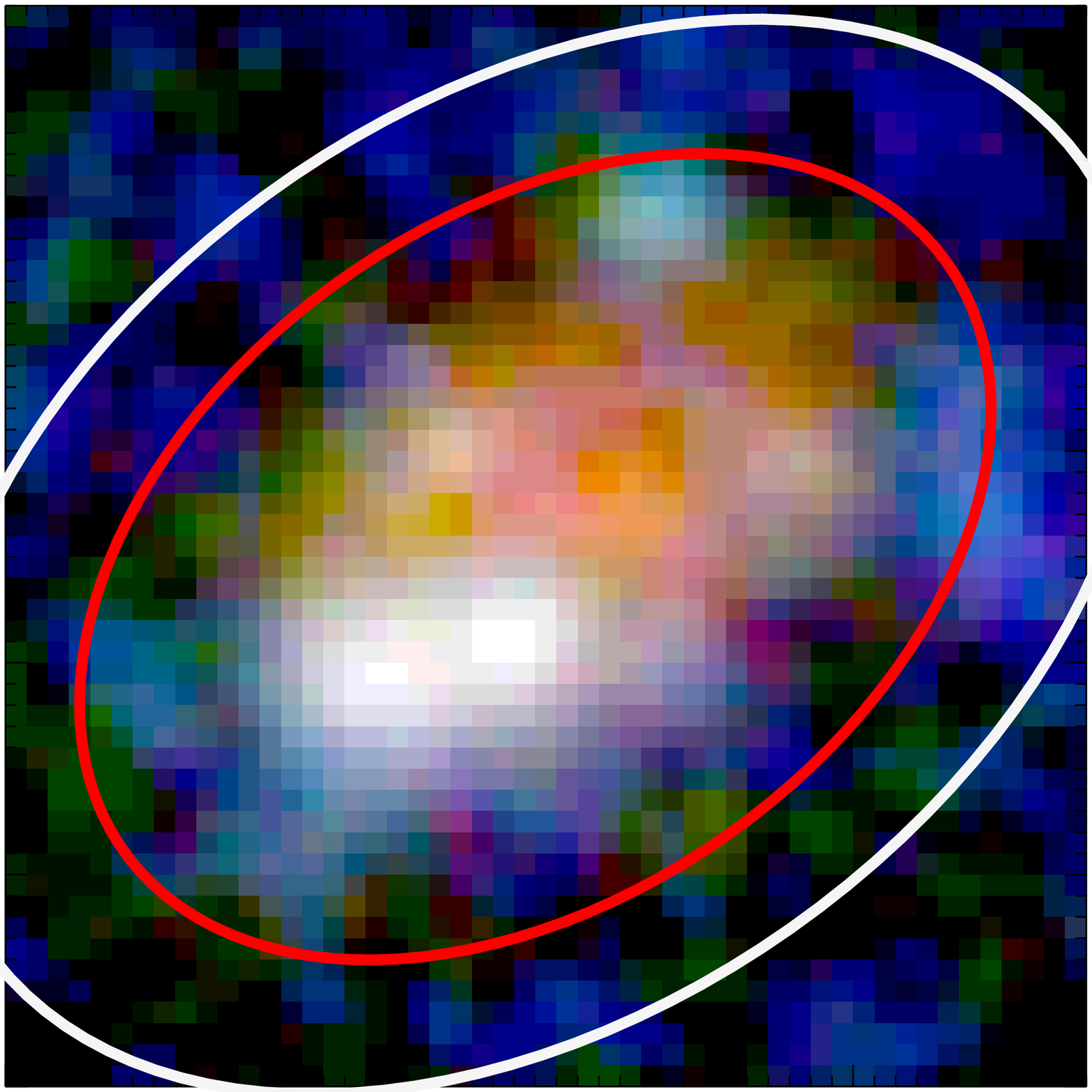} &
\includegraphics[width=1cm,height=1cm]{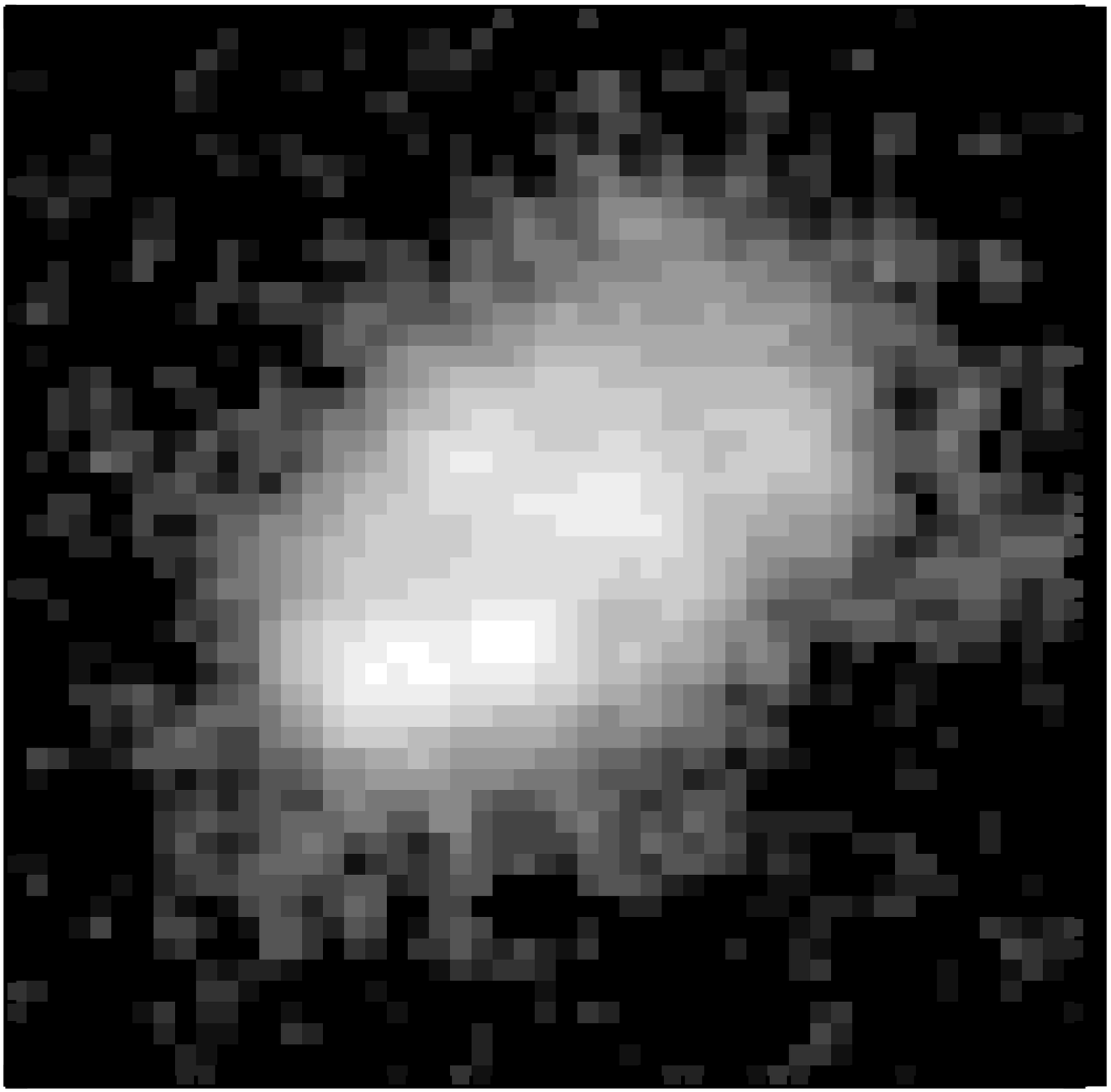} &
\includegraphics[width=1cm,height=1cm]{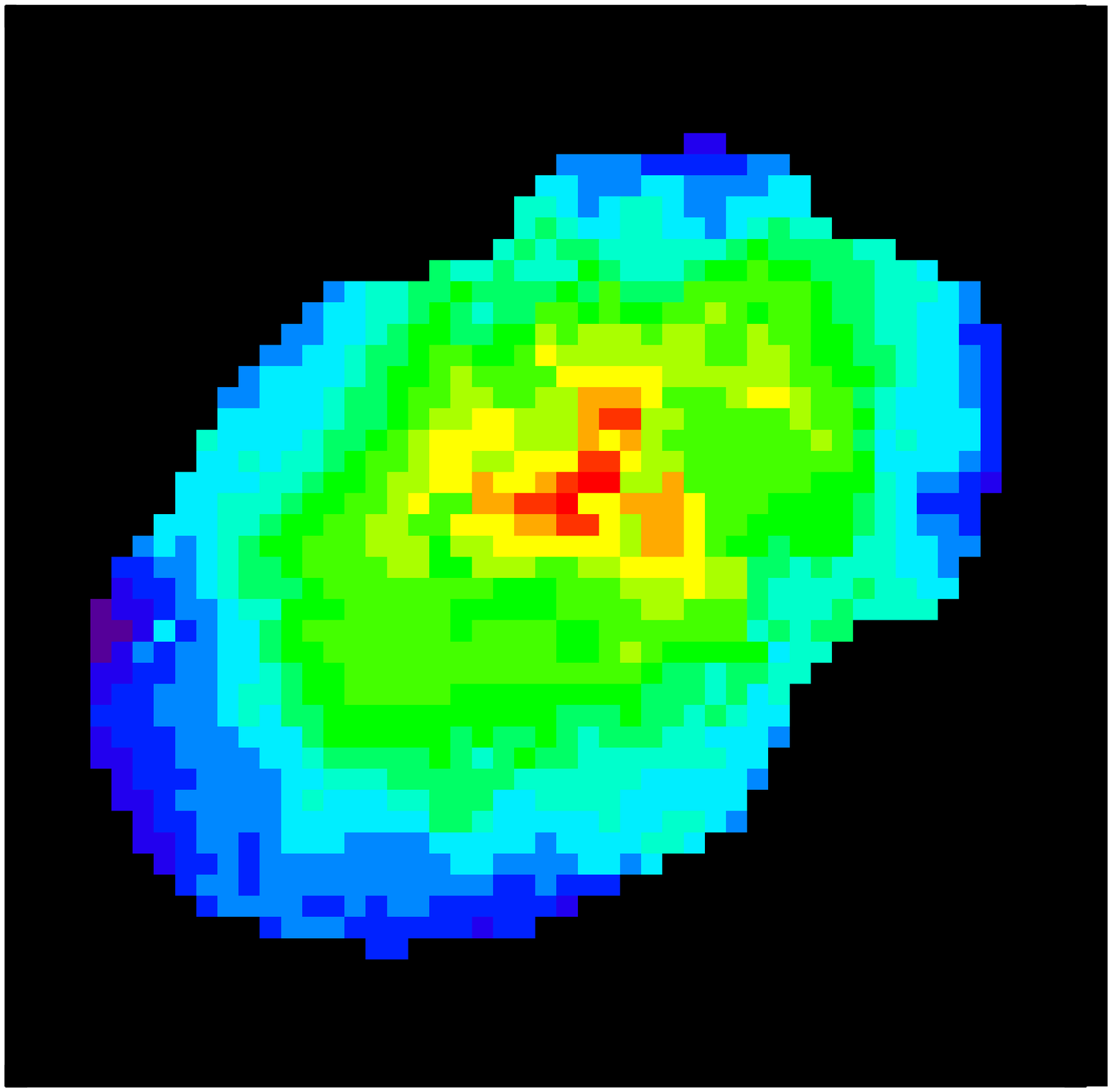} &
   53.16764 &  -27.83037 & 1.88  & -1.670 &  0.065 & 0.114 & 0.010 & Not merging  \\
 8750 &
 \includegraphics[width=1cm,height=1cm]{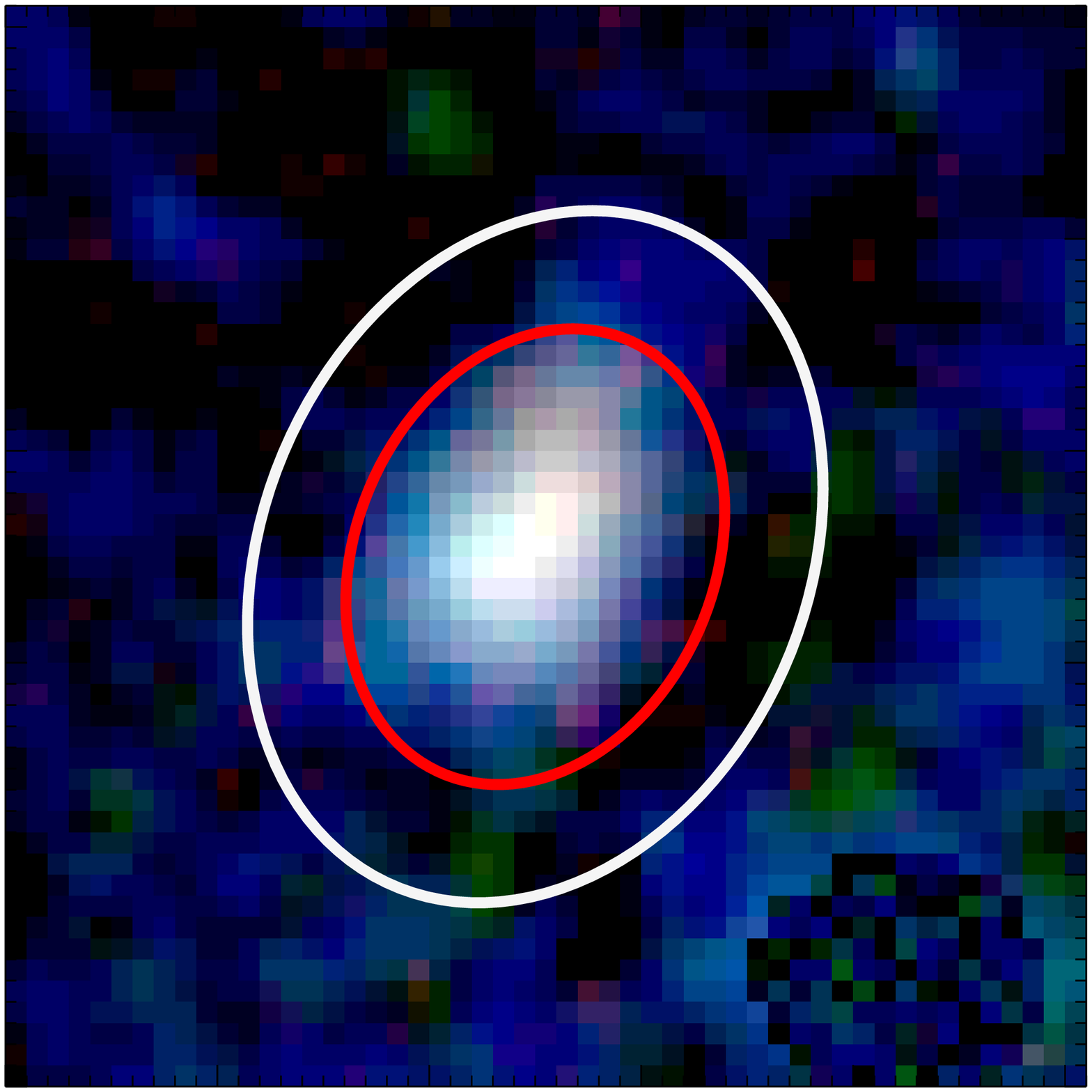} &
\includegraphics[width=1cm,height=1cm]{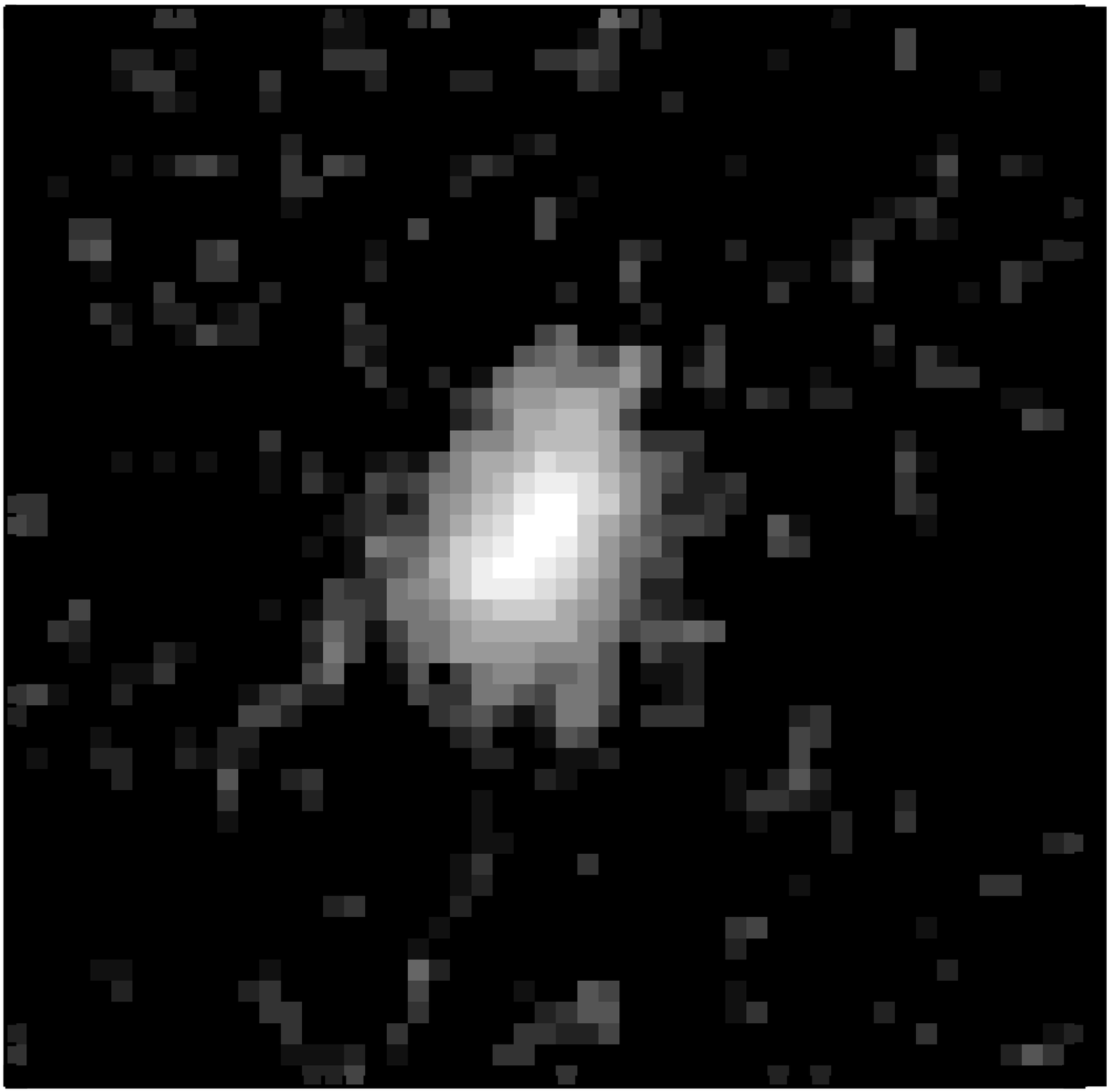} &
\includegraphics[width=1cm,height=1cm]{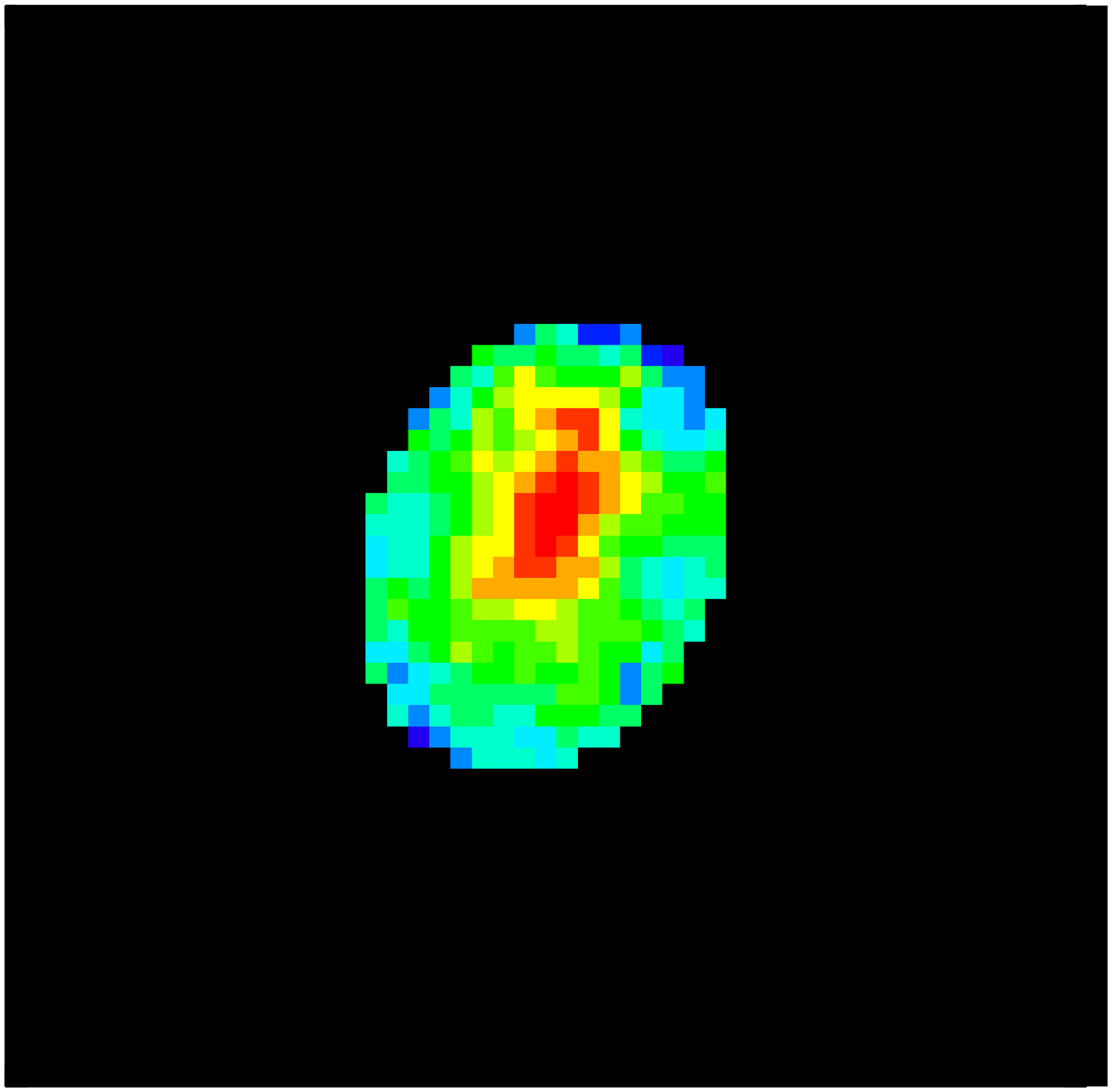} &
   53.16287 &  -27.82947 & 2.04  & -1.220 &  0.125 & 0.125 & 0.040 & Unres./Faint \\
 9295 &
 \includegraphics[width=1cm,height=1cm]{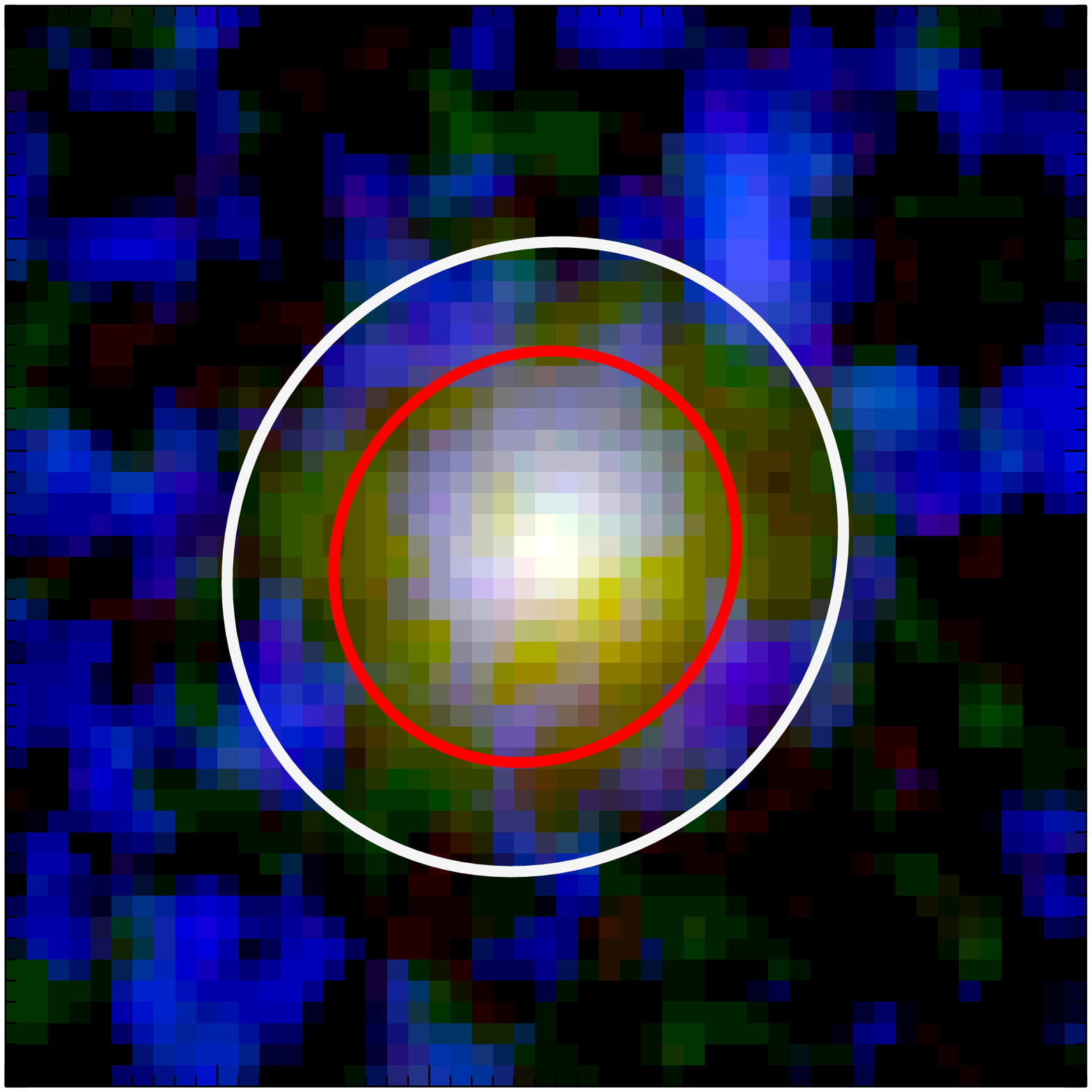} &
\includegraphics[width=1cm,height=1cm]{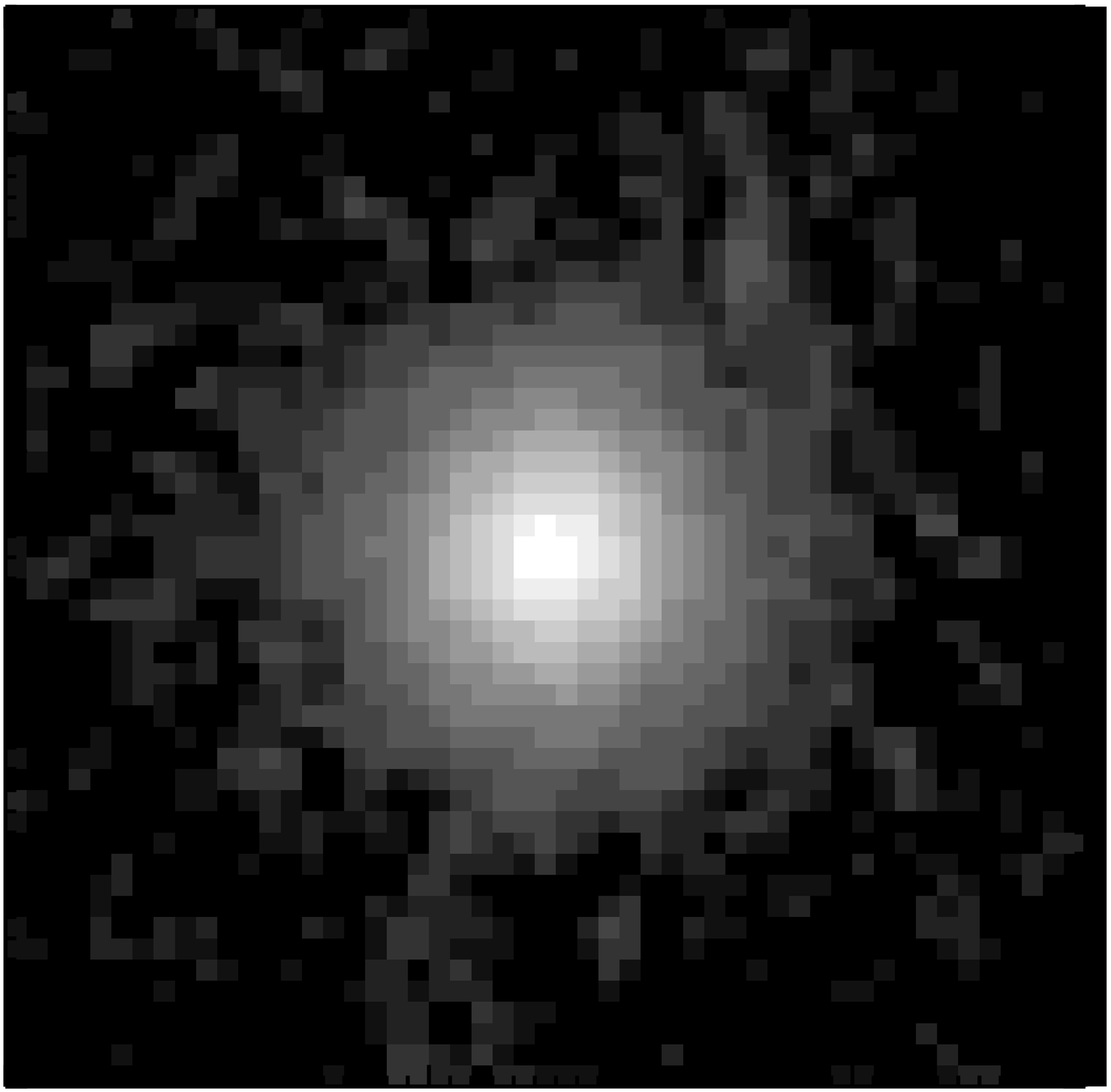} &
\includegraphics[width=1cm,height=1cm]{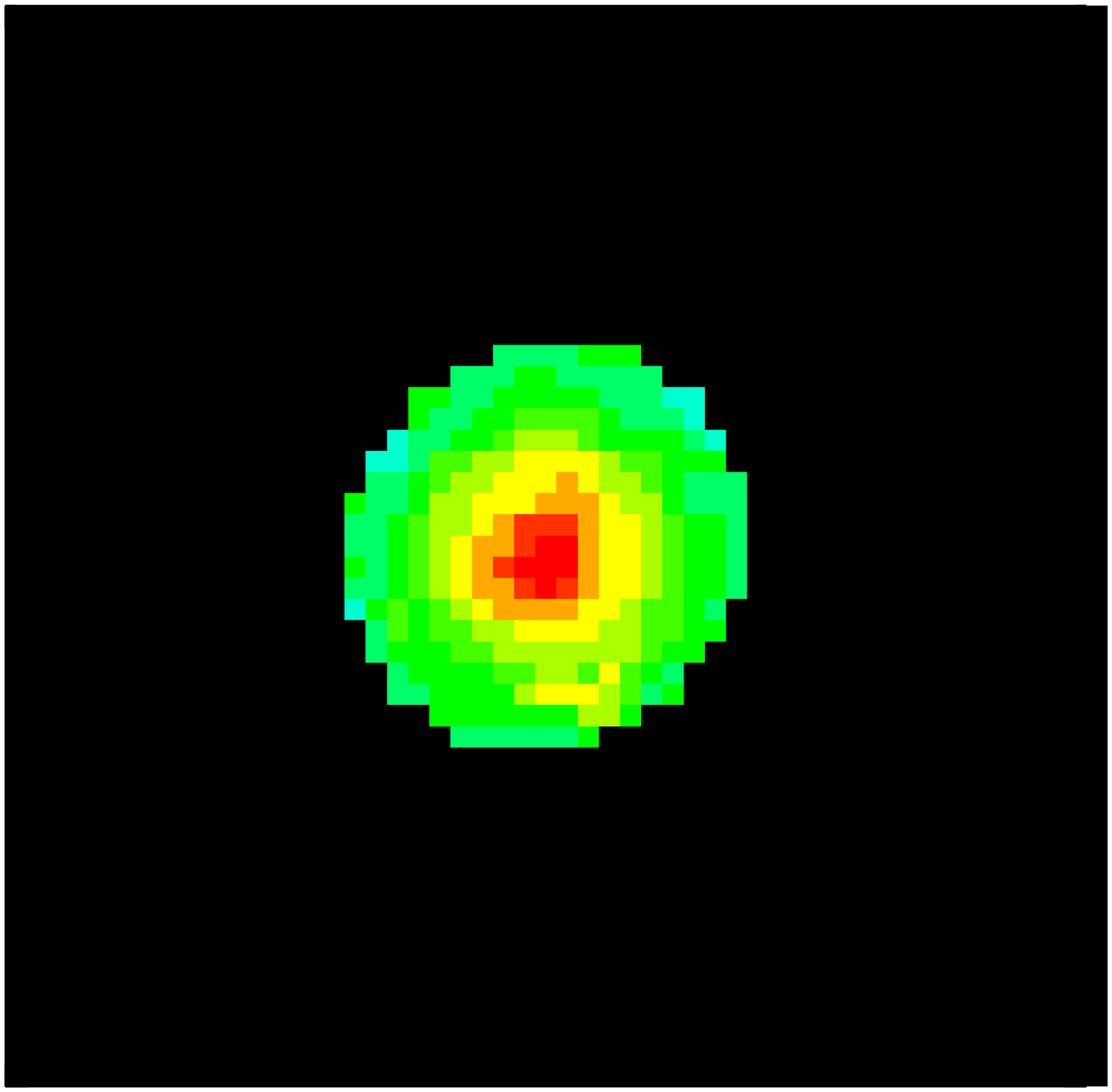} &
   53.17170 &  -27.82566 & 1.73  & -1.850 &  0.070 & 0.096 & 0.010 & Not merging  \\
 9343 &
 \includegraphics[width=1cm,height=1cm]{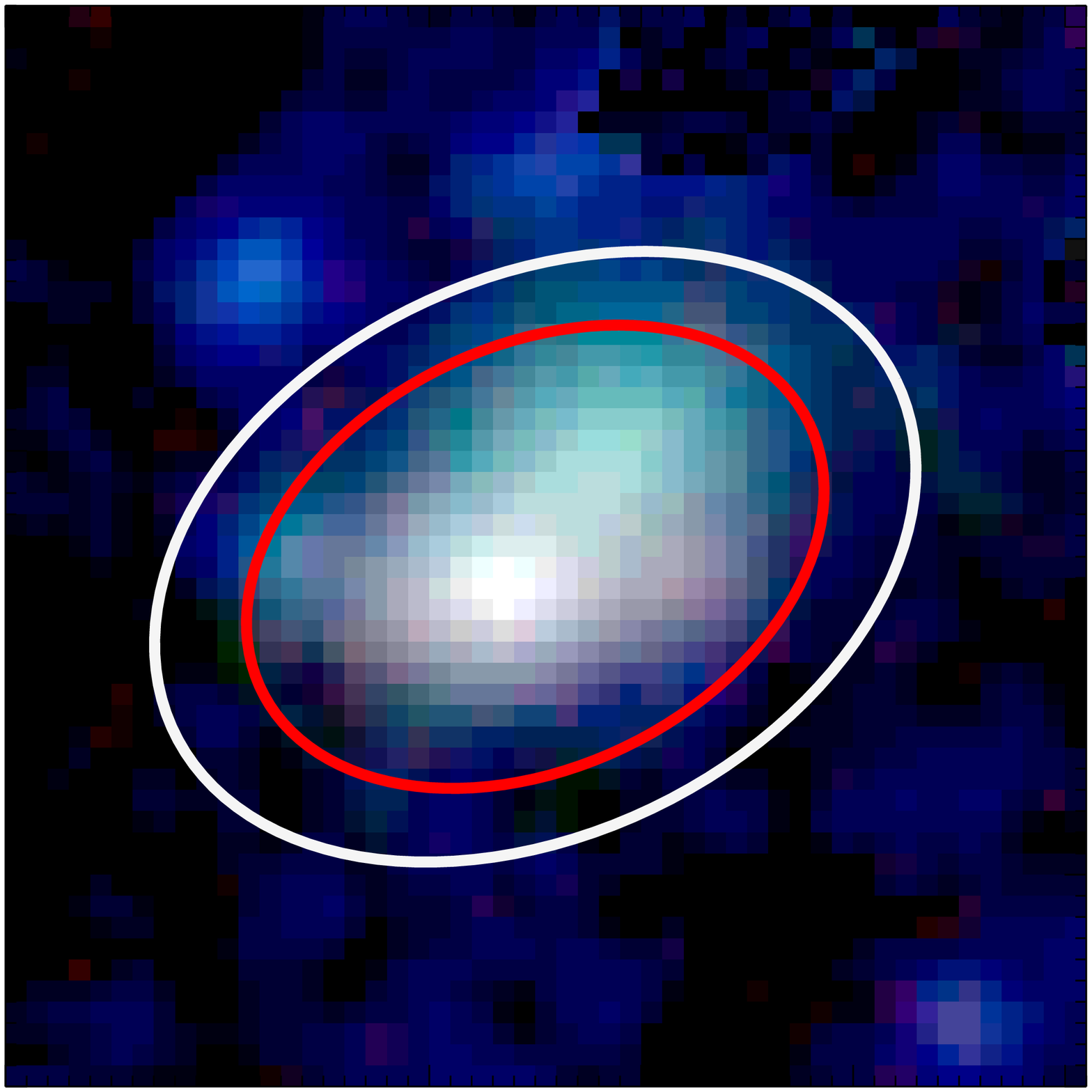} &
\includegraphics[width=1cm,height=1cm]{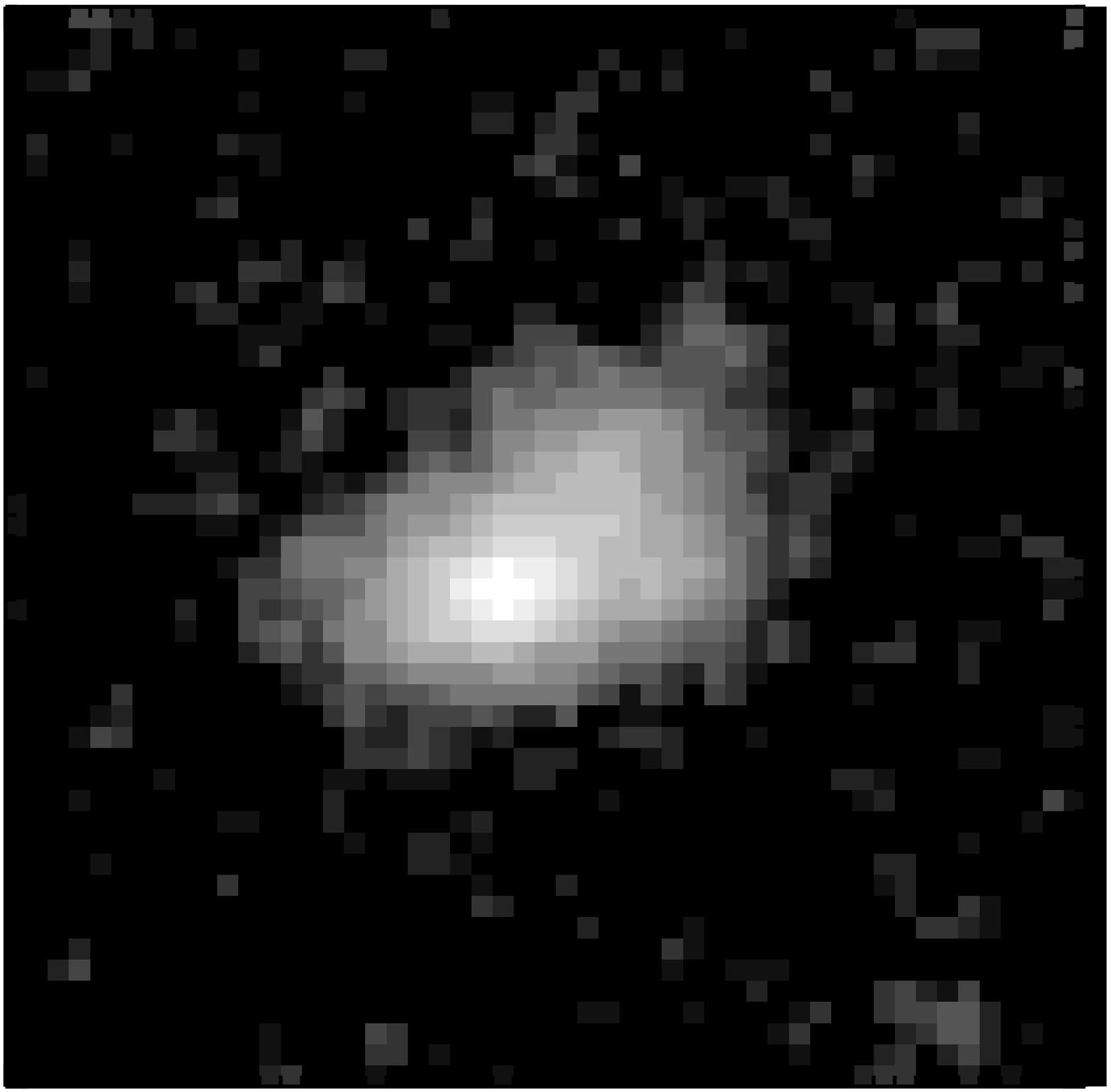} &
\includegraphics[width=1cm,height=1cm]{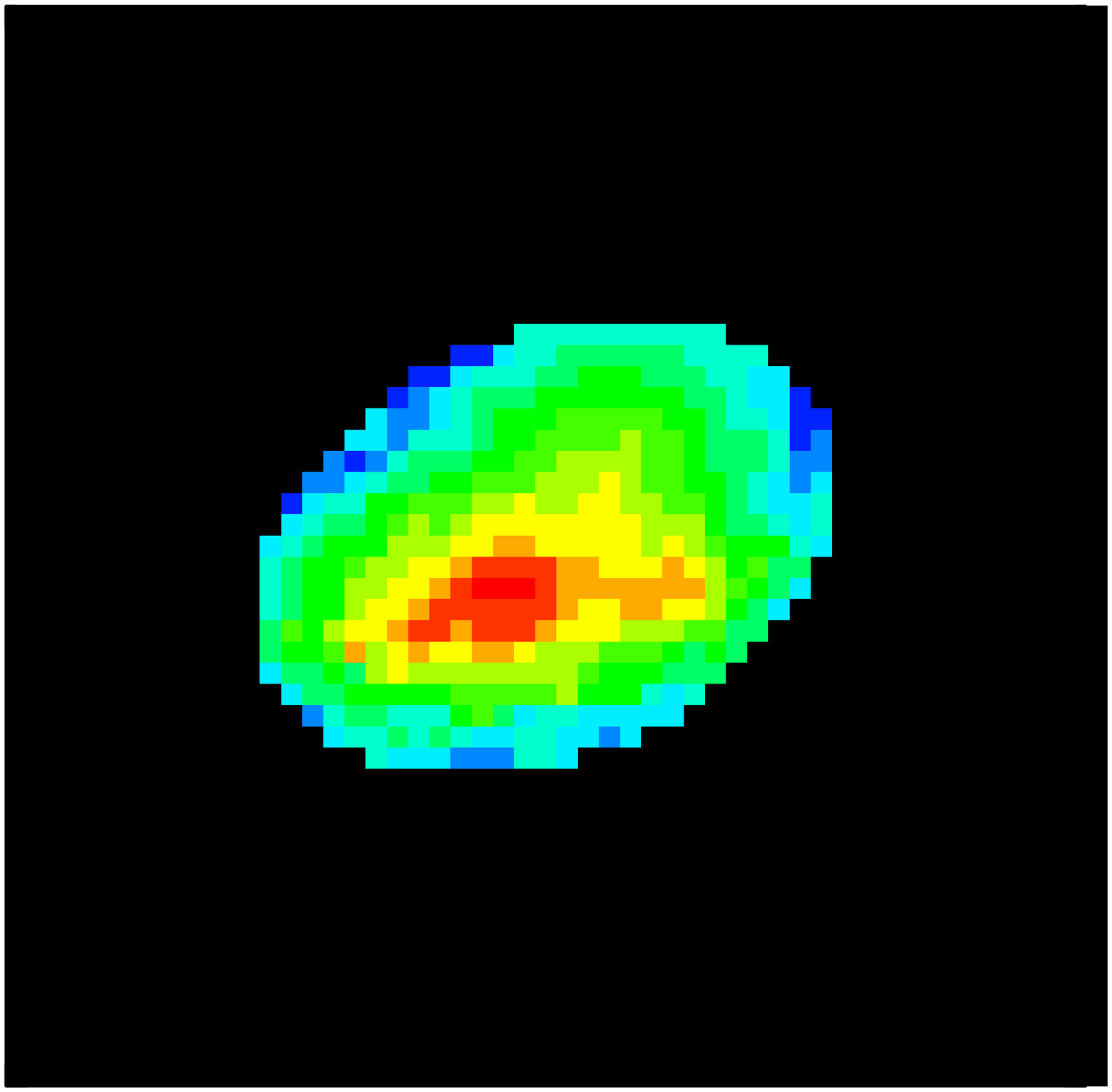} &
   53.16714 &  -27.82450 & 1.84  & -1.210 &  0.095 & 0.160 & 0.014 & Not merging  \\
 9407 &
 \includegraphics[width=1cm,height=1cm]{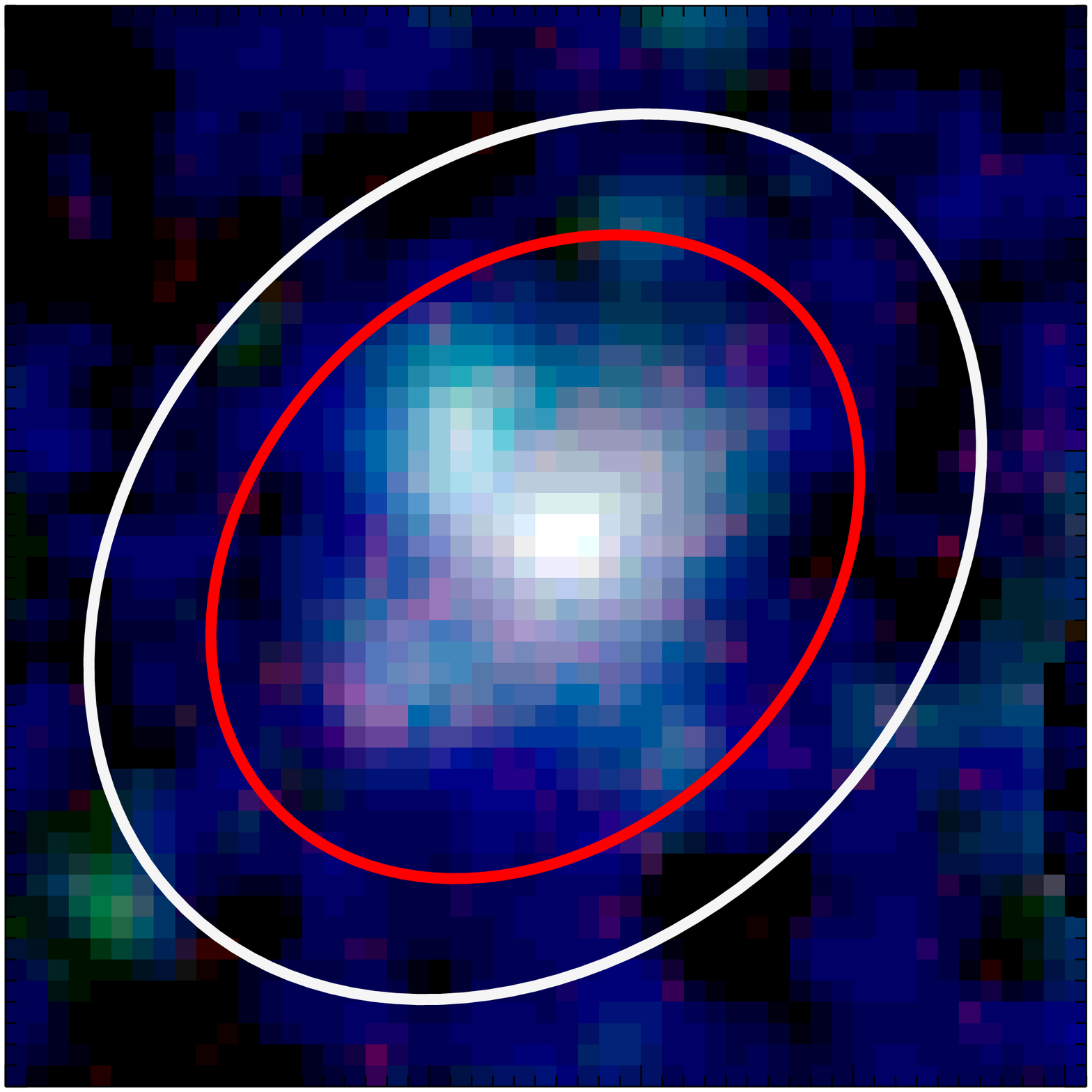} &
\includegraphics[width=1cm,height=1cm]{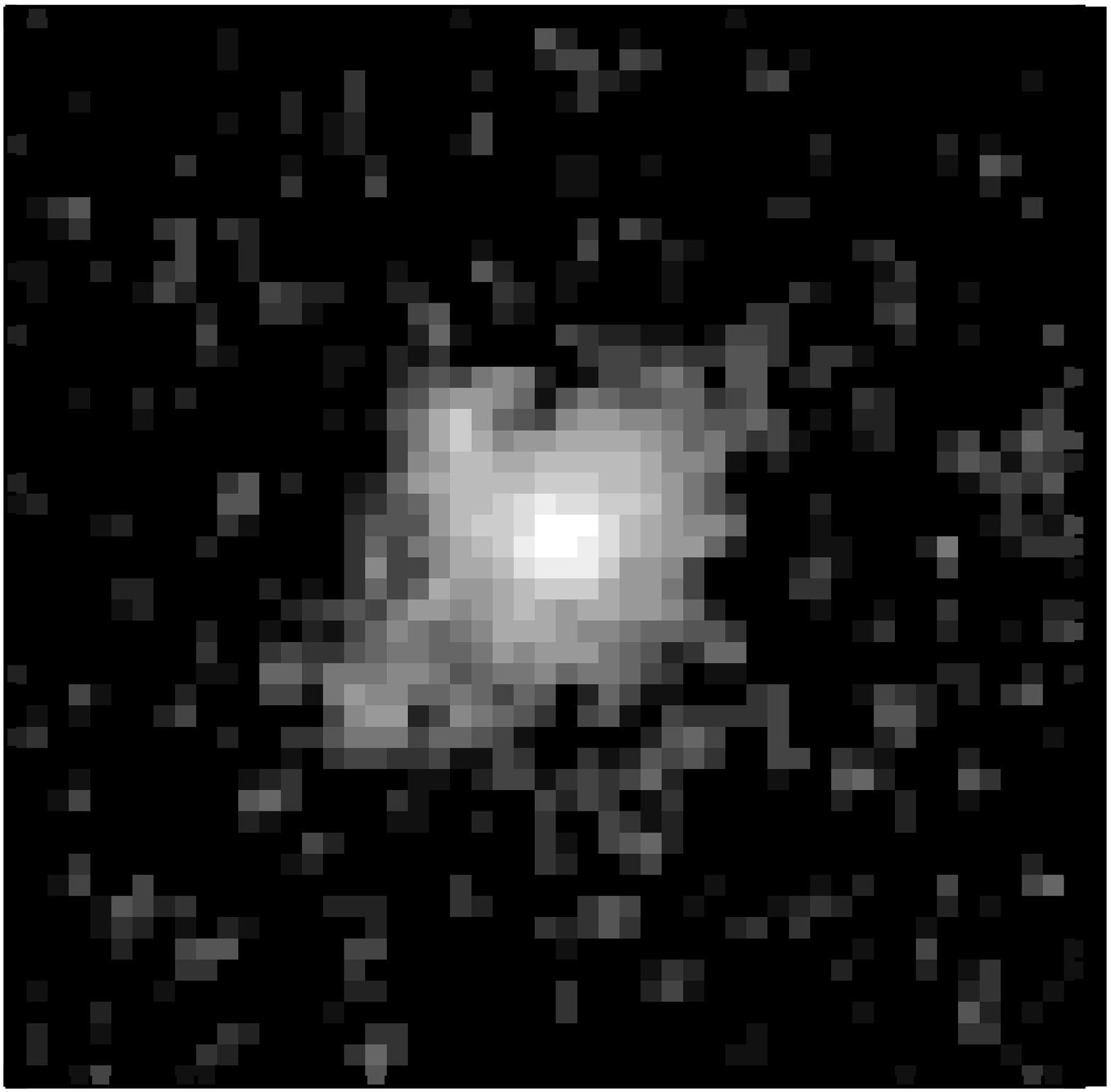} &
\includegraphics[width=1cm,height=1cm]{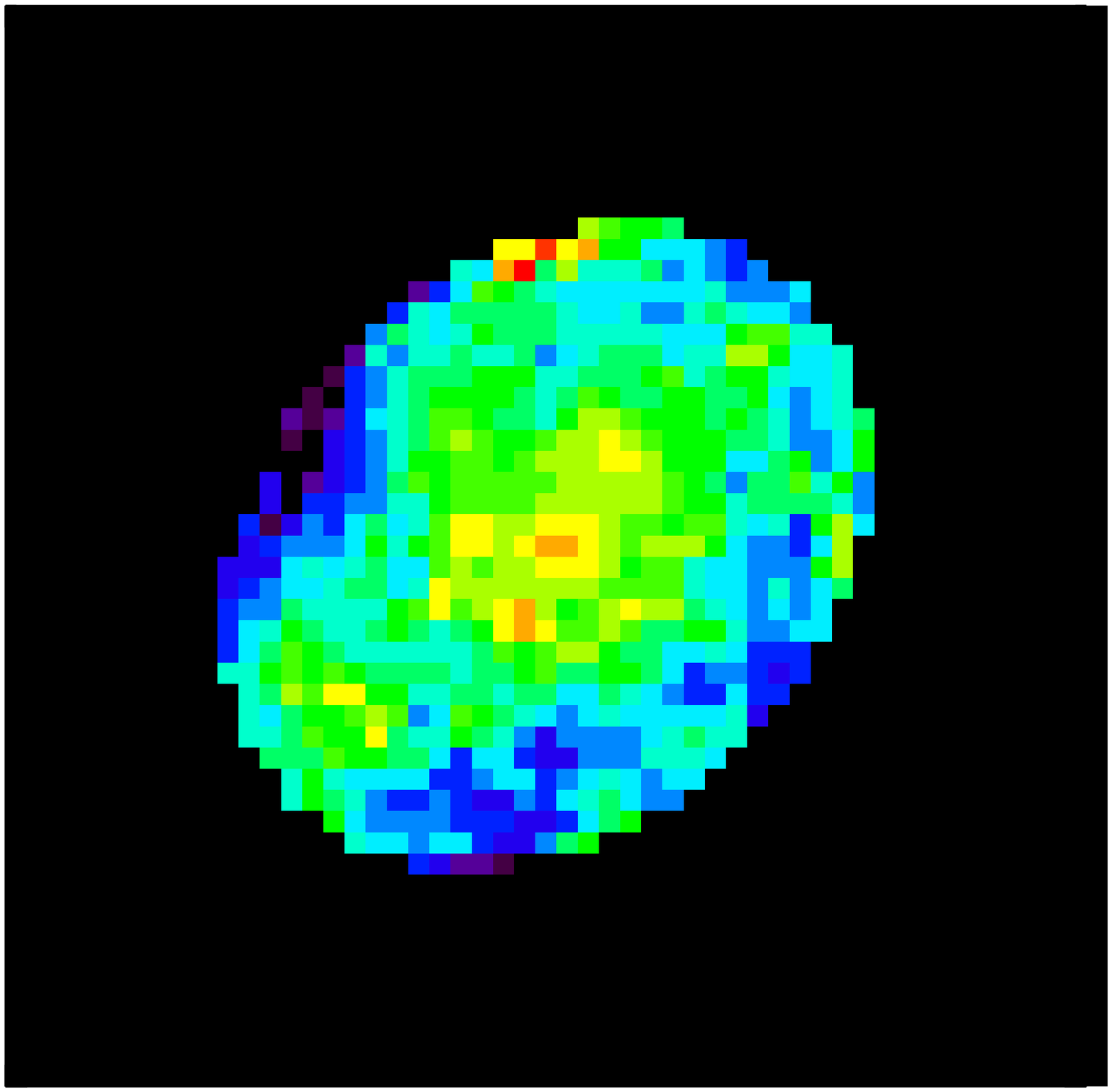} &
   53.17061 &  -27.82379 & 2.69$^{s}$ & -0.610 &  0.070 & 0.460 & 0.186 & Merger (2)   \\
 9474 &
 \includegraphics[width=1cm,height=1cm]{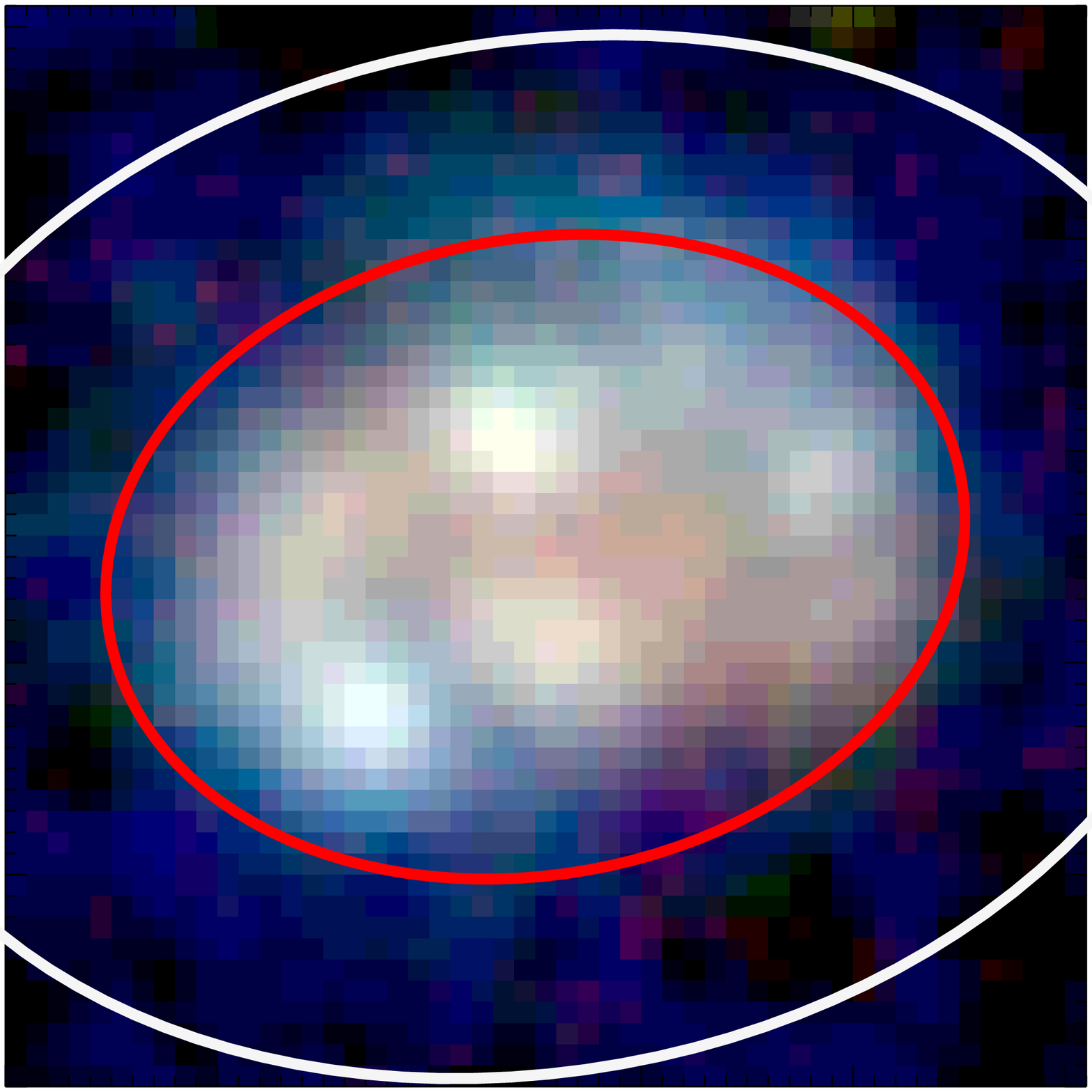} &
\includegraphics[width=1cm,height=1cm]{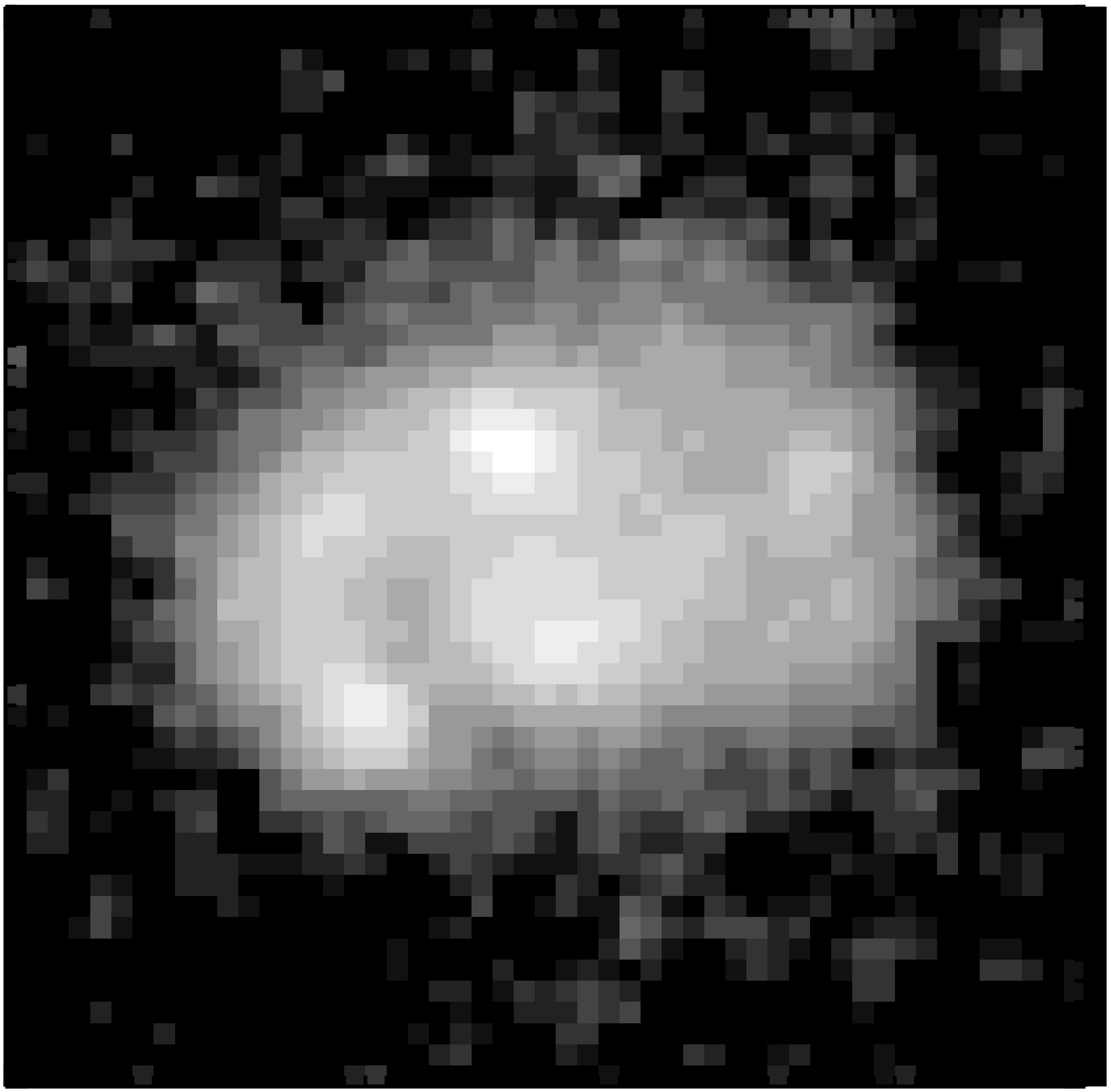} &
\includegraphics[width=1cm,height=1cm]{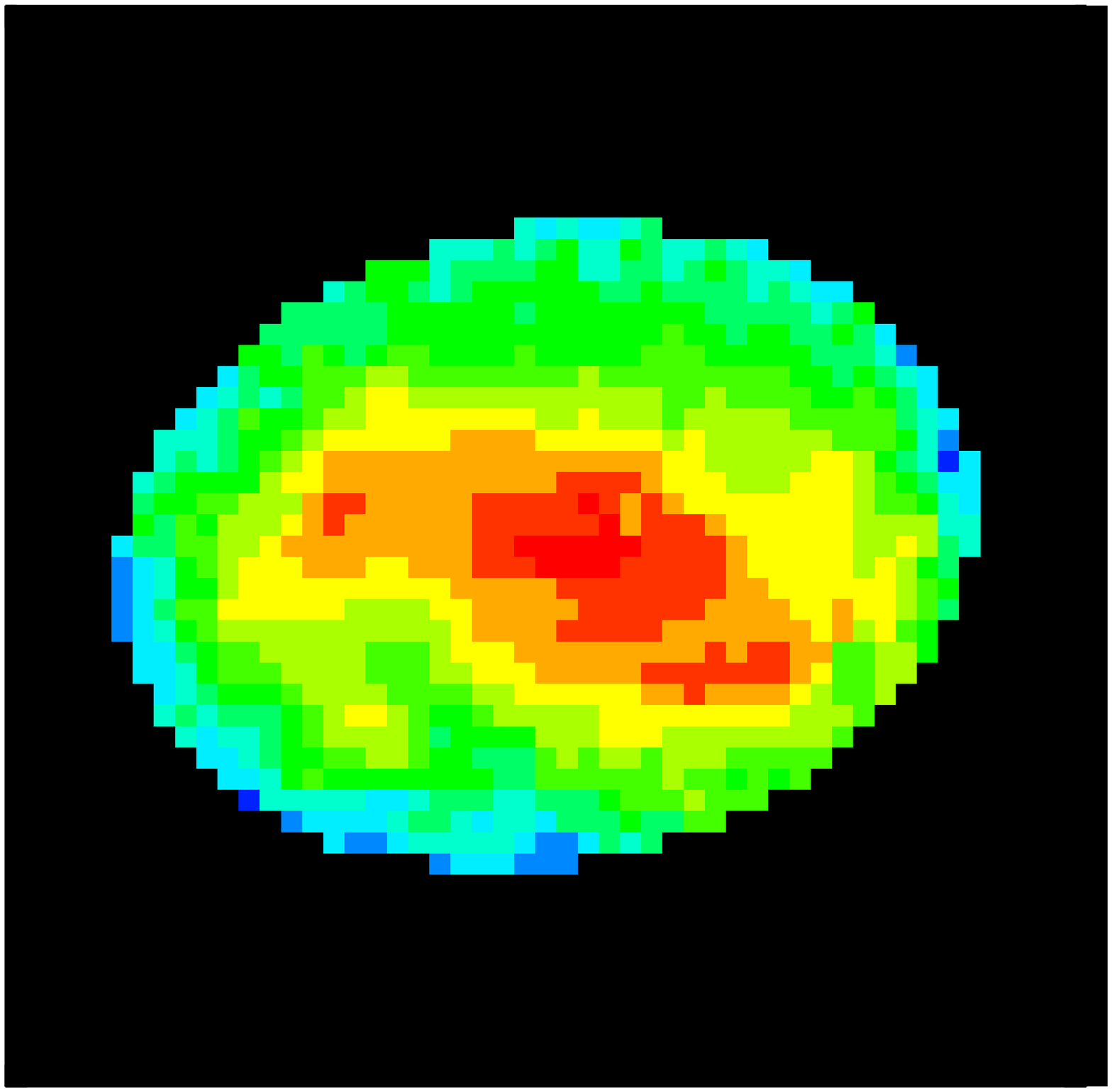} &
   53.16978 &  -27.82394 & 2.24  & -1.260 &  0.055 & 0.112 & 0.021 & Not merging  \\
 9527 &
 \includegraphics[width=1cm,height=1cm]{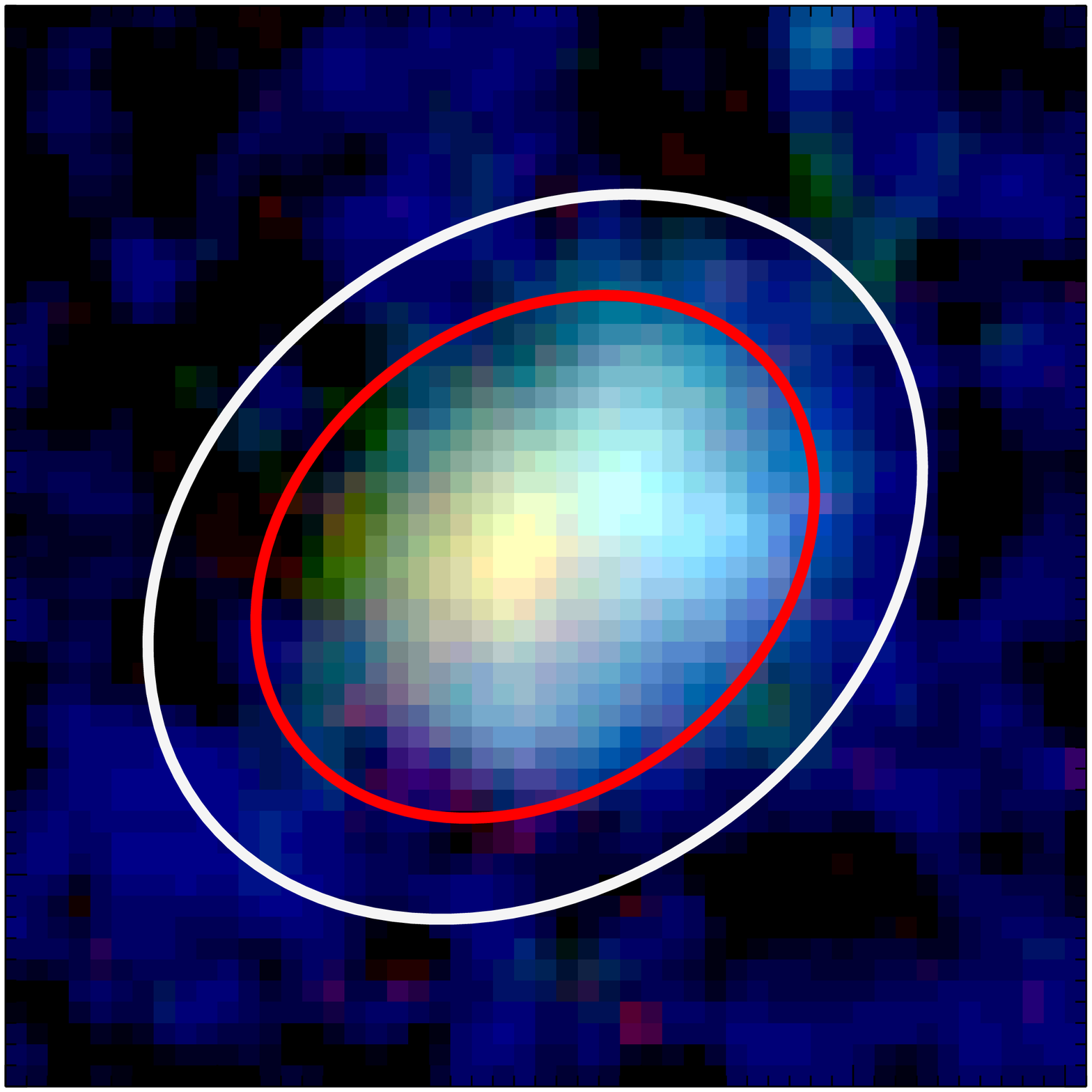} &
\includegraphics[width=1cm,height=1cm]{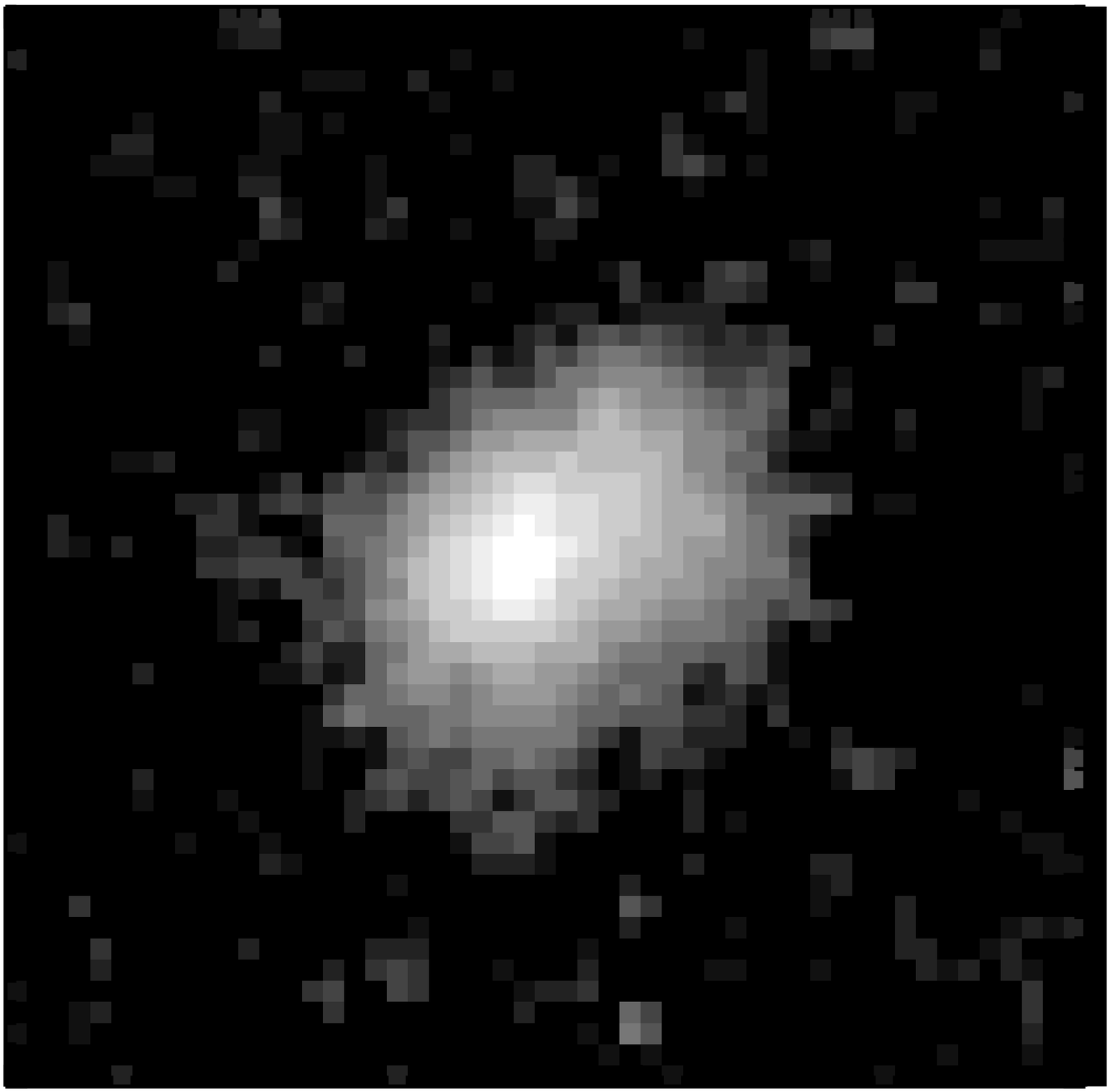} &
\includegraphics[width=1cm,height=1cm]{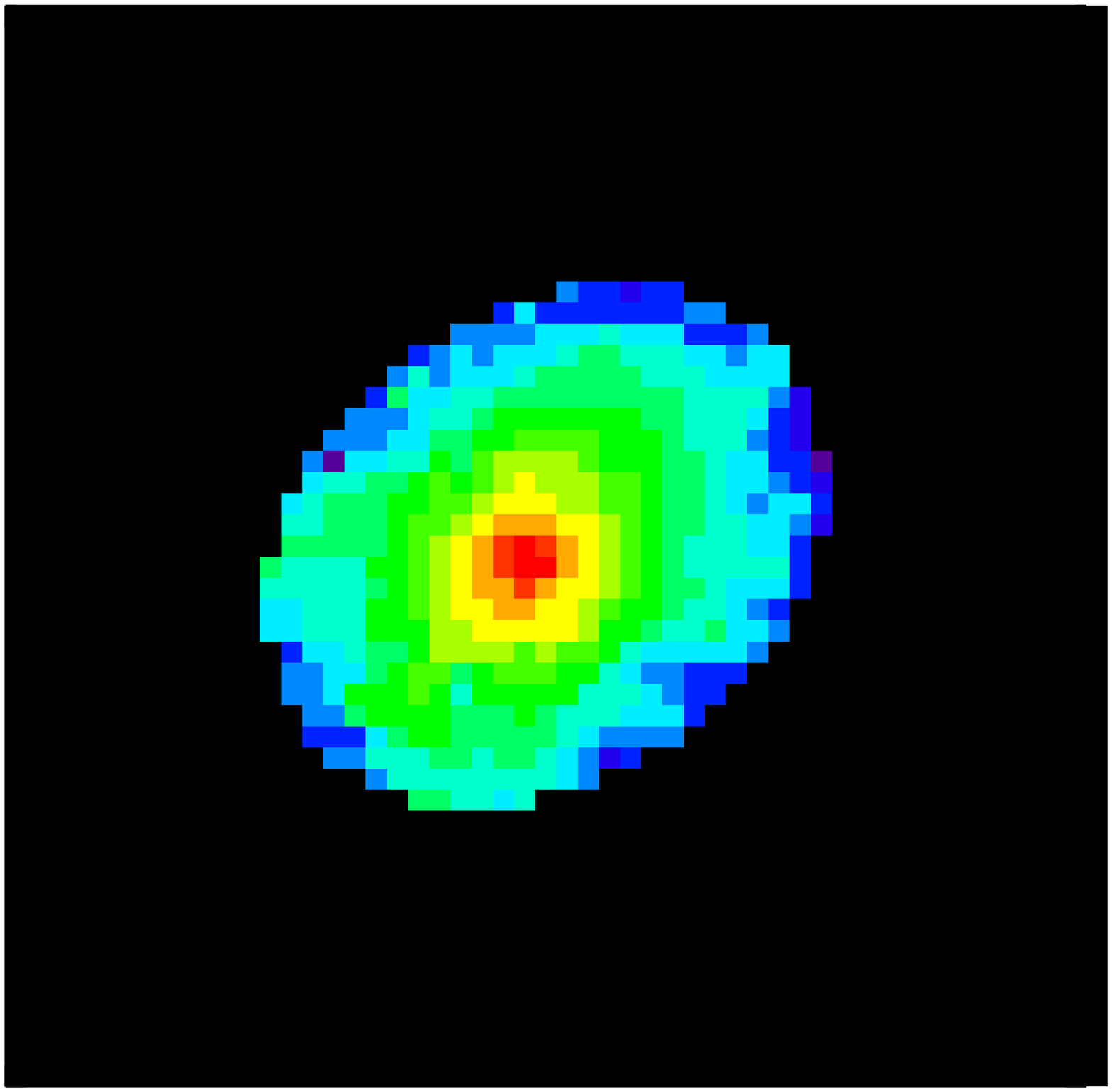} &
   53.15673 &  -27.82306 & 1.72  & -1.950 &  0.085 & 0.036 & 0.013 & Not merging  \\
 9835 &
 \includegraphics[width=1cm,height=1cm]{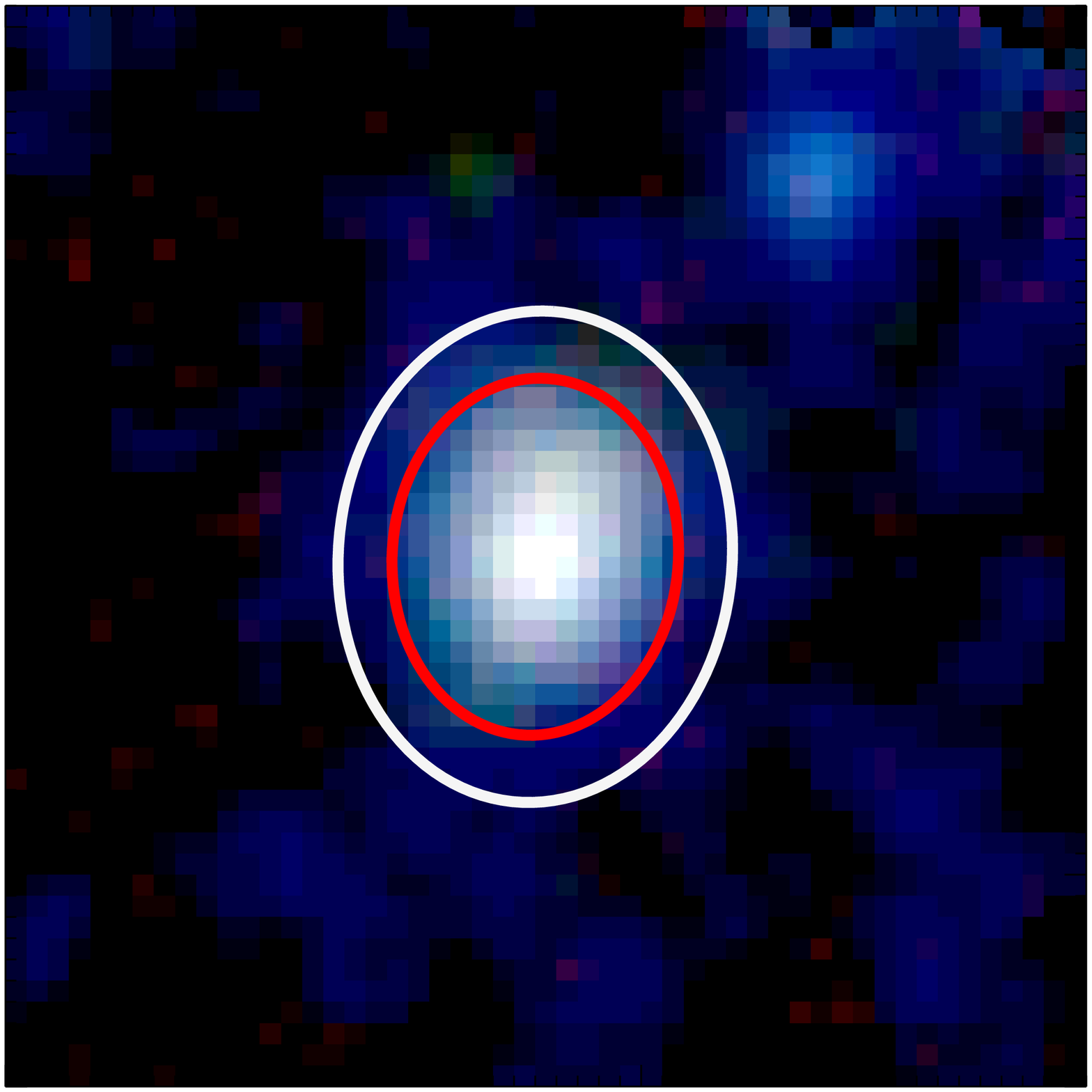} &
\includegraphics[width=1cm,height=1cm]{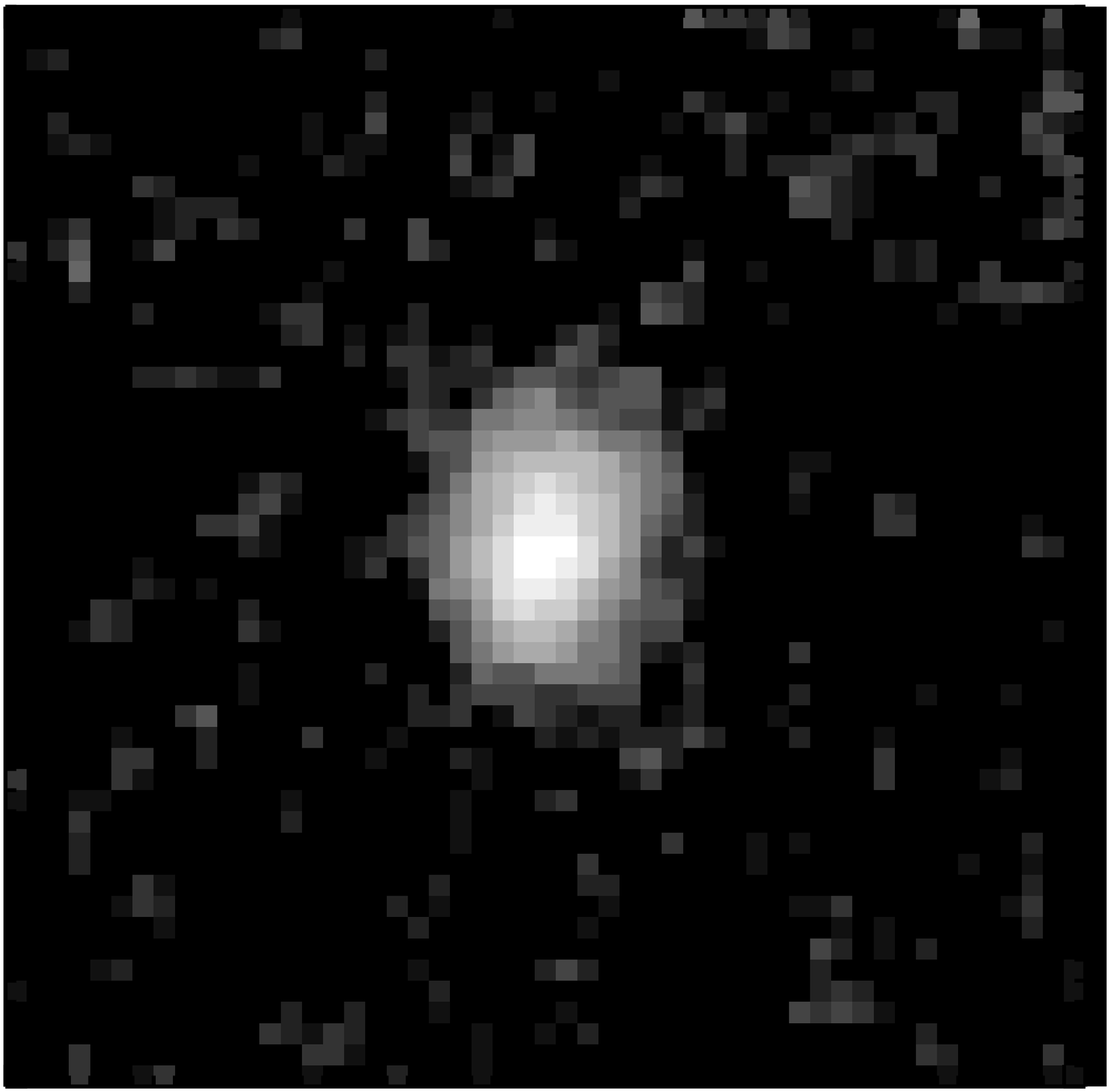} &
\includegraphics[width=1cm,height=1cm]{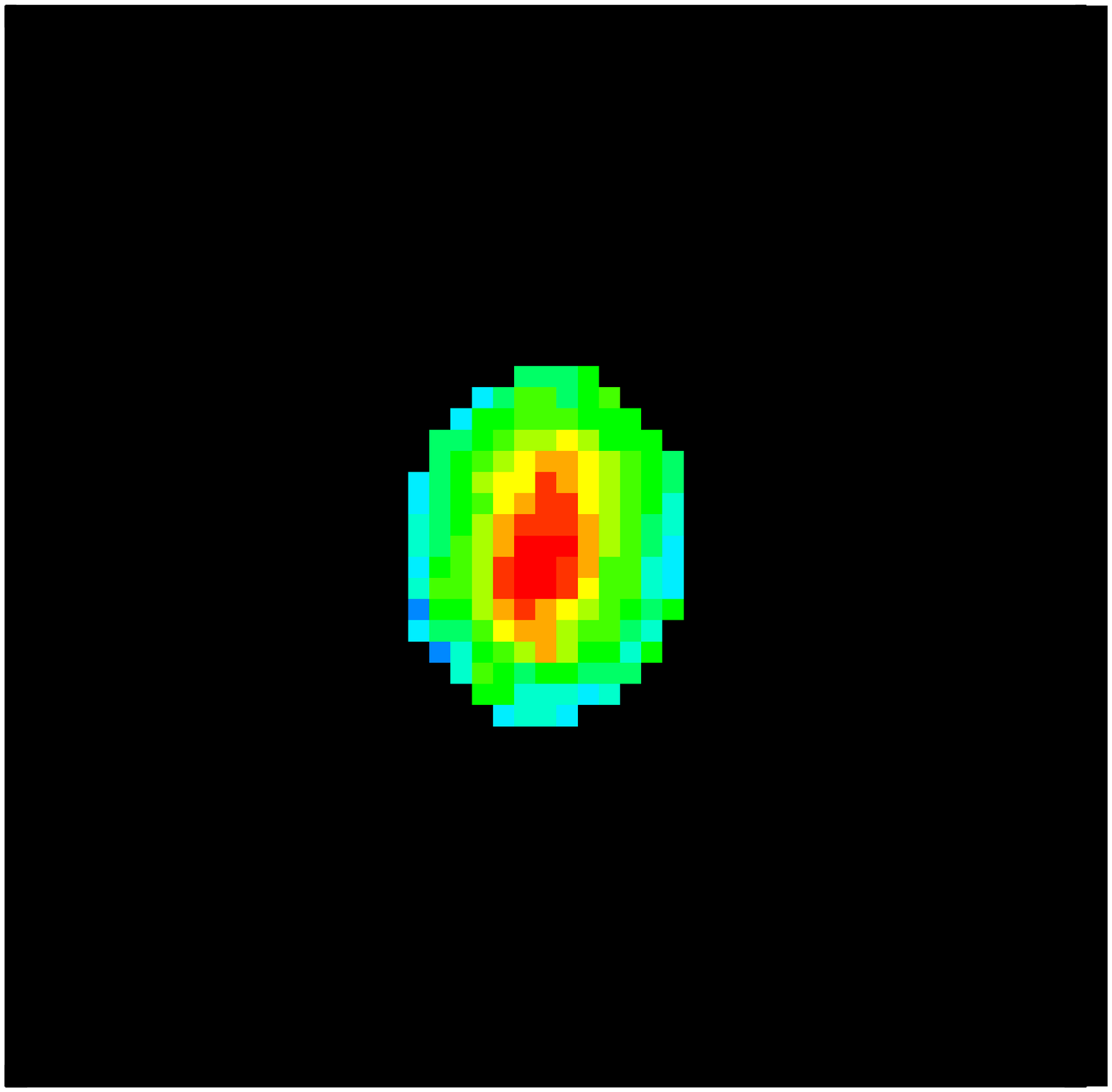} &
   53.17343 &  -27.82028 & 2.28  & -1.320 &  0.105 & 0.036 & 0.078 & Unres./Faint \\
 9987 &
 \includegraphics[width=1cm,height=1cm]{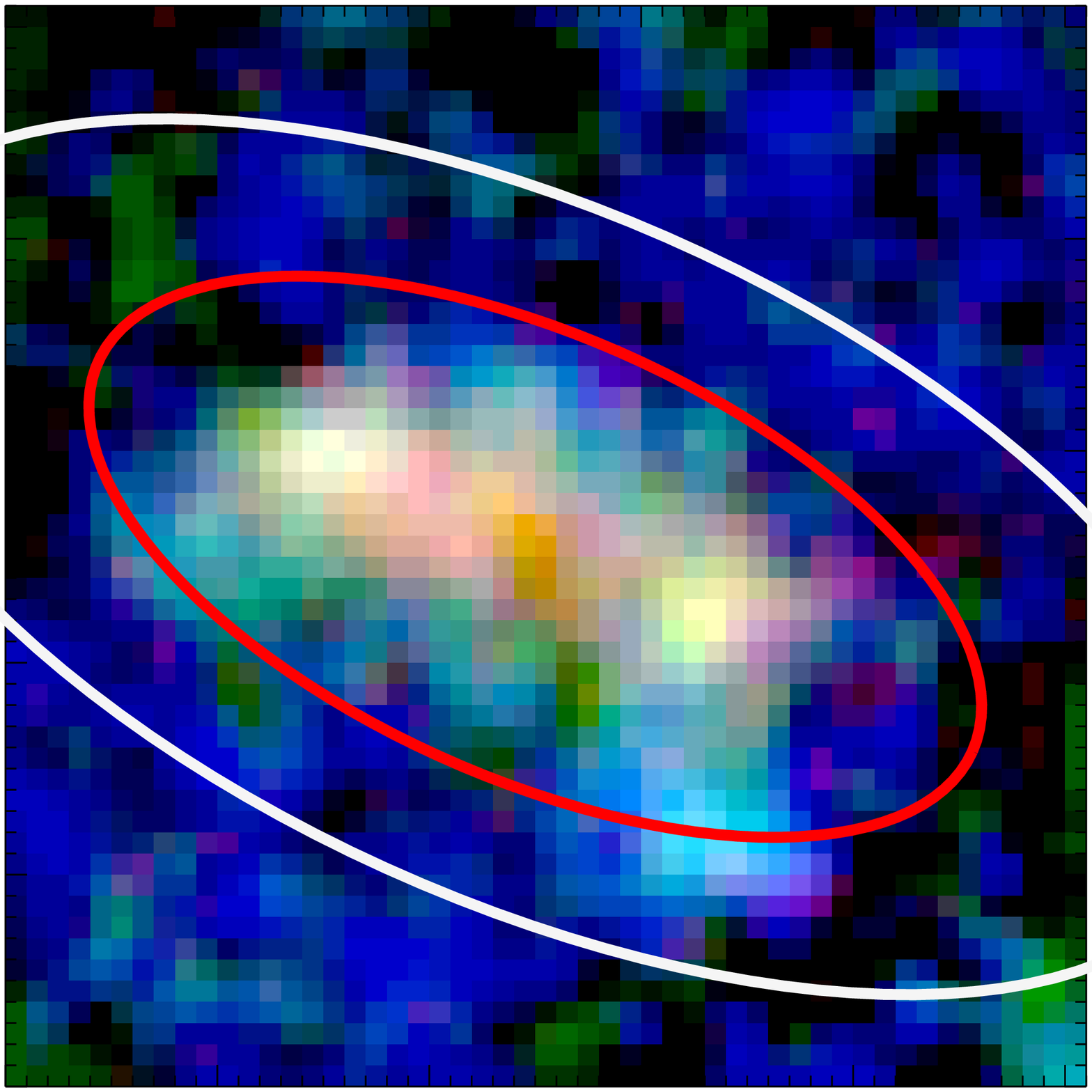} &
\includegraphics[width=1cm,height=1cm]{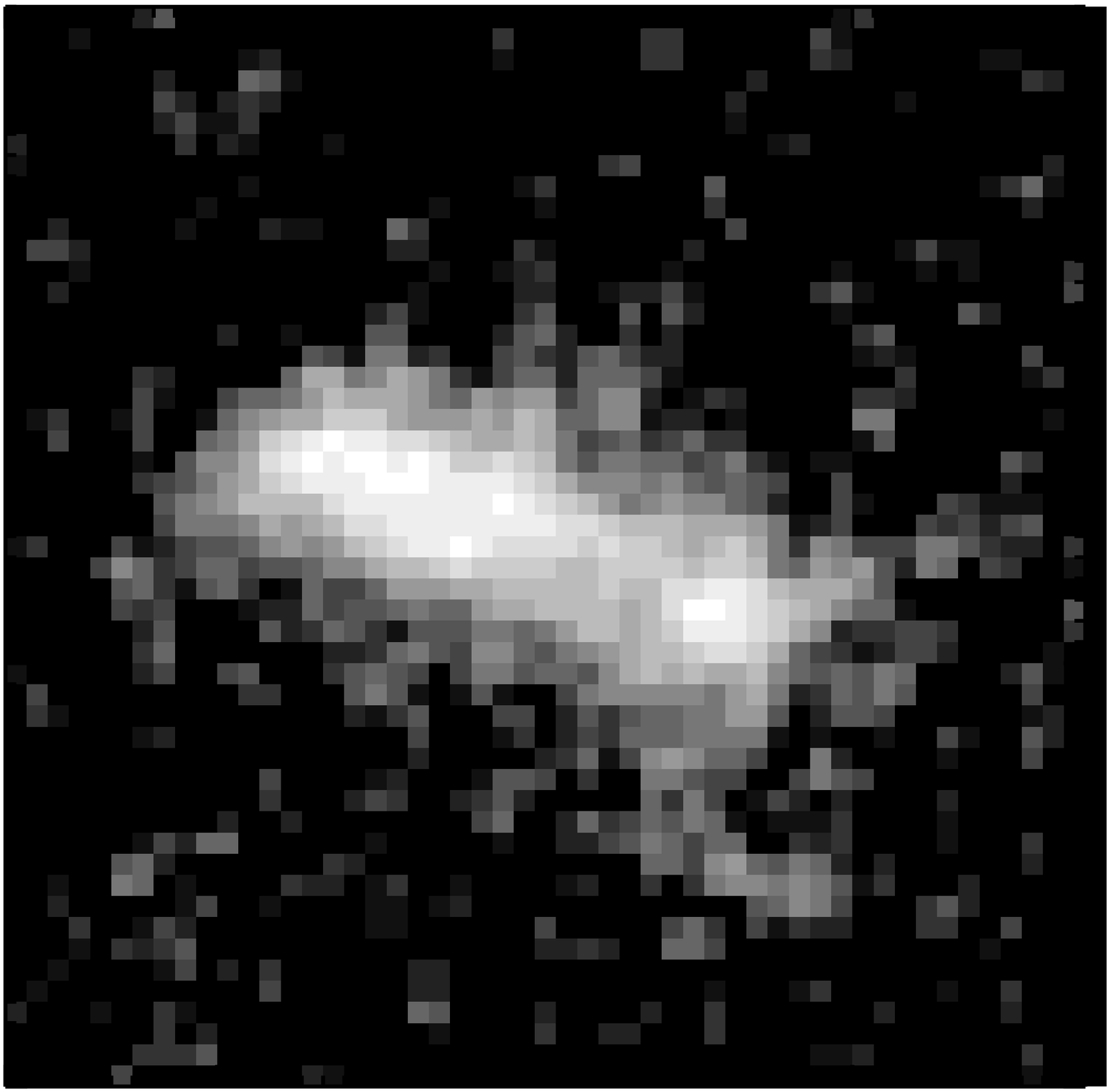} &
\includegraphics[width=1cm,height=1cm]{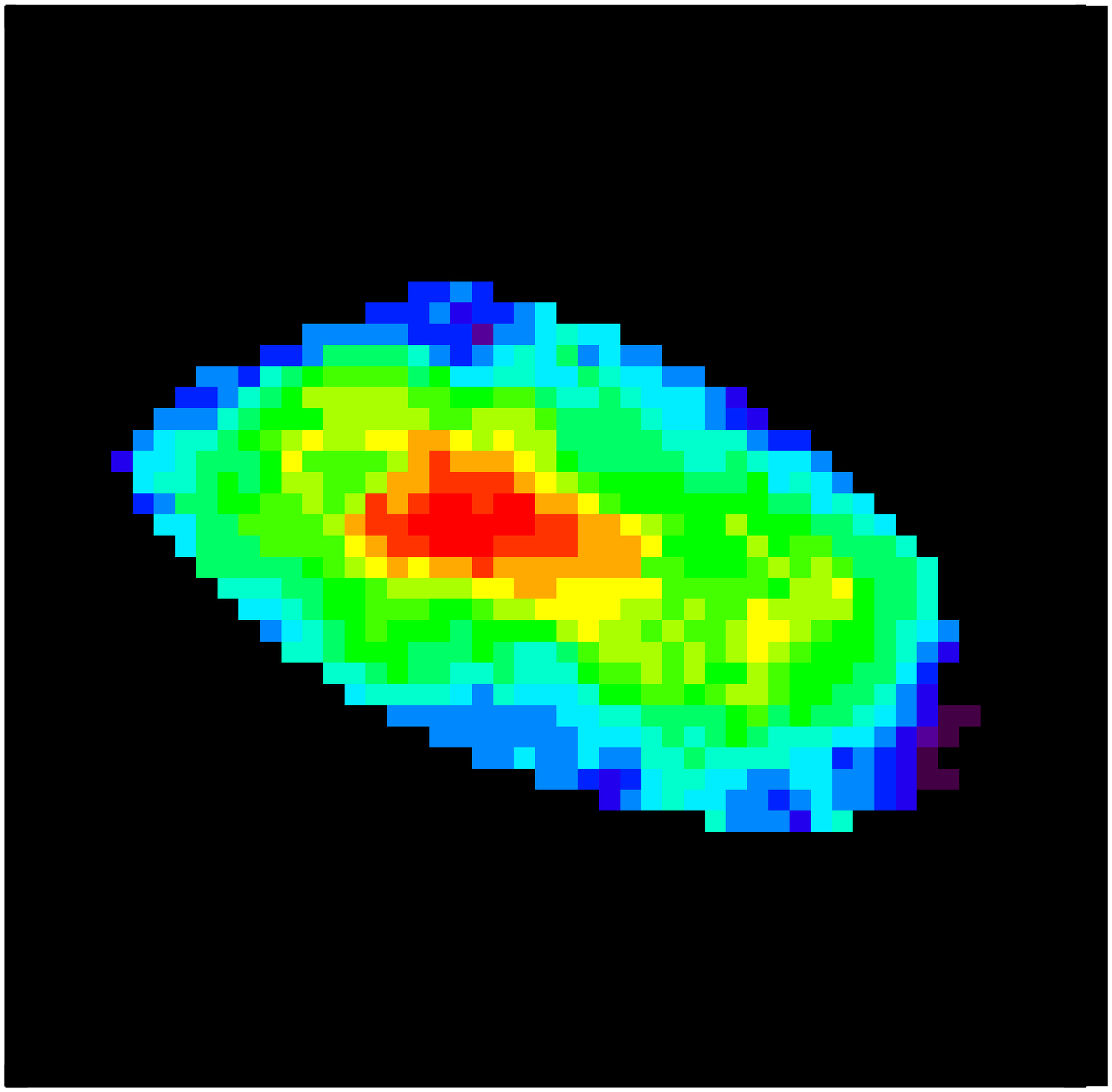} &
   53.14894 &  -27.81928 & 2.23  & -1.180 &  0.085 & 0.183 & 0.049 & Not merging  \\
\enddata
\tablecomments{
For all galaxies in the extended sample of Section \ref{sec:sampleSelection} we provide: (1) the CANDELS serial number from the \citet{Guo_et_al_2013} catalog; (2) the BzH composite image; (3) the H-band image; (4) the mass maps; (5)-(6) RA and DEC in J2000; (7) photometric or spectroscopic redshift (galaxies flagged with ``\emph{s}" have a spectroscopic redshift); (8)-(9) Asymmetry index $A_{\rm MASS}$ measured from the mass map and associated error; (10)-(11) $M_{\rm 20, MASS}$ index measured from the mass map and associated error; (12) classification according to the structural measurements performed on the mass maps: galaxies flagged as ``Merger (1)" are merger candidates selected with the criterium in Equation \ref{eq:eq1}, whereas galaxies flagged as ``Merger (2)" also satisfy the depth-dependent selection in Equation \ref{eq:eq2}. Galaxies flagged as Unres./Faint are below our limits of $r_{KRON}>5\times PSF$ and/or $H\leqslant24.5$ for a reliable mass estimate. All images are 3$^{\prime\prime}$ wide. Galaxies highlighted with a red exclamation mark are possible chance projections based on the analysis in Section \ref{sec:ProjectionEffects} and those with a ``*" symbol have photometry and mass maps strongly contaminated by neighbouring objects. The red and white ellipses in the RGB images show the Kron and Petrosian radius (used in the calculation of the structural indices from the mass maps), respectively. \\
\textbf{[The complete table can be found at: http://www.phys.susx.ac.uk/$\sim$ac625/Table1\_full.pdf]}}
\end{deluxetable*}

 \end{document}